\documentclass[10pt,journal,cspaper,compsoc]{IEEEtran}
\usepackage{amssymb}
\usepackage{bm}
\usepackage{amsmath}
\usepackage{graphicx}
\usepackage{url}
\usepackage{balance}
\usepackage{array}
\usepackage{subfigure}
\usepackage{multirow}

\newcommand{\comment}[1]{}

\newcommand {\bomega}{\mbox{\boldmath $\omega$}}
\newcommand {\bpi}{\mbox{\boldmath $\pi$}}

\newcommand{\squishlist}{
 \begin{list}{$\bullet$}
  { \setlength{\itemsep}{0pt}
     \setlength{\parsep}{3pt}
     \setlength{\topsep}{3pt}
     \setlength{\partopsep}{0pt}
     \setlength{\leftmargin}{1.5em}
     \setlength{\labelwidth}{1em}
     \setlength{\labelsep}{0.5em} } }

\newcommand{\squishend}{
  \end{list}  }

\begin{document}
%
\title{Sampling Content Distributed Over Graphs}
\author{Pinghui Wang, Junzhou Zhao,
        John C.S. Lui,~\IEEEmembership{Fellow,IEEE,}\\
        Don Towsley,~\IEEEmembership{Fellow,IEEE,}
        and~Xiaohong Guan,~\IEEEmembership{Fellow,IEEE,}
\IEEEcompsocitemizethanks{\IEEEcompsocthanksitem This work is done at The Chinese University of Hong Kong and Xi¡¯an Jiaotong University.

\IEEEcompsocthanksitem Pinghui Wang is with School
of Computer Science, McGill University, QC, Canada.
E-mail: phwang@sei.xjtu.edu.cn

\IEEEcompsocthanksitem John C.S. Lui is with the Department
of Computer Science and Engineering, The Chinese University of Hong Kong, Hong Kong.
E-mail: cslui@cse.cuhk.edu.hk

\IEEEcompsocthanksitem Don Towsley is with the Department
of Computer Science, University of Massachusetts Amherst, MA, USA.
E-mail: towsley@cs.umass.edu

\IEEEcompsocthanksitem Junzhou Zhao and Xiaohong Guan are with MOE Key Lab for Intelligent Networks and Network
Security, Xi¡¯an Jiaotong University, Xi¡¯an, Shaanxi, China.
E-mail: jzzhao@sei.xjtu.edu.cn, xhguan@xjtu.edu.cn.

\IEEEcompsocthanksitem Junzhou Zhao and Xiaohong Guan is also with the Center for Intelligent and Networked
Systems, Tsinghua National Lab for Information Science and Technology,
Tsinghua University, Beijing, China.}
\thanks{}}


\IEEEcompsoctitleabstractindextext{%
\begin{abstract}
Despite recent effort to estimate topology characteristics of
large graphs (i.e., online social networks and
peer-to-peer networks), little attention has been given to develop a
formal methodology to characterize the vast amount of content
distributed over these networks. Due to the
large scale nature of these networks,
exhaustive enumeration of this content
is computationally prohibitive.
In this paper, we show how one can
obtain content properties by sampling only a small fraction of vertices.
We first show that when sampling is naively applied,
this can produce a huge bias in content statistics
(i.e., average number of content duplications).
To remove this bias, one may use maximum likelihood
estimation to estimate content characteristics.
However our experimental results show that one needs to sample
most vertices in the graph to obtain accurate statistics using such a method.
To address this challenge, we propose two efficient estimators:
special copy estimator (SCE) and weighted copy estimator (WCE)
to measure content characteristics using available information
in sampled contents.
SCE uses the special content copy indicator to compute the estimate, while
WCE derives the estimate
based on meta-information in sampled vertices.
We perform experiments to show WCE and SCE are cost effective and
also ``{\em asymptotically unbiased}''.
Our methodology provides a new tool
for researchers to efficiently query content distributed in
large scale networks.
\end{abstract}

}

\maketitle

\IEEEdisplaynotcompsoctitleabstractindextext
\IEEEpeerreviewmaketitle

\section{Introduction} \label{sec:introduction}
\IEEEPARstart{N}{owadays} online social networks (OSNs) (i.e., Facebook and Twitter)
and P2P networks (i.e., BitTorrent) are two popular classes of Internet
applications.
Measuring content characteristics such as file duplication level
and information spreading rate on such networks become important since it
helps one develop effective
advertising strategies~\cite{SuhRetweetFactor2010},
and provides valuable information for designing content
delivery strategies, i.e., video sharing
techniques~\cite{LiSocialTube2012} to increase
video pre-fetch accuracy
by delivering videos based on users' social relationships and
interests, or to develop information seeding
techniques~\cite{MalandrinoProactiveSeeding2012} to minimize the
peak load of cellular networks by proactively pushing some
videos to OSN users.
Meanwhile measuring characteristics of OSNs' content provided by other networks also helps to us understand interactions
between different networks,
i.e.,~\cite{LiSocialTube2012} found that 80\% of videos in
Facebook come from other video service providers such as YouTube.

Due to the large sizes of these networks,
it is a challenge to measure content properties,
such as the distribution of tweets in an OSN by the number of
replies/retweets, or the distribution of videos in OSNs by external video
service providers.
To measure content properties, we formulate the problem as follows.
Define $L(\mathbf{c})$ as a generic labeling function
of content $\mathbf{c}$, with range
$\mbox{\boldmath $L$}=\{l_0,...,l_K\}$.
We present methods to estimate the content distribution $\bomega=(\omega_0,\ldots,\omega_{K})$, where
$\omega_k$ ($0\le k \le K$) is the fraction of content with label $l_k$.
For example, $L(\mathbf{c})$ can be defined as the number of comments
of post $\mathbf{c}$ in OSNs, and then $\omega_k$ is the fraction of posts with $k$
comments, where $\mbox{\boldmath $L$}=\{0, 1, 2, \ldots \}$.
In P2P networks, $L(\mathbf{c})$ can denote the number of replicas
of file $\mathbf{c}$; then $\omega_k$ is the fraction of files possessing $k$
replicas, where $\mbox{\boldmath $L$}=\{1, 2, \ldots \}$.
Similarly $L(\mathbf{c})$ can also be the file type of $c$ in P2P networks,
with $\mbox{\boldmath $L$}=\{l_0\!=\!``video", l_1\!=\!``musci",
l_2\!=\!``text", l_3\!=\!``others"\}$.
Then $\omega_k$ ($k=0,1,2,3$) is the fraction of files of type $l_k$.

\begin{figure}[htb]
\begin{center}
\includegraphics[width=0.3\textwidth]{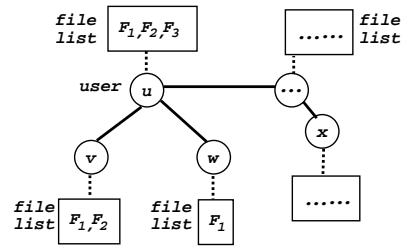}
\caption{An example of sampling files in a P2P network.}\label{fig:exampleIntro}
\end{center}
\end{figure}

Due to the size of these networks, the direct enumeration
is computationally prohibitive and one must consider
using sampling methods to estimate $\bomega$.
Unfortunately, previous graph sampling work developed for
estimating degree or workplace distributions~\cite{Gjoka2010,Ribeiro2010}
does not directly apply in our context.
This is because content and vertex are intrinsically different since
content may be {\em duplicated}.
To illustrate this, consider a simple example of a
P2P network as shown in Fig.~\ref{fig:exampleIntro}.
Assume file $F_1$ is cached by $10,000$ users,
$F_2$ is cached by two users $u$ and $v$,
and $F_3$ is cached by user $u$.
When sampling is applied,
clearly $F_1$ is more likely to be observed than
$F_2$ and $F_3$.
Therefore, estimation algorithms for topological metrics such as degree distribution cannot be blindly applied. They need to be modified to deal with biases introduced because of the nature of content characterization.

To the best of our knowledge, our work is the first analytical and
qualitative study on the problem of estimating characteristics of content distributed
over large graphs.
We propose methods to
accurately estimate the content distribution $\bomega$.
Our contributions are:\\
\noindent
$\bullet$ We show that when sampling is naively applied, there can be
huge bias in content statistics $\bomega$. One can remove this bias using the maximum likelihood
estimation. However our experimental
results show that one needs to sample most vertices in the graph
in order to obtain accurate statistics.

\noindent
$\bullet$ We present two efficient methods to estimate the content distribution $\bomega$ using available information in sampled content based on two different assumptions. The first method assumes that we can determine whether a collected content copy is a source or not, which is true for
most OSNs that have a label in each content copy to indicate whether it is a source or a duplicated copy.
%
For example, a tweet in microblog networks can be
classified as an original tweet or
a retweet. Therefore, one can utilize a
special content copy set consisting of all original tweets
to characterize distribution of tweets in the network.
To measure $\bomega$ of content in these networks,
we propose a special copy estimator (SCE)
based on collected source content copies.
We will show that SCE uses only a fraction of sampled content copies.
Moreover for networks such as P2P networks that cannot classify video copies as
original and not original, we propose another weighted
copy estimator (WCE). It assumes that each content copy records the
number of copies its content holds.
This feature is true for many OSNs~\cite{Renren,SinaMicroblog,Xiami} and
P2P networks~\cite{Yang04deploymentof}.
Our experiments show that WCE and SCE are
asymptotically unbiased, and WCE is much more accurate
estimator than SCE.

\noindent
$\bullet$ We also use WCE to estimate graph structure statistics for OSNs such as Sina microblog~\cite{SinaMicroblog} and Xiami~\cite{Xiami} where users maintain graph property summaries of their neighbors. For example,
by crawling the profile of a user in Sina microblog,
we can obtain its neighbors' properties such
as the number of followers, the number of following,
the number of tweets, etc.
This allows us to collect more information than previous graph sampling
methods under the same sampling cost. Since a user's graph property summary can
be viewed as content maintained by itself and its neighbors,
we apply WCE to estimate graph statistics.
Our experiments show that WCE can obtain the same level of accuracy of
graph properties with a {\em much less} sampling cost as
compared with previous works.

This paper is organized as follows.
In Section~\ref{sec:struturesampling} we summarize the most popular graph sampling techniques.
Section~\ref{sec:contentampling} presents several new methods for measuring characteristics of content in the graph.
The performance evaluation and testing results are presented in Section~\ref{sec:results}. Section~\ref{sec:application} presents real applications on Twitter and Sina microblog websites.
Related work is given in Section~\ref{sec:related},
and conclusion is given in Section~\ref{sec:conclusions}.

\section{Graph Sampling Methods} \label{sec:struturesampling}

In this section we present
some graph sampling methods that are underlying techniques for sampling content discussed in the later Section.
For ease of presentation, we assume the underlying graph
is undirected.
One way to convert a directed graph into an undirected graph
is by ignoring the direction of edges.
Consider an undirected graph $G=(V,E)$, where $V$ is the set of vertices and
$E$ is the set of undirected edges. Breadth-First-Search (BFS) is one
of most popular graph sampling techniques.
However it introduces a large bias
towards high-degree vertices that is unknown and difficult to
remove in general graphs~\cite{Achlioptas2005, Kurant2010}.
Therefore we do not consider BFS in this paper.
In what follows, we present popular graph sampling methods:
{\em Uniform Vertex Sampling} (UNI) and
{\em Random Walk} (RW), Metropolis-Hasting RW
(MHRW)~\cite{Stutzbach2009}, and Frontier Sampling (FS)~\cite{Ribeiro2010}.
Unless we state otherwise,
we denote $\bpi=(\pi_v\!:\! v\in V)$ as the probability distribution for the underlying sampling method, where $\pi_v$ is the probability that vertex $v$ is sampled at each sampling step.

\subsection {Uniform Vertex Sampling (UNI)}
UNI randomly samples vertices from the
vertex set $V$ uniformly and independently with replacement.
Not all network graphs support UNI but some do.
For example, one can view Wikipedia as a graph
and Wikipedia provides a query API to obtain a randomly sampled vertex
(wiki page) from its entire vertices.
Therefore, at each step, UNI samples each vertex $v$ with
the same probability, so we have
\[
\pi_v^{\text{UNI}}=\frac{1}{|V|}, \quad v\in V.
\]

For networks such as Facebook, MySpace, Flickr~\cite{Flickr},
Renren, Sina microblog, and Xiami, one can sample users (vertices)
as users have numeric IDs between the minimum and the maximum ID values.
Unfortunately, ID values of users in many networks
(e.g. Flickr, Facebook, Sina microblog, and MySpace)
are not sequentially assigned, and the ID space
is sparsely populated~\cite{RibeiroNetSci2010,Ribeiro2012}.
Hence, a randomly generated ID may not correspond to a valid user,
so considerable computational effort in generating a random ID will be wasted.
Therefore, UNI should only be applied to those graphs whose
user ID values are densely packed.

\subsection {Random Walk (RW)}
RW has been extensively studied in the graph theory
literature~\cite{Lovasz1993}. From an initial vertex, a walker selects a
neighbor at random as the next-hop vertex.
The walker moves to this neighbor and repeats the process.
Denote $\mathcal{N}(u)$ as the set of neighbors of any vertex $u$,
\mbox{$\text{deg}(u)=|\mathcal{N}(u)|$} is the degree of $u$.
Formally, RW can be viewed as a Markov chain with transition
matrix $P^{\text{RW}}=[P^{\text{RW}}_{u,v}]$, $u,v\in V$,
where $P^{\text{RW}}_{u,v}$ is defined as the probability of vertex $v$
being selected as the next-hop vertex given that its current vertex is $u$,
we have:
\begin{equation*}
P^{\text{RW}}_{u,v}=\left\{
\begin{array}{ll}
  \frac{1}{\text{deg}(u)} &\text{if } v\in \mathcal{N}(u), \\
  0 & \text{otherwise.}
\end{array}
\right.
\end{equation*}
The stationary distribution
$\bpi^{\text{RW}}$ of this Markov chain is
\[
\pi_v^{\text{RW}}=\frac{\text{deg}(v)}{2|E|}, \quad v\in V.
\]
For a connected and non-bipartite graph $G$,
the probability of being at a vertex $v\in V$ converges to
the above stationary distribution~\cite{Lovasz1993}.
Note that $\bpi^{\text{RW}}$ is biased toward vertices with high degree.
However, this bias can be
corrected~\cite{Heckathorn2002,Salganik2004}.

\subsection {Metropolis-Hastings Random Walk (MHRW)}
MHRW~\cite{Zhong2006,Stutzbach2009,Gjoka2010} provides another way to modifies RW using Metropolis-Hasting technique~\cite{Chib1995, Hastings1970, Metropolis1953}, which aims to collect vertices uniformly. To generate a sequence of random samples from a desired stationary distribution $\pi^{\text{MHRW}}$, the Metropolis-Hastings technique is a Markov chain Monte Carlo method based on modifying the transition matrix of RW as
\begin{equation*}
P_{u,v}^{\text{MHRW}}=\left\{
\begin{array}{ll}
  P^{\text{RW}}_{u,v}\min\left (\frac{\pi^{\text{MHRW}}_v P^{\text{RW}}_{v,u}}{\pi^{\text{MHRW}}_u P^{\text{RW}}_{u,v}},1\right) &\text{if } v\in \mathcal{N}(u),\\
  1-\sum_{w\neq u} P_{u,w}^{MHRW} &\text{if }v=u,\\
  0 & \text{otherwise}.
\end{array}
\right.
\end{equation*}
%
%
For a MHRW with target distribution $\pi^{\text{MHRW}}=\pi^{\text{UNI}}$, it works as follows:
At each step, MHRW selects a neighbor $v$ of current vertex $u$ at random and then accept the move randomly with probability $\min\left ( \frac{\text{deg}(u)}{\text{deg}(v)},1\right)$. Otherwise, MHRW still remains at $u$. Essentially, MHRW removes the bias of RW at each step by rejecting moves towards high degree vertices with a certain probability.

\subsection {Frontier Sampling (FS)}
FS~\cite{Ribeiro2010} is a centrally coordinated sampling which performs $T$ dependent RWs in graph $G$. Compared to a single RW, FS is less likely to get stuck in a loosely connected component of $G$.
Denote $\vec{L} = (v_1,\ldots, v_T)$ as the vector with $T$ vertices. Each $v_i$ ($1\le i\le T$) is initialized with a random vertex uniformly selected from $V$. At each step, FS selects a vertex $u\in \vec{L}$ with probability $\frac{\text{deg}(u)}{\sum_{\forall v\in \vec{L}} \text{deg}(v)}$, and uniformly selects a node $w$ from $\mathcal{N}(u)$, the neighbors of $u$. Thus $w$ is uniformly selected from the vertices connected to the vertices in $\vec{L} = (v_1,\ldots, v_T)$. Then FS replaces $u$ by $w$ in $\vec{L}$ and add $w$ to sequence of sampled vertices. If $G$ is a connected and non-bipartite graph, the probability that a vertex $v$ is sampled by FS converges to the following distribution
\[
\pi_v^{\text{FS}}=\frac{\text{deg}(v)}{2|E|}, \quad v\in V.
\]

\subsection{Estimator}
Previous work has considered how to estimate topology properties,
e.g., degree distribution, via sampling methods.
Define $L'(v)$ to be the vertex label of vertex $v$ under study, with range $\mbox{\boldmath $L'$}=\{l_0',...,l_{K'}'\}$.
Denote vertex label density $\mbox{\boldmath $\tau$}=(\tau_0,\ldots,\tau_{K'})$, where $\tau_k$ ($0\le k \le K'$) is the fraction of vertices with label $l_k'$.
For example, when $L'(v)$ is defined as the degree of vertex $v$, and $\mbox{\boldmath $\tau$}$ is the degree distribution.
To estimate $\mbox{\boldmath $\tau$}$,
the stationary distribution $\bpi$ is needed to correct the bias induced
by the underlying sampling method.
Since the values of $|V|$ and $|E|$ are usually unknown,
unbiasing the error is not straightforward. Instead,
one may use a non-normalized stationary
distribution $\hat\bpi=(\hat\pi_v: v\in V)$ to reweight sampled
vertices $s_i$ ($1\le i\le n$), where $\hat\pi_v$ is computed as
\begin{equation}\label{eq:knownscale}
\hat\pi_v=\left\{
\begin{array}{ll}
  1 &\text{for UNI and MHRW},\\
  \text{deg}(v) &\text{for RW and FS}.\\
\end{array}
\right.
\end{equation}
Let $\mathbf{1}(\mathbf{P})$ define the indicator function that equals one when predicate $\mathbf{P}$ is true, and zero otherwise.
Finally $\tau_k$ is estimated as follows
\[
\hat\tau_k=\frac{1}{S}\sum_{i=1}^n \frac{\mathbf{1}(L'(s_i)=l_k')}{\hat\pi_{s_i}}, \quad 0\le k \le K',
\]
where $S=\sum_{i=1}^n {\hat\pi}_{s_i}^{-1}$.

In summary, RW and FS are biased to sample vertices with high degree vertices. These biases can be later corrected, giving us smaller estimation errors for the characteristics of high degree vertices.
The accuracy of RW and MHRW is compared
in~\cite{Rasti2009,Gjoka2010}.
RW is shown to be consistently more accurate than MHRW.
Compared with RW and MHRW, FS requires UNI sampling for its initial settings, but is more accurate for sampling loosely connected and disconnect graphs~\cite{Ribeiro2010}.

\section{Content Sampling Methods} \label{sec:contentampling}

Denote by $\mathbf{C}=\{\mathbf{c}_1, \ldots, \mathbf{c}_H\}$
the set of all content under study, where $H$ is total number of
distinct contents in $G$. In this section, we study how to characterize
the content distribution
$\bomega=(\omega_0,\ldots,\omega_{K})$ defined in
Section~\ref{sec:introduction}, that is
\[
\omega_k=\frac{\sum_{\mathbf{c}\in \mathbf{C}} \mathbf{1}(L(\mathbf{c})=l_k)}{H}, \quad k=0,\ldots,K,
\]
where $L(\mathbf{c})$ is the label of content $\mathbf{c}$ with range $\{l_0,...,l_K\}$.
We let $c^{*}$ denote a special copy, such as the original source of content $\mathbf{c}$.
Note that each content $\mathbf{c}$ has one and only one special copy $c^{*}$.
For content $\mathbf{c} \!\in \! \mathbf{C}$,
let $\{c^{(1)},\ldots,c^{(f(\mathbf{c}))}\}$ denote
the set of its copies appearing in graph $G$ including its special copy $c^{*}$, where $f(\mathbf{c})$ is the number of copies of $\mathbf{c}$.
For a content copy $c$, let $v(c)$ be the vertex that maintains $c$.
Unless we state otherwise, in what follows the notation $\mathbf{c}$ is used to depict content and $c$ is used to depict a copy of content $\mathbf{c}$.
Meanwhile, we define $L(c)=L(\mathbf{c})$ and $f(c)=f(\mathbf{c})$.
Let $C_{\mathbf{S}}=\{c_1^{*}, \ldots, c_H^{*}\}$ be a
{\em special content copy set}.
For some graphs, sampling methods can check whether a
sampled content copy is special or not and generate such a set $C_{\mathbf{S}}$.
For example,
a tweet in the Sina microblog can be classified into an original tweet or
a retweet, therefore we can generate $C_{\mathbf{S}}$ consisting
of all original tweets.

We assume that $n$ vertices $s_i$ ($1\le i\le n$) are obtained by a graph sampling method that samples a vertex randomly from $V$ according to probability distribution $\bpi$ at each sampling step.
Denote by $C(s_i)$ the set of the content copies maintained by vertex $s_i$.
Denote by $\mathbf{C_D}$ the set of content that has at
least one copy maintained by sampled vertices $s_i$ ($1\le i\le n$).
In this section, we study how to characterize the content
distribution $\bomega$ based on $C(s_i)$, $1\le i\le n$.

We present four estimators:
(1) {\em distinct content estimator} ({\em DCE}),
(2) {\em maximum likelihood estimator} ({\em MLE}),
(3) {\em special copy estimator} ({\em SCE}), and
(4) {\em weighted copy estimator} ({\em WCE}).
DCE estimates $\bomega$ directly based on the collected content $\mathbf{C_D}$. Later we will show that content in $\mathbf{C_D}$
is not uniformly sampled from $\mathbf{C}$. Therefore estimates of $\bomega$ obtained by DCE are biased. MLE uses duplication level information of copies in $C(s_i)$ ($1\le i\le n$) to remove the bias of DCE.
However, we will show via experiment
that MLE needs to sample most vertices in graph $G$ in order to obtain
accurate statistics. SCE and WCE use meta
information in sampled content copies to remove sampling biases for estimating $\bomega$.
SCE estimates $\bomega$ based on collected content copies in $C_\mathbf{S}$, which assumes that we can determine whether a collected content copy is special or not.
WCE utilizes all collected content to estimate $\bomega$ based on the assumption that each copy of any content $\mathbf{c}$ records the value of $f(\mathbf{c})$.
A list of notations used is shown in Table~\ref{tab:notations}.
\begin{table}[htb]
	\centering
	\caption{Table of notations}
	\label{tab:notations}
	\begin{tabular}{||c|l||}
		\hline
$s_i, 1\le i\le n$&sampled vertices\\ \hline
$\bpi=(\pi_v: v\in V)$&vertex sampling probability distribution\\ \hline
$\mathbf{C}=\{\mathbf{c}_1, \ldots, \mathbf{c}_H\}$&set of all content appearing in graph $G$\\ \hline
$L(\mathbf{c})$&label of content $\mathbf{c}\in \mathbf{C}$\\ \hline
$\{l_0,...,l_K\}$&range of label function $L(\mathbf{c})$\\ \hline
\multirow{2}{*}{$\bomega=(\omega_0,\ldots,\omega_{K})$}&distribution of content by\\
&the content label\\ \hline
\multirow{2}{*}{$f(\mathbf{c})$}&number of copies that content $\mathbf{c}$\\
&possesses\\ \hline
$\{c^{(1)},\ldots,c^{(f(\mathbf{c}))}\}$&all copies of content $\mathbf{c}$\\ \hline
$c^{*}$&the special copy of content $\mathbf{c}$\\ \hline
$C_\mathbf{S}=\{c_1^{*}, \ldots, c_H^{*}\}$&special content copy set\\ \hline
$c$&a copy of content $\mathbf{c}$\\ \hline
$v(c)$ & vertex that maintains content copy $c$\\ \hline
$L(c)$, $f(c)$& $L(c)=L(\mathbf{c})$, $f(c)=f(\mathbf{c})$\\ \hline
$C(v),v\in V$&content copies maintained by vertex $v$\\ \hline
\multirow{2}{*}{$\mathbf{C_D}$}&set of content that has a copy\\
& maintained by sampled vertices $s_i$\\
		\hline
	\end{tabular}
\vspace{-0.5em}
\end{table}

\subsection{Distinct Content Estimator (DCE)}
DCE directly estimates $\bomega$ using all distinct collected content $\mathbf{C_D}$ as follows
\[
\hat\omega_k^{\text{DCE}}=\frac{1}{|\mathbf{C_D}|}\sum_{\mathbf{c}\in \mathbf{C_D}} \mathbf{1}(L(\mathbf{c})=l_k), \quad 0\le k \le K.
\]
Content $\mathbf{c}\in \mathbf{C}$ is maintained by vertices $v(c^{(j)})$ ($1\le j\le f(\mathbf{c})$) and vertex $v(c^{(j)})$ is sampled with probability $\pi_{v(c^{(j)})}$ at each sampling step,
therefore the probability that one copy of content $\mathbf{c}$ is collected by randomly sampling a vertex is
$\sum_{j=0}^{f(\mathbf{c})} \pi_{v(c^{(j)})}$.
Note that this probability depends both on the graph sampling method and
the number of copies of $\mathbf{c}$, therefore content in $\mathbf{C_D}$
is {\em not} uniformly sampled from $\mathbf{C}$.
Even when the UNI sampling method is used,
where each vertex $u$ is sampled with the
same probability $\pi_u=\frac{1}{|V|}$,
the probability that $\mathbf{C_D}$ contains $\mathbf{c}$ is
proportional to $\frac{f(\mathbf{c})}{|V|}$.
This clearly shows that $\hat\omega_k^{\text{DCE}}$ using UNI is still biased.

\subsection{Maximum Likelihood Estimator (MLE)}
In what follows, we use the maximum likelihood estimation method
to remove the bias of DCE.
Due to page limit, We only present the MLE for
graph sampling method UNI with $\pi_v=\frac{1}{|V|}$, $v\in V$.
Suppose that the graph size is known
(this can be estimated by sampling methods proposed in~\cite{Katzir2011}),
$n < |V|$ vertices are sampled, and then each copy of $\mathbf{c}$ is sampled with the same probability
$p=\frac{n}{|V|}$. For simplicity, we assume that content are distributed over networks uniformly at random.
Let $M$ be the maximum number of copies that content has.
Denote $P_{i,j}$ as the probability that $i$ copies are sampled for
content which has $j$ copies,
where $ 1 \le i \le j \le M$.
Let $q=1-p$, we have
$P_{i,j}=\frac{\binom{j}{i} p^i q^{j-i}}{1 - q^j}$.

When the content label under study is the number of copies associated
with content. For randomly sampled content, let
$\alpha_i$ ($1\le i\le M$) be the probability that it has $i$ copies sampled.
Note that $\alpha_i$ can be estimated based on collected content copies.
In what follows, we propose a method to
estimate $\bomega$ based on the relationship of $\alpha_i$ and $\bomega$. the likelihood function of
$\alpha_i$ is
\begin{equation}\label{eq:alphai}
\mathcal{L}(\alpha_{i}|\bomega)
=\sum_{j=i}^{M} \omega_j P_{i,j}.
\end{equation}
This is similar to packet sampling based flow size distribution estimation studied in~\cite{Duffield2003}, where each packet is sampled with probability $p$. Here a flow refers to a group of packets with the same source and destination, and the flow size is the number of packets that it contains. In our context content corresponds to a flow, and its copies to packets in the flow.
Therefore we can develop a maximum likelihood estimate $\hat\omega_k^{\text{MLE}}$ of $\omega_k$ ($1\le k\le M$) similar to the method proposed in~\cite{Duffield2003}.

When the content label under study is insensitive to the
number of duplicates.
We use the following approach to derive the MLE.
Meanwhile it is available in each content copy, which is not a latent property such as the number of copies content has.
Define $\beta_{k,j}$ ($0\le k\le K$, $1\le j\le M$) as the fraction of the number of content with label $l_k$ and $j$ copies over the number of  content with label $l_k$.
For randomly sampled content, let
$\alpha_{k,i}$ ($1\le i\le M$) be the probability that its content label is $l_k$ and has $i$ copies sampled.
Then the likelihood function of $\alpha_{k,i}$ is
\[
\mathcal{L}(\alpha_{k,i}|\bomega)=\sum_{j=i}^M \beta_{k,j} P_{i,j}.
\]
$\alpha_{k,i}$ can be estimated based on collected content copies.
Then similar to~(\ref{eq:alphai}), we can develop a maximum likelihood estimate $\hat\beta_{k,j}$ of $\beta_{k,j}$, $1\le j\le M$.
Since
\[
\alpha_k=\omega_k \sum_{i=1}^M \sum_{j=i}^M \beta_{k,j} P_{i,j},
\]
Then we have the following estimator of $\omega_k$
\[
\hat\omega_k^{\text{MLE}}=\frac{\hat\alpha_k} {S^{\text{MLE}}\sum_{i=1}^M \sum_{j=i}^M \hat\beta_{k,j} P_{i,j}}, \quad 0\le k\le K,
\]
where $\hat\alpha_k$ is the fraction of sampled content with
label $k$, and
$S^{\text{MLE}}=\sum_{k=0}^K \frac{\hat\alpha_k}
     {\sum_{i=1}^M \sum_{j=i}^M \hat\beta_{k,j} P_{i,j}}$.
In a later section, we will show that to calculate $\hat\omega_k^{\text{MLE}}$,
one has to sample a large number of vertices in $G$. It is consistent with results observed in~\cite{Murai2012}.

\subsection{Special Copy Estimator (SCE)}
SCE estimates $\bomega$ only using collected special content copies, which are content copies in set $C_\mathbf{S}$.
For content $\mathbf{c}$, $v(c^{*})$ is the vertex maintained
its special copy $c^{*}$.
Then the probability that $c^{*}$ is collected by sampling a random vertex is
$\pi_{v(c^{*})}$. Similar to the estimator given in
Section~\ref{sec:struturesampling}, we use
$\hat\bpi$ defined in Eq.~(\ref{eq:knownscale}) to estimate
$\omega_k$ ($0\le k \le K$),
\begin{equation}
\hat\omega_k^{\text{SCE}}=\frac{1}{S^{\text{SCE}}}\sum_{i=1}^n \sum_{c\in C(s_i)} \frac{\mathbf{1}(L(c)=l_k) \mathbf{1}(c \in C_\mathbf{S})}
{\hat\pi_{s_i}},
\end{equation}
where $S^{\text{SCE}}=\sum_{i=1}^n \sum_{c\in C(s_i)}
\frac{\mathbf{1}(c \in C_\mathbf{S})}{\hat\pi_{s_i}}$.
It is important to point out that
$\hat\omega_k^{\text{SCE}}$ is an {\em asymptotically unbiased estimator} of
$\omega_k$.
For each vertex $v\in V$, Eq.~(\ref{eq:knownscale}) shows that $\pi_v/\hat\pi_v$ has the same value, denoted as $S_{\pi}$.
We have the following equation for each $k=0,\ldots, K$ and $i=1, \ldots, n$
\begin{equation*}
\begin{split}
&\text{E}\left[\sum_{c\in C(s_i)} \frac{\mathbf{1}(L(c)=l_k) \mathbf{1}(c \in C_\mathbf{S})}{\hat\pi_{s_i}}\right]\\
&= \sum_{v\in V}\pi_v \sum_{c\in C(v)} \frac{\mathbf{1}(L(c)=l_k) \mathbf{1}(c \in C_\mathbf{S})}{\hat\pi_v}\\
&= S_{\pi}\sum_{v\in V}\sum_{c\in C(v)}\mathbf{1}(L(c)=l_k) \mathbf{1}(c \in C_\mathbf{S})\\
&= S_{\pi}\sum_{c\in C_\mathbf{S}}\mathbf{1}(L(c)=l_k)
= S_{\pi} H\omega_k.
\end{split}
\end{equation*}
Applying the law of large numbers, we have
\begin{equation*}
\lim_{n\rightarrow \infty}\frac{1}{n}
  \sum_{i=1}^n \sum_{c\in C(s_i)} \!\!\!
\frac{\mathbf{1}(L(c)=l_k) \mathbf{1}(c \in C_\mathbf{S})}{\hat\pi_{s_i}}
\,\,\xrightarrow{a.s.} \,\, S_{\pi}H\omega_k,
\end{equation*}
where ``$a.s.$" denotes ``almost sure" converge, i.e., the event happens with probability one.
Similarly, we have $\lim_{n\rightarrow \infty} \frac{S^{\text{SCE}}}{n} \xrightarrow{a.s.} S_{\pi}H$. Therefore $\hat\omega_k^{\text{SCE}}$ is an
asymptotically unbiased estimator of $\omega_k$.

\subsection{Weighted Copy Estimator (WCE)}
WCE estimates $\bomega$ using all collected content copies $C(s_i)$ ($1\le i\le n$).
This estimator is useful for networks (i.e., Sina microblog
or Renren) in which
each copy of any content $\mathbf{c}$ records the value of $f(\mathbf{c})$,
the number of copies
$\mathbf{c}$ has in the network.
For content $\mathbf{c}$, vertex $v(c^{(j)})$ maintains the
copy $c^{(j)}$ ($1\le j\le f(\mathbf{c})$),
and the vertex is sampled with probability $\pi_{v(c^{(j)})}$.
Meanwhile a random vertex maintains a copy of $\mathbf{c}$ with probability proportional to $f(\mathbf{c})$. Therefore we assign a weight $\frac{1}{\hat\pi_{v(c^{(j)})} f(\mathbf{c})}$ for $c^{(j)}$ to remove the sampling bias.
Finally $\omega_k$ ($0\le k \le K$) is estimated as follows
\begin{equation}
\hat\omega_k^{\text{WCE}}=\frac{1}{S^{\text{WCE}}}\sum_{i=1}^n \sum_{c\in C(s_i)}\frac{\mathbf{1}(L(c)=l_k)}{\hat\pi_{s_i} f(c)},
\label{eq:estimatorwce}
\end{equation}
where $S^{\text{WCE}}=\sum_{i=1}^n
\sum_{c\in C(s_i)}\frac{1}{\hat\pi_{s_i} f(c)}$.
Note that $\hat\omega_k^{\text{WCE}}$ is an
{\em asymptotically unbiased estimator} of $\omega_k$. To see that,
we have the following equation for each $k=0,\ldots, K$ and $i=1, \ldots, n$
\begin{equation*}
\begin{split}
&\text{E}\left[\sum_{c\in C(s_i)}\frac{\mathbf{1}(L(c)=l_k)}{\hat\pi_{s_i} f(c)}\right]\\
&=\sum_{v\in V}\pi_v \sum_{c\in C(v)}\frac{\mathbf{1}(L(c)=l_k)}{\hat\pi_v f(c)}\\
&= S_{\pi}\sum_{v\in V}\sum_{c\in C(v)}\frac{\mathbf{1}(L(c)=l_k)}{f(c)}\\
&= S_{\pi}\sum_{\mathbf{c}\in \mathbf{C}} \sum_{j=1}^{f(\mathbf{c})} \frac{\mathbf{1}(L(c^{(j)})=l_k)}{f(c^{(j)})}\\
&= S_{\pi}\sum_{\mathbf{c}\in \mathbf{C}} \mathbf{1}(L(\mathbf{c})=l_k)
= S_{\pi}H\omega_k.
\end{split}
\end{equation*}
Then we have
\[
\lim_{n\rightarrow \infty}\frac{1}{n}\sum_{i=1}^n \sum_{c\in C(s_i)}\frac{\mathbf{1}(L(c)=l_k)}{\hat\pi_{s_i} f(c)}\xrightarrow{a.s.} S_{\pi}H\omega_k.
\]
Similarly, we have $\lim_{n\rightarrow \infty}
\frac{S^{\text{WCE}}}{n} \xrightarrow{a.s.} S_{\pi}H$.
Therefore $\hat\omega_k^{\text{WCE}}$ is an
asymptotically unbiased estimator of $\omega_k$.

\begin{figure}[htb]
\begin{center}
\includegraphics[width=0.45\textwidth]{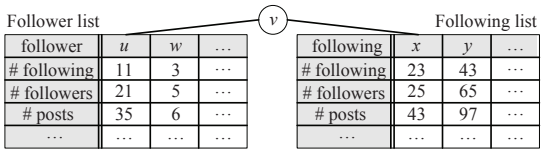}
\caption{Graph properties maintained by vertex $v$.}\label{fig:graphexample}
\end{center}
\end{figure}

We also note that,
compared with previous sampling methods\cite{Ribeiro2010,Gjoka2010},
WCE is a more {\em cost effective method} to estimate graph structure statistics
for OSNs (i.e., Sina Microblog and Xiami) which carry such meta information.
As shown in Fig.~\ref{fig:graphexample}, the webpage
of a user in Sina Microblog
maintains a summary for each of its neighbors
(both followers and following), which includes graph properties
such as the number of followers,
the number of following, and the number of posts.
Hence, one can obtain properties of any vertex $v$
{\em and} all its neighbors by simply sampling $v$.
So compared with previous works for measuring structure characteristics,
we can obtain more accurate estimates by utilizing
the meta information of sampled vertices. 
It is important to point that when we use this meta information,
we are biased toward vertices with a large number of neighbors
(even when using UNI).  Therefore, we need a way to unbias this error.

Denote $\text{outdeg}(v)$ as the number of vertices that vertex $v$ follows, and by $\text{indeg}(v)$ is the number of vertices that follow $v$.
To remove the sampling bias for observing high degree vertices' graph properties, we use WCE to estimate vertex label density $\mbox{\boldmath $\tau$}=(\tau_0,\ldots,\tau_{K'})$ defined in Section~\ref{sec:struturesampling}, where $\tau_k$ ($0\le k \le K'$) is the fraction of vertices with vertex label $l_k'$.
The property summary of each vertex $v$ can be viewed as content with
$f'(v)=\text{indeg}(v)+\text{outdeg}(v)+1$ copies maintained by
the followers, following of $v$ and $v$ itself.
For a collected vertex $v$, define its associated vertices $C'(v)$ as the collection of its following, its followers, and itself. Note that $C'(v)$ might contain duplicate elements since a vertex can be both the following and follower of $v$.
We use WCE to estimate $\tau_k$ ($0\le k \le K'$) as follows
\begin{equation}
\hat\tau_k^{\text{WCE}}=\frac{1}{S'}\sum_{i=1}^n \sum_{v\in C'(s_i)}\frac{\mathbf{1}(L'(v)=l_k')}{\hat\pi_{s_i} f'(v)},
\end{equation}
where $S'\!\!=\!\!\sum_{i=1}^n \sum_{v\in C'(s_i)}\frac{1}{\hat\pi_{s_i} f'(v)}$.
Note that $\hat\tau_k^{\text{WCE}}$ is an
asymptotically unbiased estimator of $\tau_k$.
 To see that,
we have the following equation for each $k=0,\ldots, K'$ and $i=1, \ldots, n$
\begin{equation*}
\begin{split}
&\text{E}\left[\sum_{v\in C'(s_i)}\frac{\mathbf{1}(L'(v)=l'_k)}{\hat\pi_{s_i} f'(v)}\right]\\
&=\sum_{u\in V}\pi_u \sum_{v\in C'(u)}\frac{\mathbf{1}(L'(v)=l'_k)}{\hat\pi_v f'(v)}\\
&= S_{\pi}\sum_{u\in V}\sum_{v\in C'(u)}\frac{\mathbf{1}(L'(v)=l'_k)}{f'(v)}\\
&= S_{\pi}\sum_{u\in V} \mathbf{1}(L'(v)=l'_k)
= S_{\pi}|V|\tau_k.
\end{split}
\end{equation*}
Then we have
\[
\lim_{n\rightarrow \infty}\frac{1}{n}\sum_{i=1}^n \sum_{v\in C'(s_i)}\frac{\mathbf{1}(L'(v)=l'_k)}{\hat\pi_{s_i} f'(v)}\xrightarrow{a.s.} S_{\pi}|V|\tau_k.
\]
Similarly, we have $\lim_{n\rightarrow \infty}
\frac{S'}{n} \xrightarrow{a.s.} S_{\pi}|V|$.
Therefore $\hat\tau_k^{\text{WCE}}$ is an
asymptotically unbiased estimator of $\tau_k$.

\section{Data Evaluation} \label{sec:results}

Our experiments are performed on a variety of real world networks,
which are summarized in Table~\ref{tab:datasets}. Xiami is a popular website
devoted to music streaming and music recommendations.
Similar to Twitter, Xiami
builds a social network based on follower and following relationships.
Each user has a numeric ID that is sequentially assigned. We crawled
its entire network graph and have made
the dataset publicly available\footnote{http://www.cse.cuhk.edu.hk/\%7ecslui/data}.
Flickr and YouTube are popular photo sharing and video sharing websites.
In these websites, a user can subscribe to other user updates such as blogs and photos.
These networks can be represented by a direct graph, with
vertices representing users and a directed edge from $u$ to $v$
represents that user $u$ subscribes to user $v$.
Further details of these datasets can be found in~\cite{MisloveIMC2007}.

\begin{table}[htb]
\begin{center}
\caption{Overview of directed graph datasets used in our simulations.\label{tab:datasets}}
\begin{tabular}{||c|ccc||}
\hline
{\bf Graph}&{\bf Xiami}&{\bf YouTube}&{\bf Flickr}\\
\hline \hline
vertices&1,753,690&1,138,499&1,715,255\\
edges &16,019,106&2,990,443&15,555,041\\
directed-edges &16,574,010&4,945,382&22,613,981\\
vertices (LCC)&1,748,010&1,134,890&1,624,992\\
edges (LCC)&16,015,779&2,987,624&15,476,835\\
directed-edges (LCC)&16,568,449&4,942,035&22,477,015\\
\hline
\end{tabular}
\end{center}
``directed-edges" refers to the number of directed edges in a
directed graph, ``edges" refers to the number of edges in an
undirected graph, and ``LCC" refers to the largest connected component
of a given graph.
\vspace{-0.5em}
\end{table}

Using real graph topologies which are publicly available, we generate
benchmark datasets for our simulation experiments by manually generating
content and distributing them over these graphs.
In the following experiments we generate $10^7$ distinct content and distribute each content $\mathbf{c}$ using four different content
distribution schemes (CDSs): CDS I, CDS II, CDS III, and CDS IV, which model information distribution mechanisms for undirected and directed graphs.

CDS I and II distribute content with a target content distribution by the number of copies.
Define the truncated Pareto distribution as $\phi_k=\frac{\alpha}{\gamma k^{\alpha+1}}$, $k=1,\ldots,W$, where $\alpha > 0$ and $\gamma=\sum_{k=1}^W \frac{\alpha}{k^{\alpha+1}}$ and $W$ is the maximum number of duplication. The number of copies for each content $\mathbf{c}$ is randomly
selected from set $\{1,\ldots,W\}$ according to the truncated Pareto distribution with parameter $\alpha$ and $W$ for CDS I and II. Then copies of $\mathbf{c}$ are distributed as follows

\noindent
$\bullet$ {\bf CDS I:} We distribute each content copy to a
randomly selected vertex in $G_d$ (one of the directed graph
in Table~\ref{tab:datasets}).

\noindent
$\bullet$ {\bf  CDS II:} When content has $k$ copies,
we first randomly select a vertex $v$ that can reach at least $k-1$
other vertices. In here, two vertices are reachable if there is at least one
path between them in the undirected graph $G$, which is derived from $G_d$
by ignoring the direction of edges.  Then we assign the special (or original)
copy of this content to $v$, and assign $k-1$ duplicated
copies to the top $k-1$ nearest vertices in $G$ which are reachable from $v$.

CDS III and CDS IV distribute each content $\mathbf{c}$ using the
independent cascade model~\cite{Goldenberg2001}, that is

\noindent
$\bullet$ {\bf CDS III:} We distribute $\mathbf{c}$ over the
associated undirected graph $G$.
We first distribute the special copy of $\mathbf{c}$ to a randomly
selected vertex $v$.
Then we distribute copies of
$\mathbf{c}$ to other vertices iteratively.
When a new vertex first receives a copy of $\mathbf{c}$, it is given a single chance to
distribute a copy of $\mathbf{c}$ to each of its neighbors currently without $\mathbf{c}$ with
probability $p_S$.

\noindent
$\bullet$ {\bf CDS IV:}
We distribute $\mathbf{c}$ similar to CDS III but on the
{\em direct graph} $G_d$.
The difference is that when a new vertex first receives a copy of $\mathbf{c}$,
it is given a single chance to distribute a copy of $\mathbf{c}$ to each of its incoming neighbors (followers) currently without $\mathbf{c}$ with probability $p_S$.

Here we assume that each copy of content $\mathbf{c}$ records the number of copies possessed by $\mathbf{c}$ finally.
In the following experiments we evaluate the performance of our methods for estimating $\bomega$, content distribution by the number of copies.
Let
\begin{equation*}
NMSE(\hat{\omega}_j)=\frac{\sqrt{\text{E}[(\hat{\omega}_j-\omega_j)^2]}}{\omega_j} \, , j=1,2,\dots \, ,
\end{equation*}
be a metric that measures the relative error of the estimate $\hat{\omega}_j$
with respect to its true value $\omega_j$.
In our experiment, we average
the estimates and calculate their NMSE over 1,000 runs.
Let $B$ denote the sampling budget, which is the number of distinct sampled vertices per run.
In the following experiments, we set default parameters as follows: sampling budget $B=0.01|V|$,
uniform vertex sampling cost $c=1$, the number of random walkers $T=1000$ for FS.
RW, MHRW and FS are evaluated on the LCC of graphs. UNI and RWJ are evaluated on the entire graphs.
To simplify notation, graph sampling method A combined with content estimator B
is denoted as method A\_B.

\begin{figure}[htb]
\center
\subfigure[$W=20$]{
\includegraphics[width=0.4\textwidth]{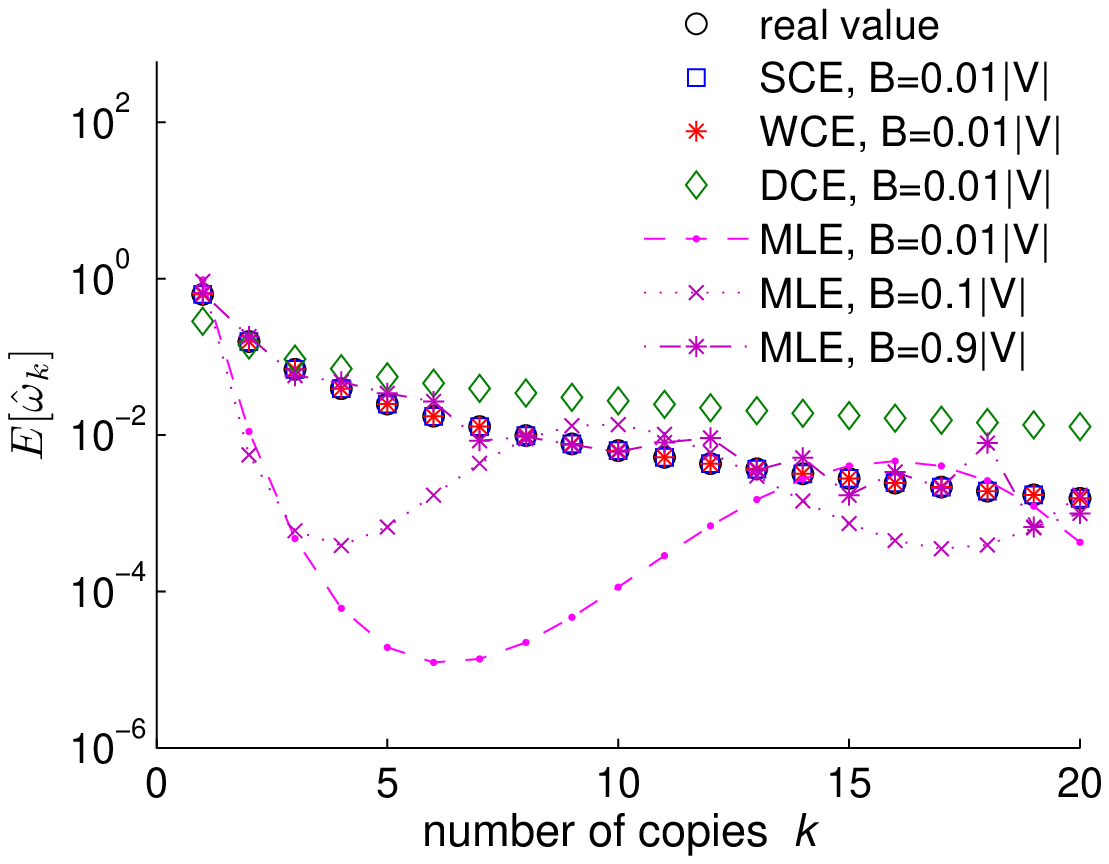}}
\subfigure[$W=100$]{
\includegraphics[width=0.4\textwidth]{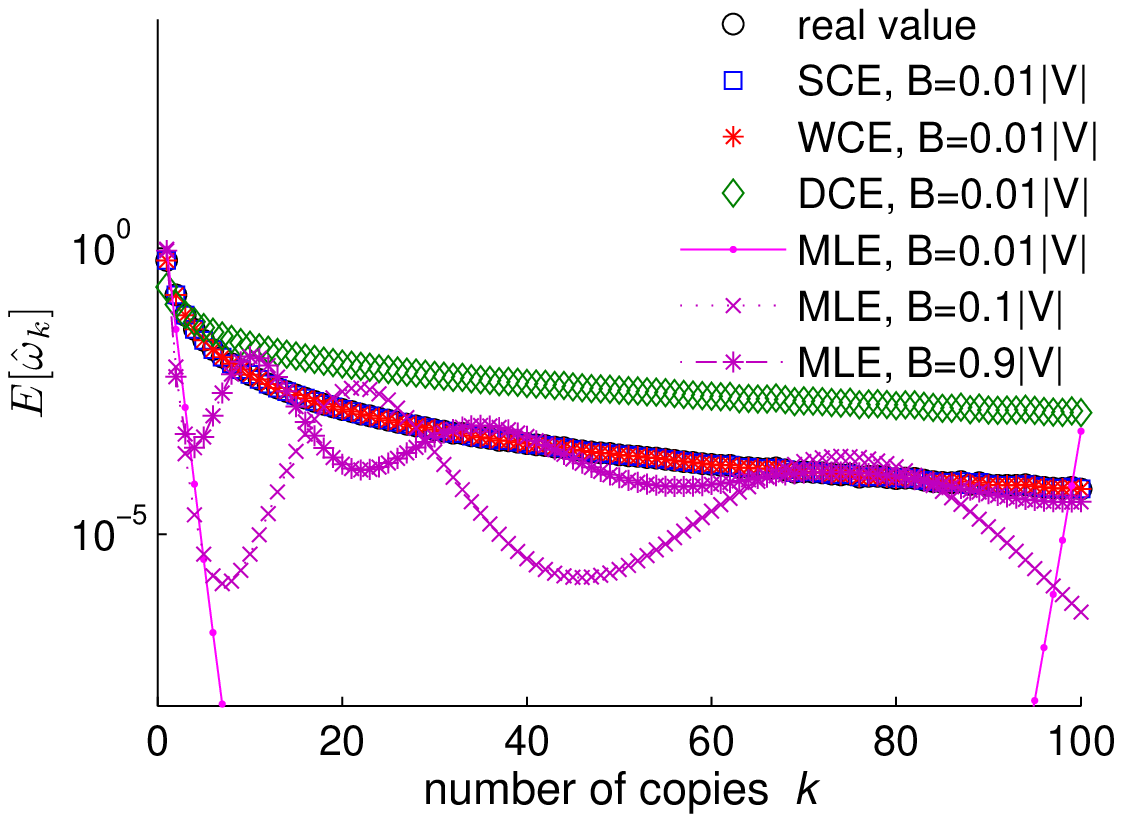}}
\caption{(Xiami) Average of content distribution estimates for different estimators.}\label{fig:MLEcmp}
\end{figure}

Fig~\ref{fig:MLEcmp} shows the average of content distribution estimates of 1,000 runs
for methods DCE, MLE, SCE and WCE.
where the graph sampling method is UNI, content distribution scheme is CDS I,
with $\alpha=1$, $W=\{20,50\}$.
We observe that DCE is {\em highly biased}, while SCE and WCE are unbiased.
MLE needs to sample most vertices to reduce biases especially for large $W$.
Note that it SCE and WCE practically coincide with the correct values.

\begin{figure*}[htb]
\center
\subfigure[CDS I]{
\includegraphics[width=0.23\textwidth]{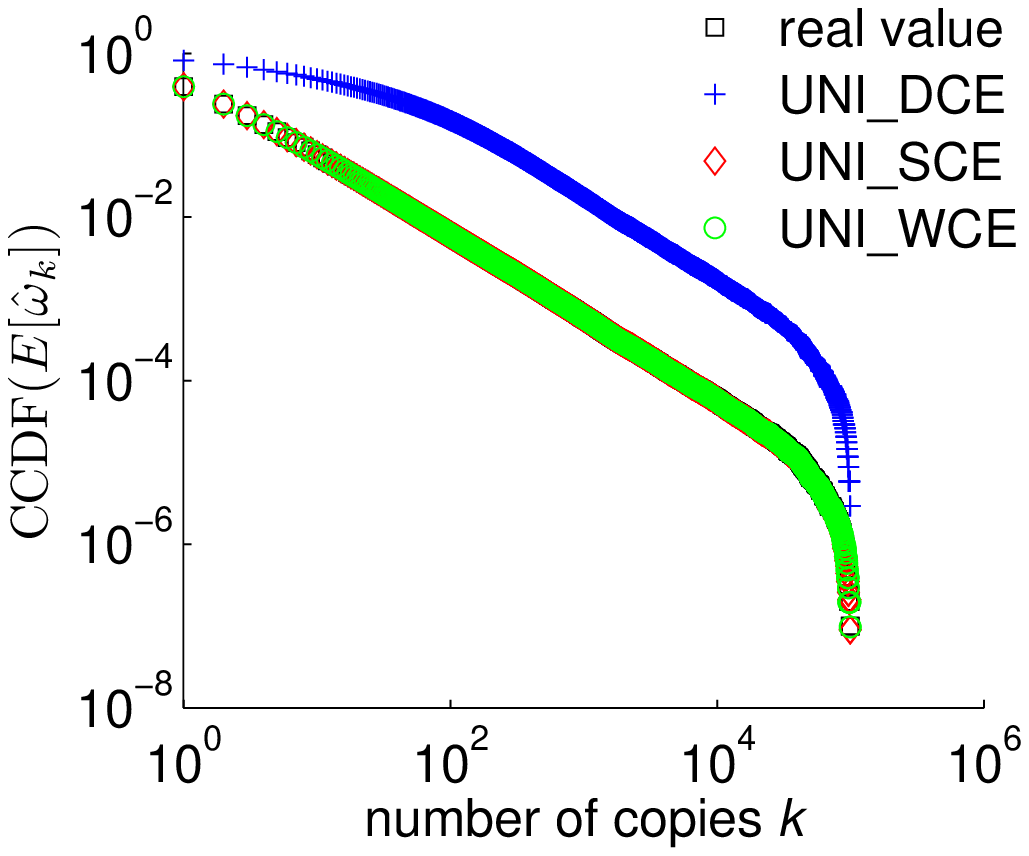}}
\subfigure[CDS II]{
\includegraphics[width=0.23\textwidth]{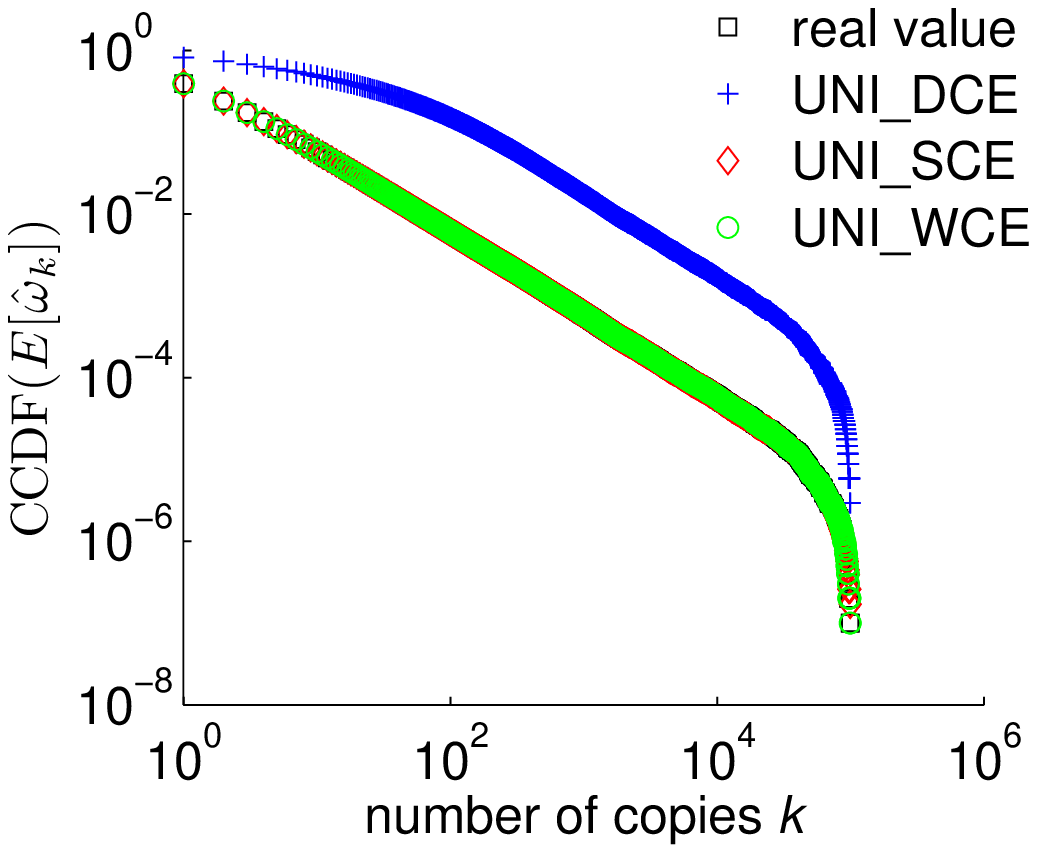}}
\subfigure[CDS III]{
\includegraphics[width=0.23\textwidth]{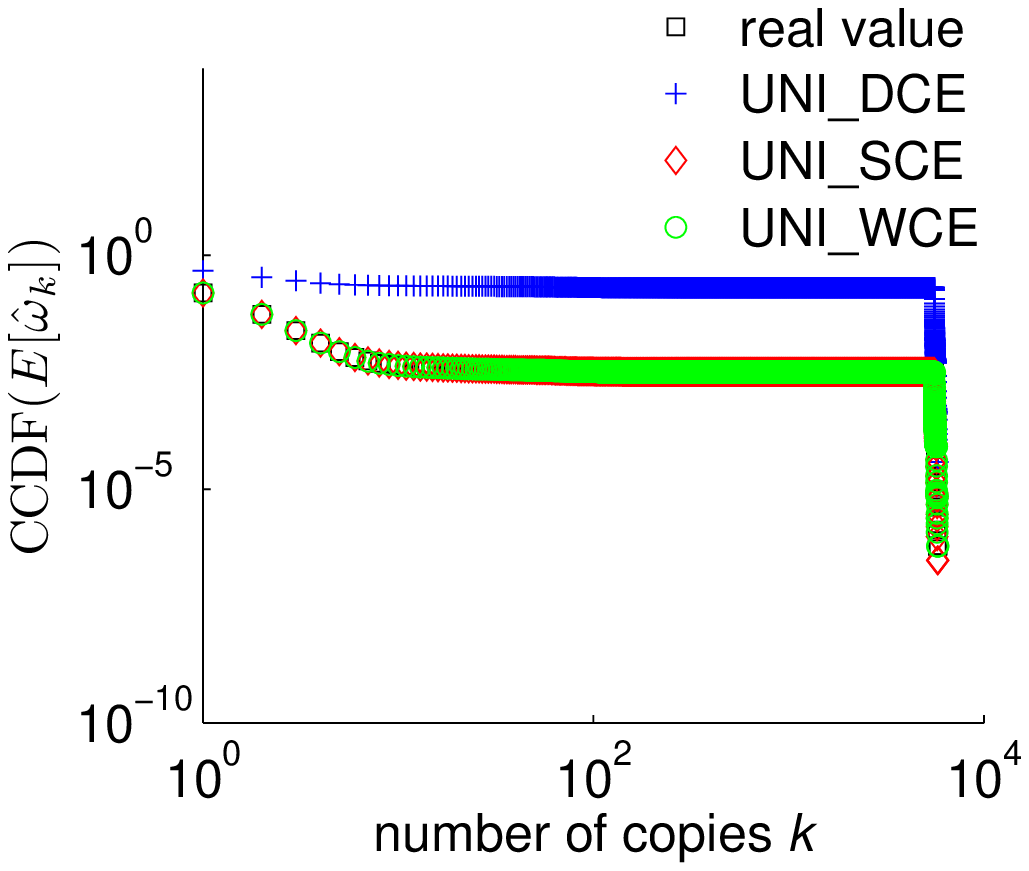}}
\subfigure[CDS IV]{
\includegraphics[width=0.23\textwidth]{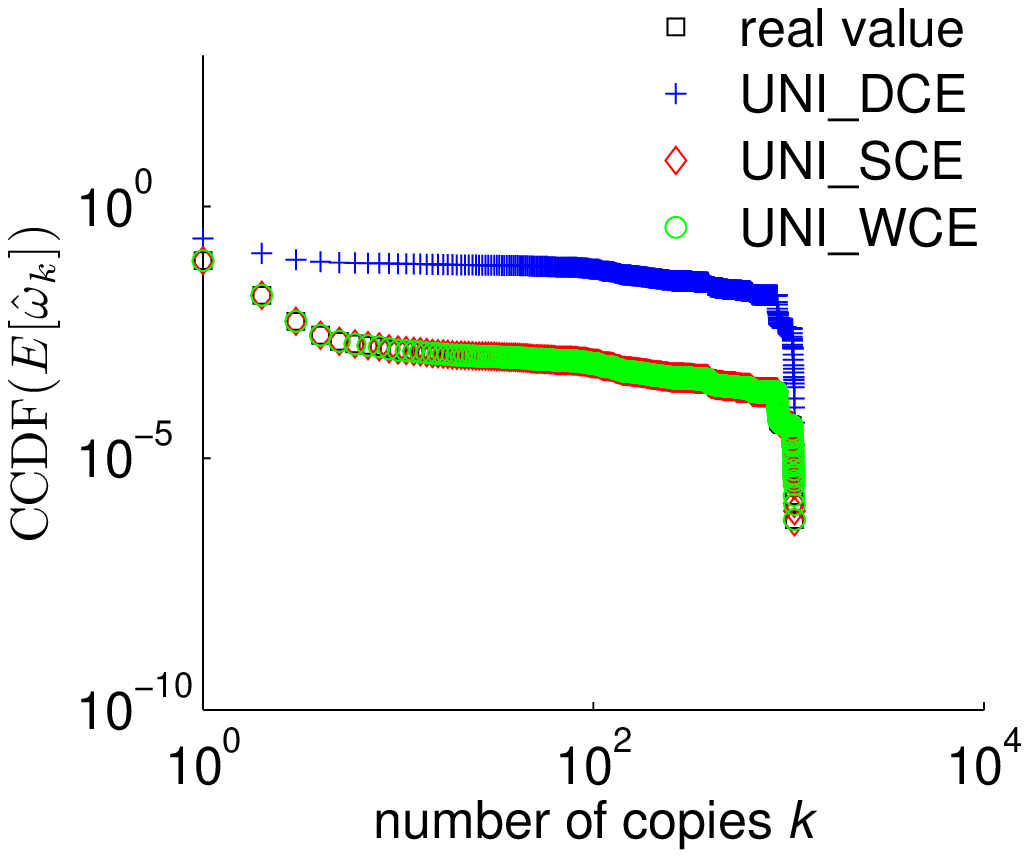}}
\subfigure[UNI, CDS I]{
\includegraphics[width=0.23\textwidth]{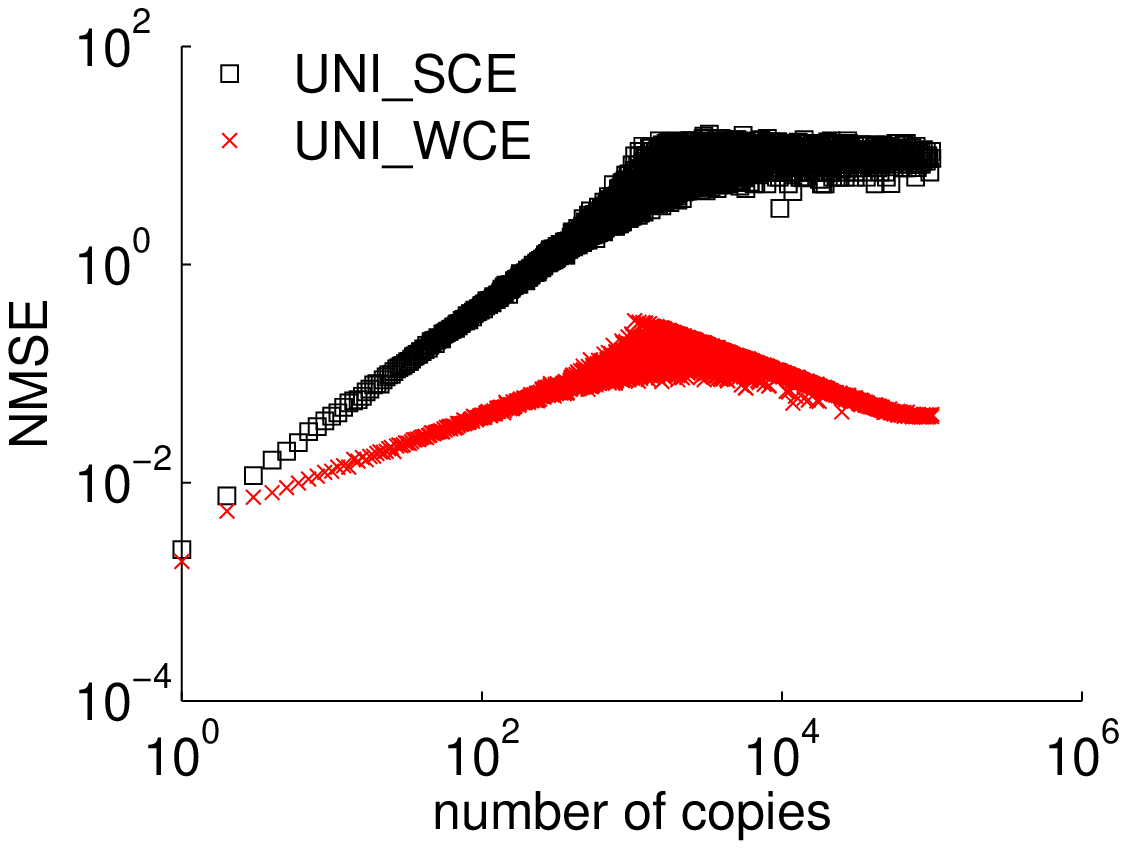}}
\subfigure[RW, CDS I]{
\includegraphics[width=0.23\textwidth]{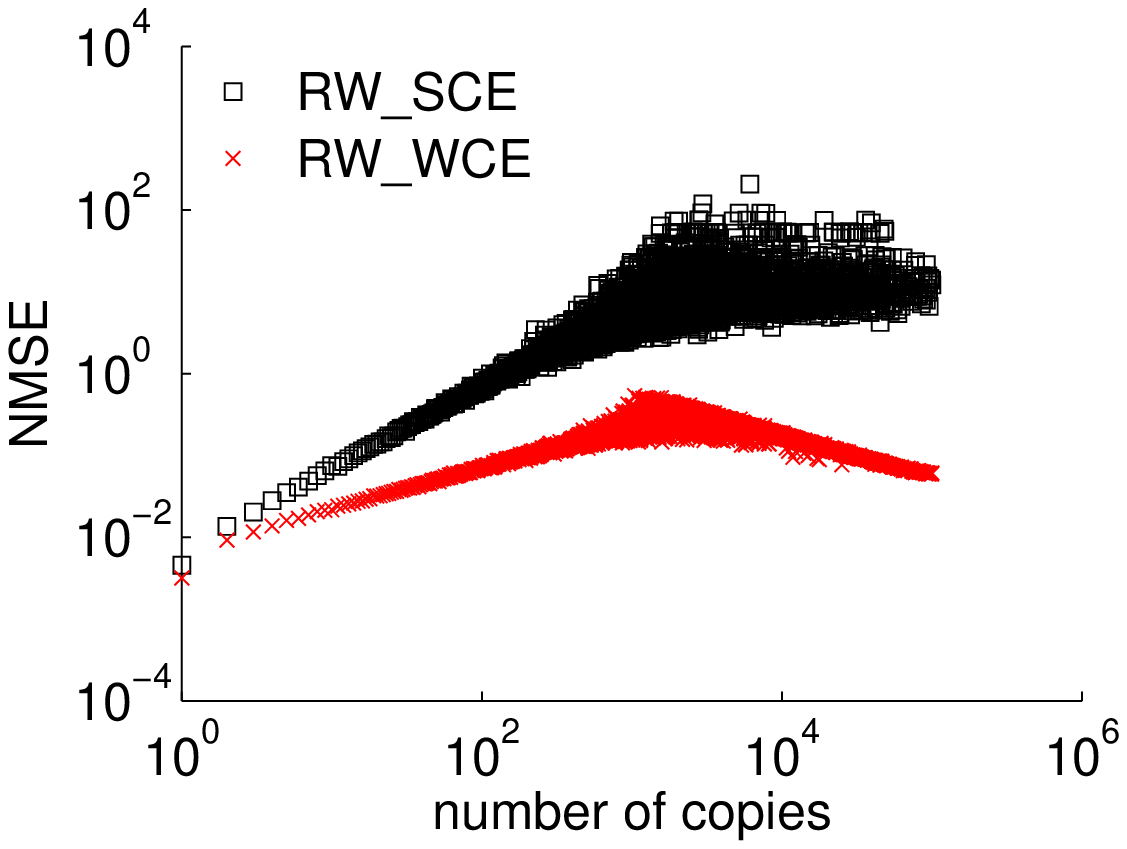}}
\subfigure[MHRW, CDS I]{
\includegraphics[width=0.23\textwidth]{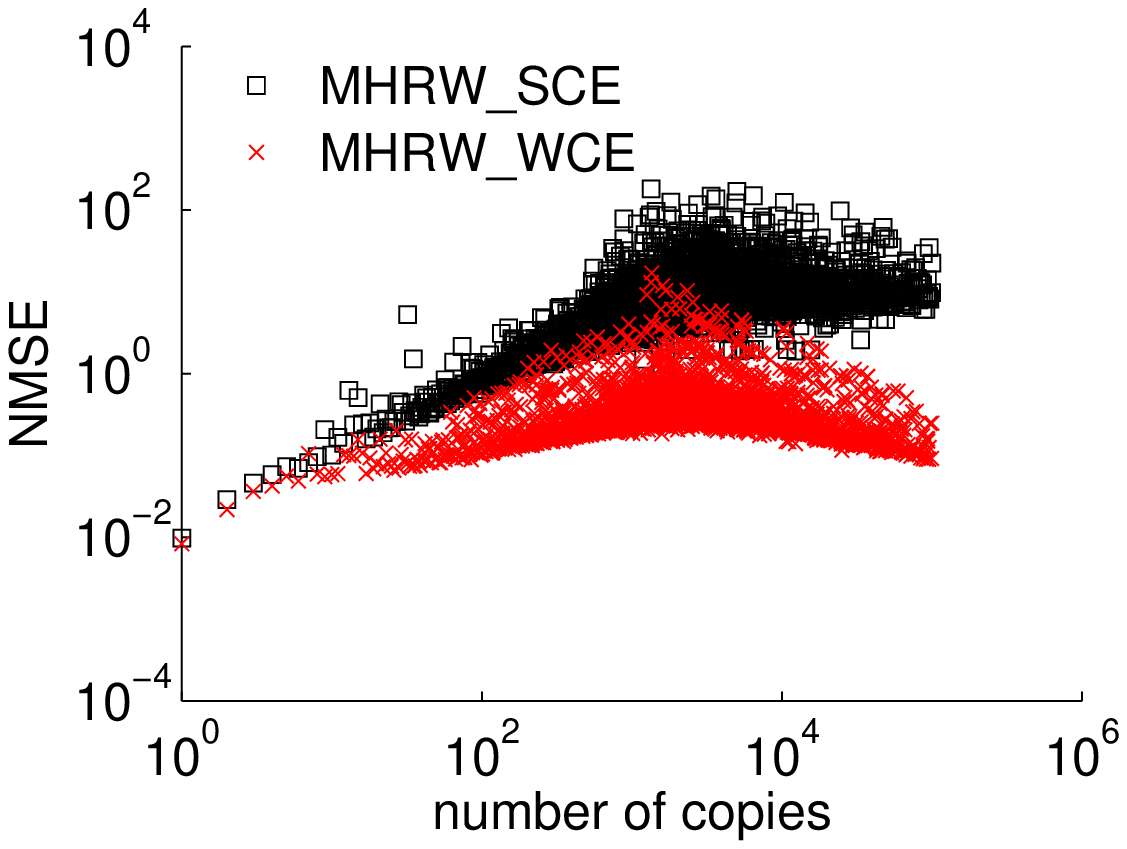}}
\subfigure[FS, CDS I]{
\includegraphics[width=0.23\textwidth]{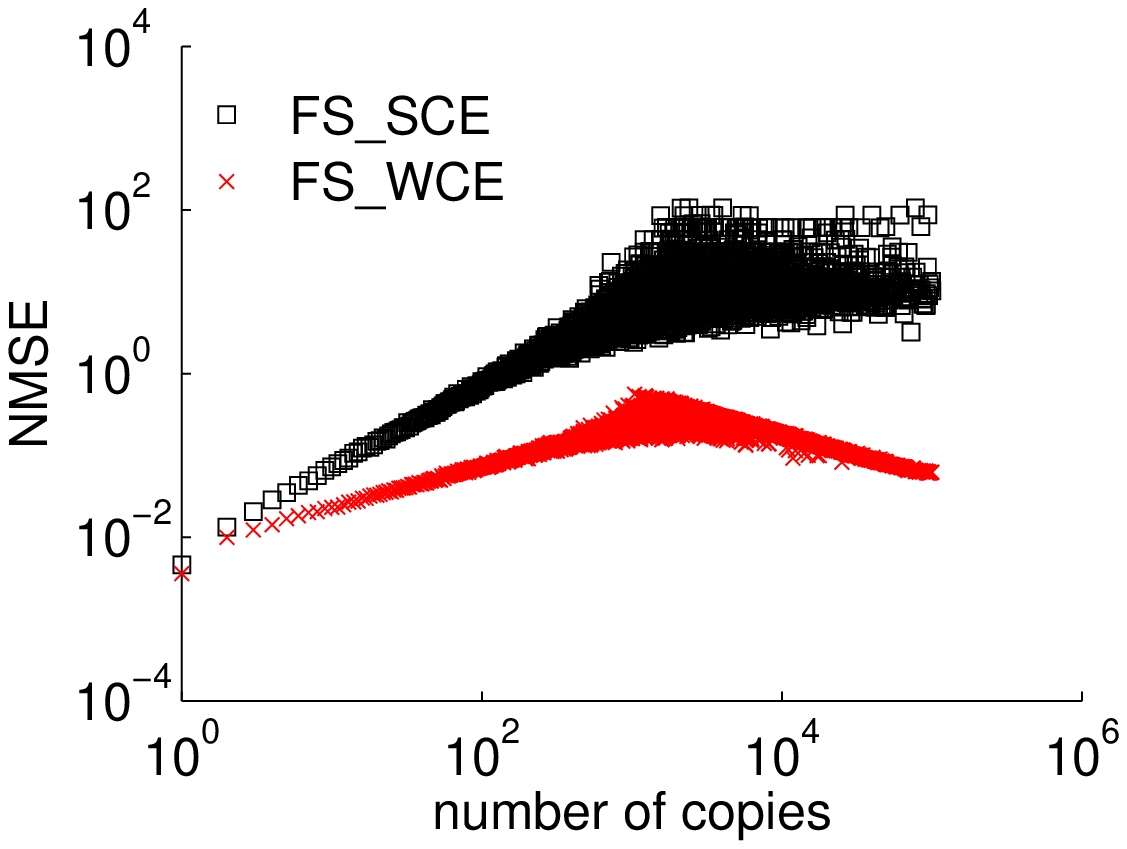}}
\subfigure[UNI, CDS II]{
\includegraphics[width=0.23\textwidth]{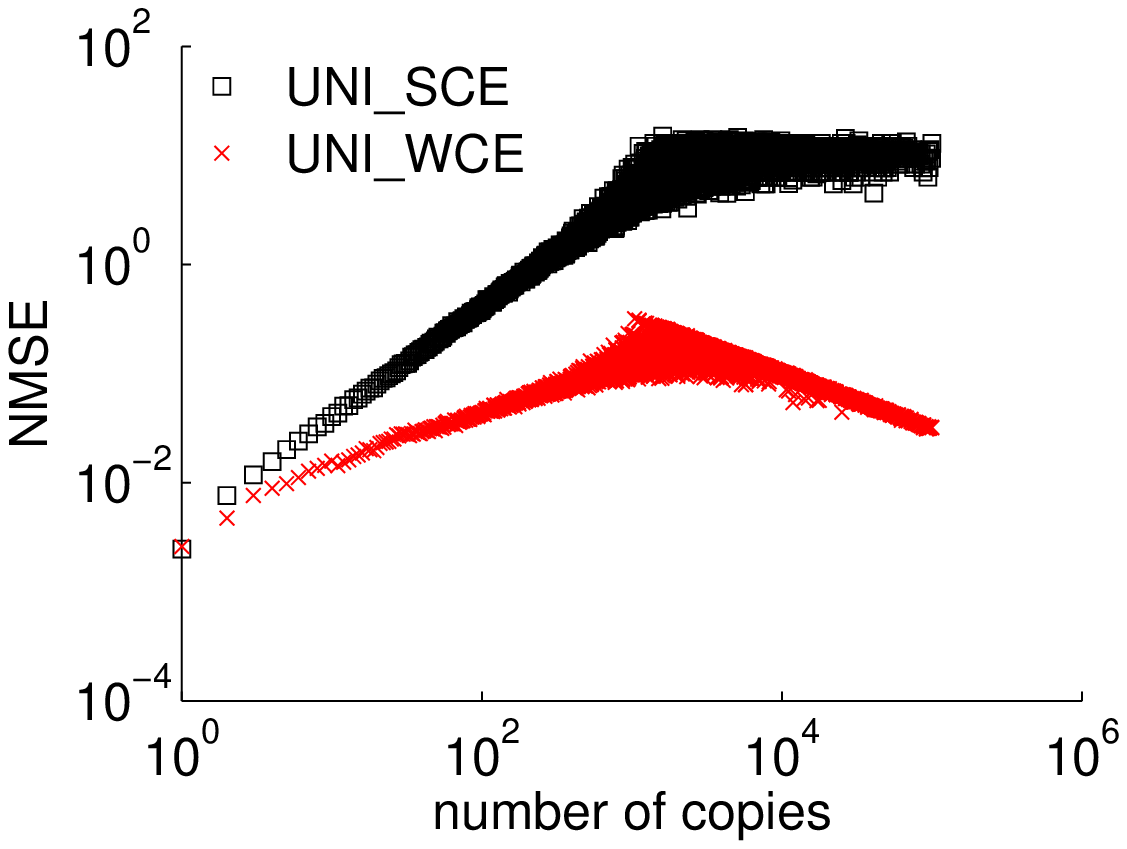}}
\subfigure[RW, CDS II]{
\includegraphics[width=0.23\textwidth]{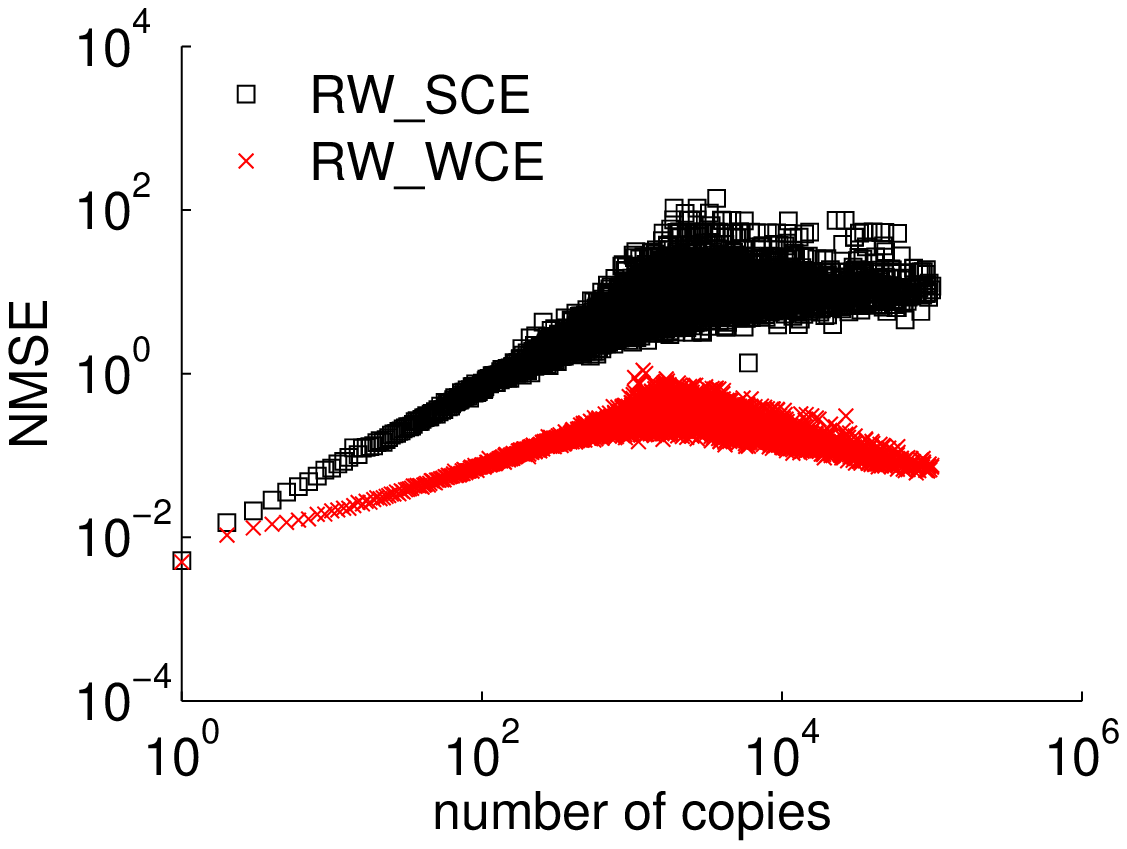}}
\subfigure[MHRW, CDS II]{
\includegraphics[width=0.23\textwidth]{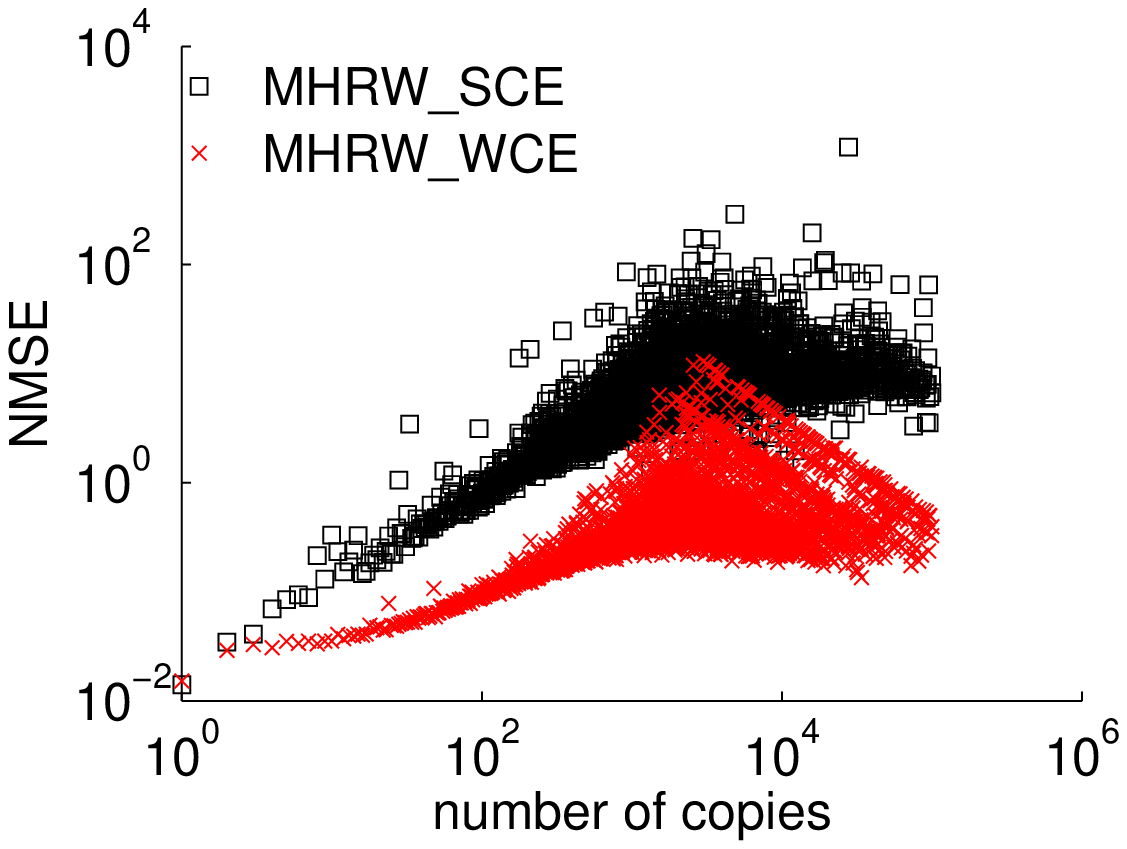}}
\subfigure[FS, CDS II]{
\includegraphics[width=0.23\textwidth]{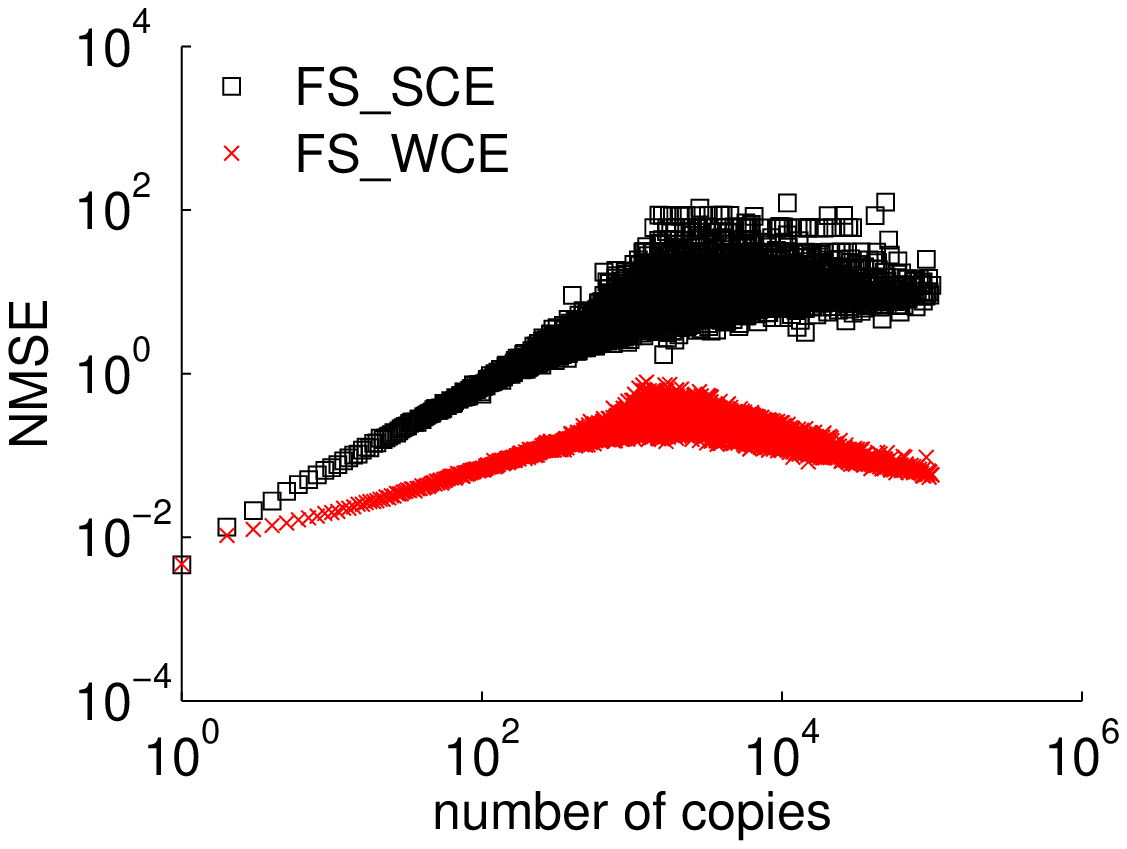}}
\subfigure[UNI, CDS III]{
\includegraphics[width=0.23\textwidth]{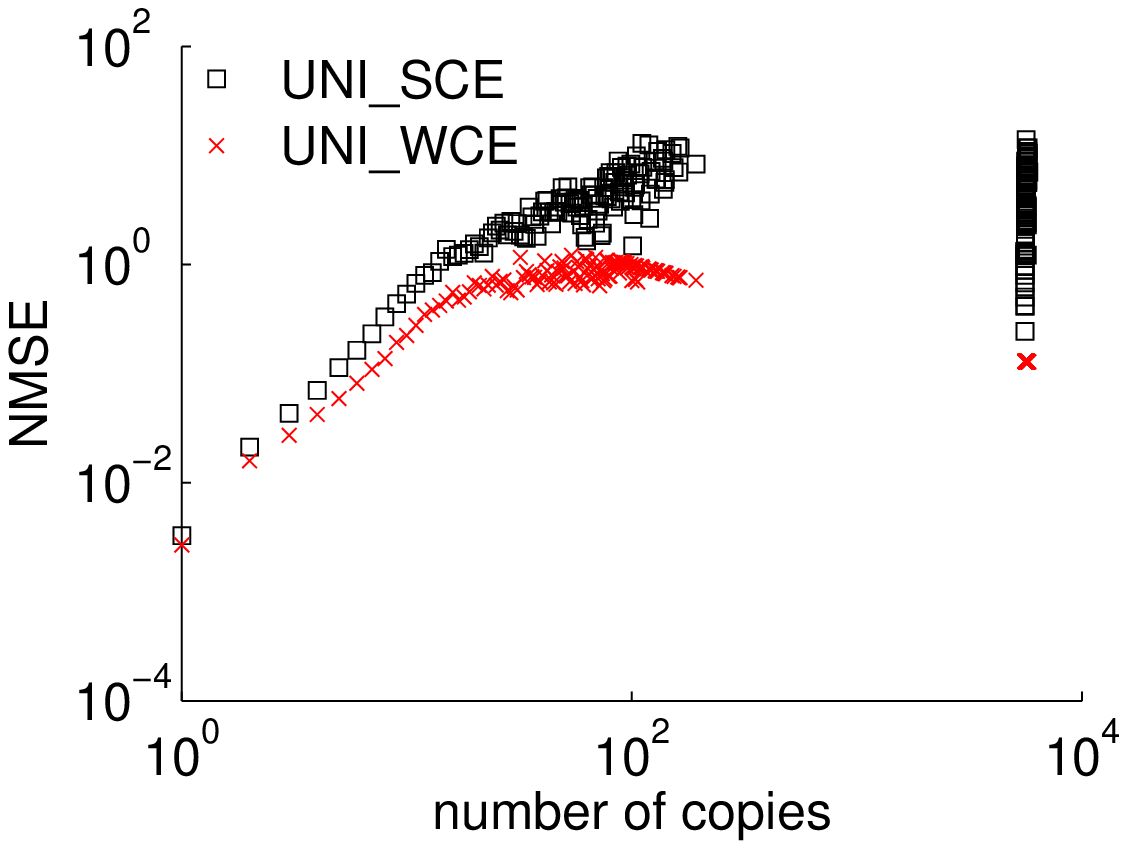}}
\subfigure[RW, CDS III]{
\includegraphics[width=0.23\textwidth]{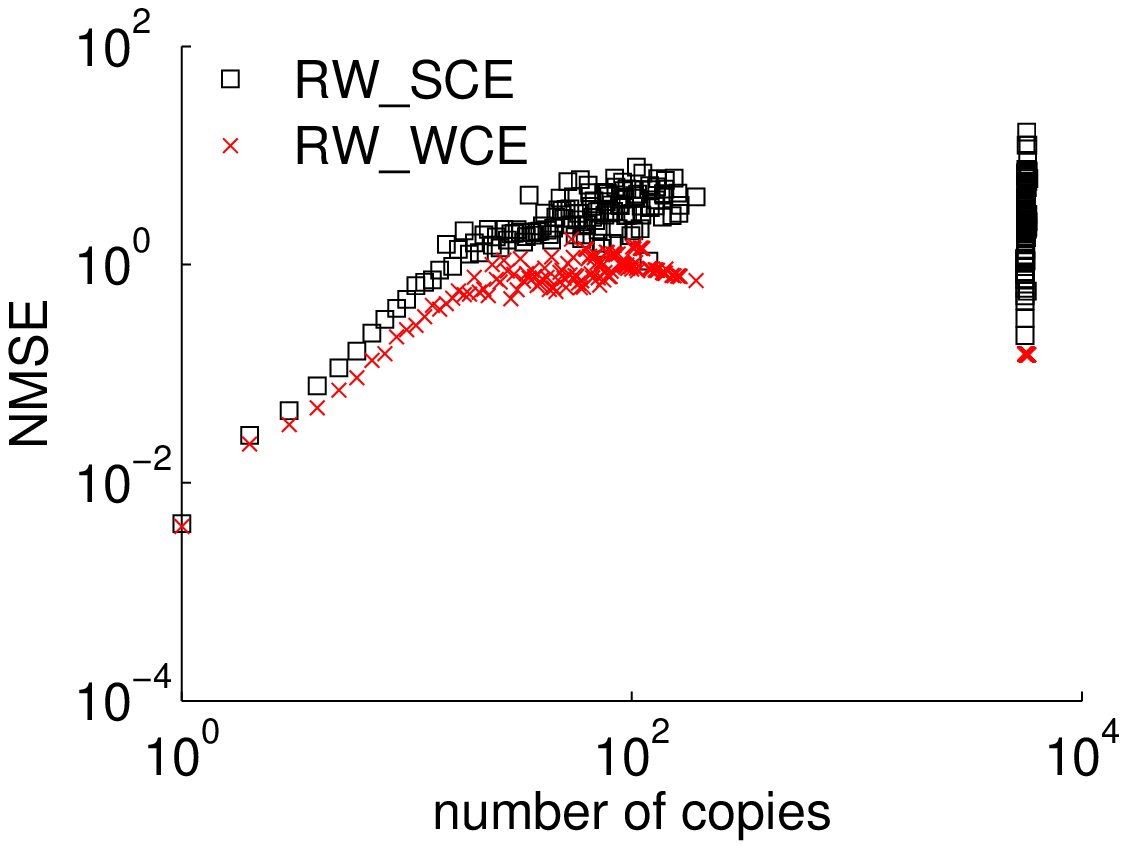}}
\subfigure[MHRW, CDS III]{
\includegraphics[width=0.23\textwidth]{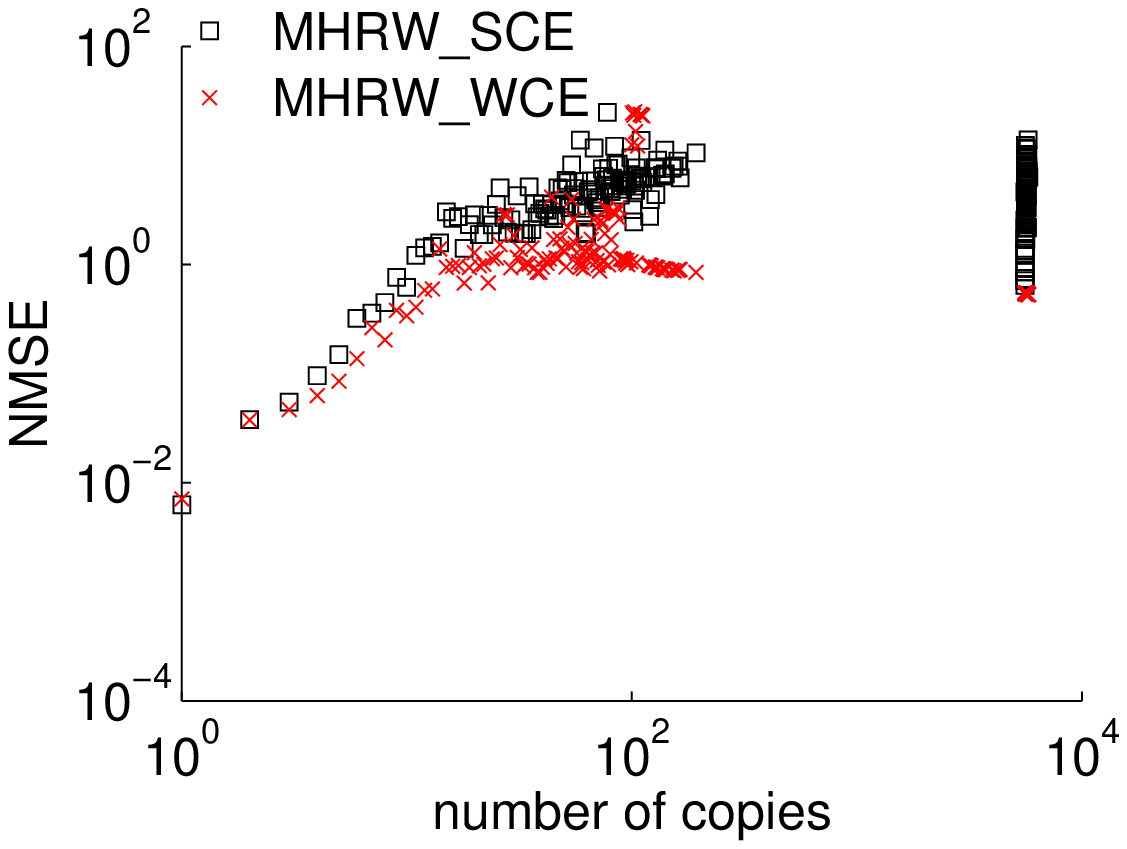}}
\subfigure[FS, CDS III]{
\includegraphics[width=0.23\textwidth]{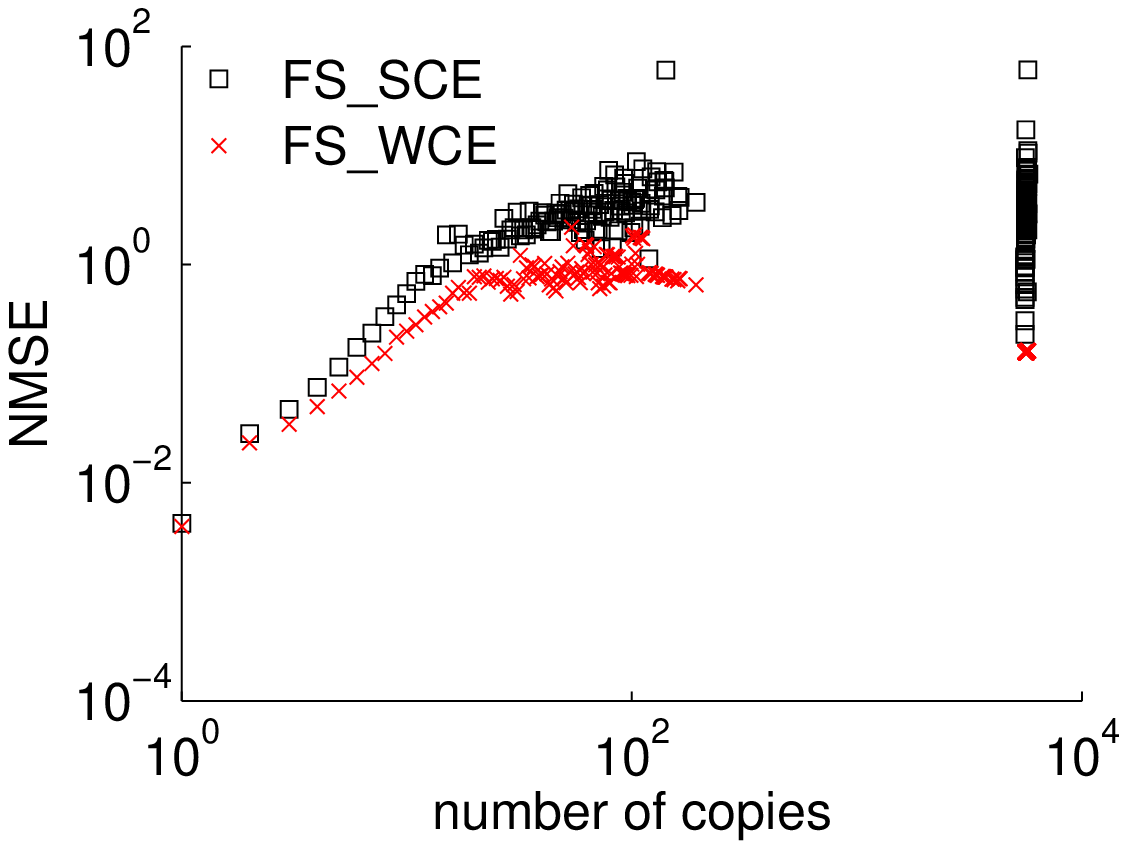}}
\subfigure[UNI, CDS IV]{
\includegraphics[width=0.23\textwidth]{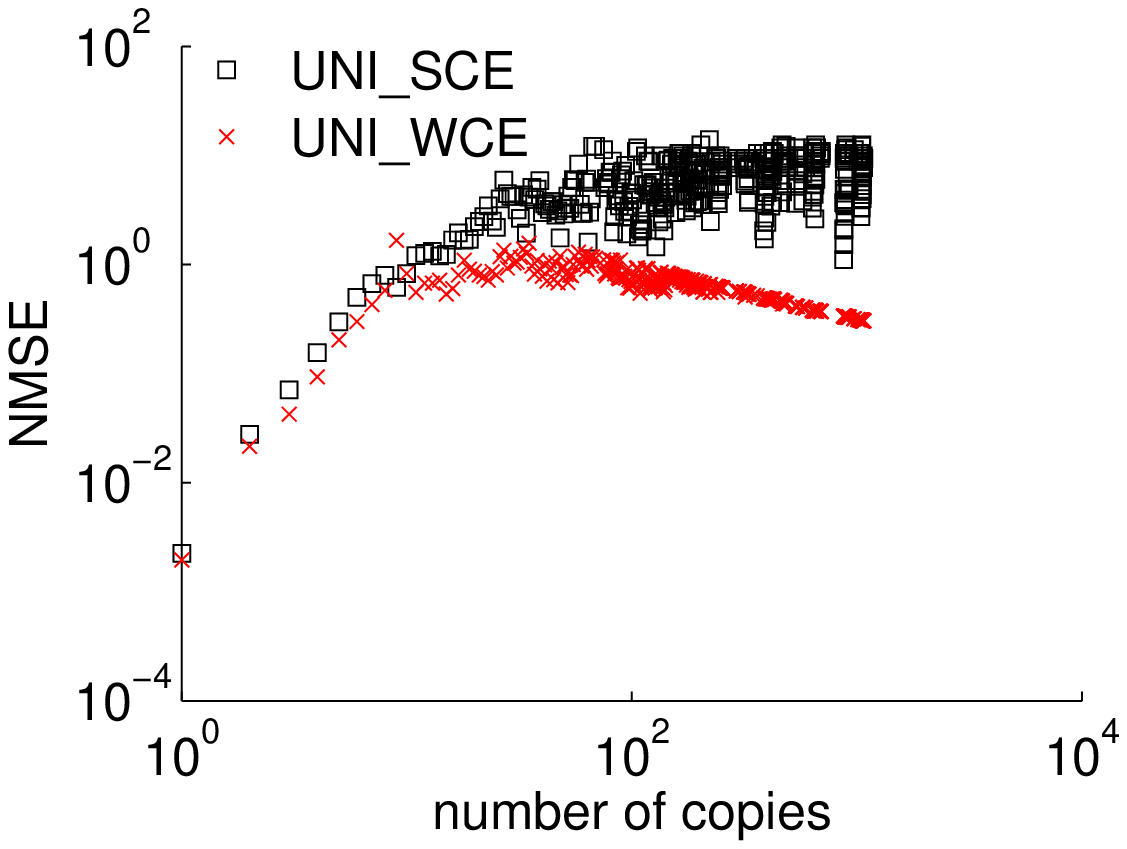}}
\subfigure[RW, CDS IV]{
\includegraphics[width=0.23\textwidth]{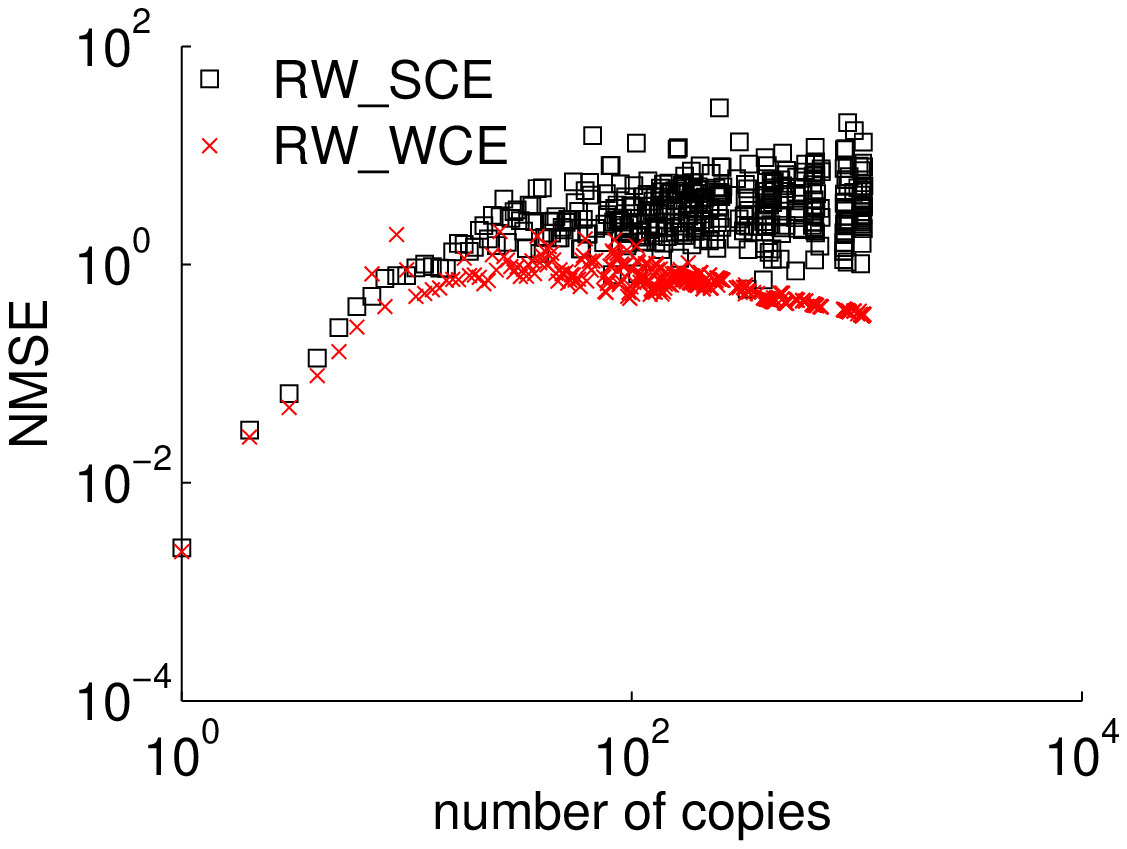}}
\subfigure[MHRW, CDS IV]{
\includegraphics[width=0.23\textwidth]{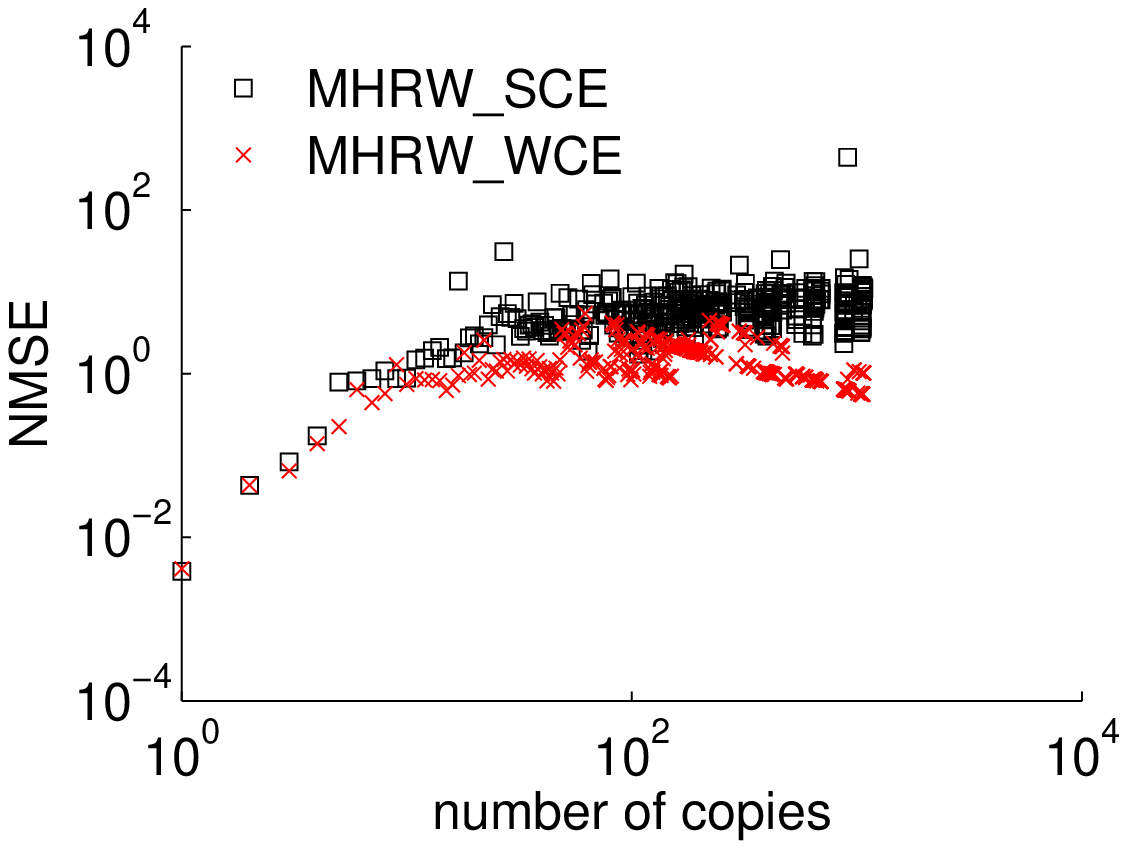}}
\subfigure[FS, CDS IV]{
\includegraphics[width=0.23\textwidth]{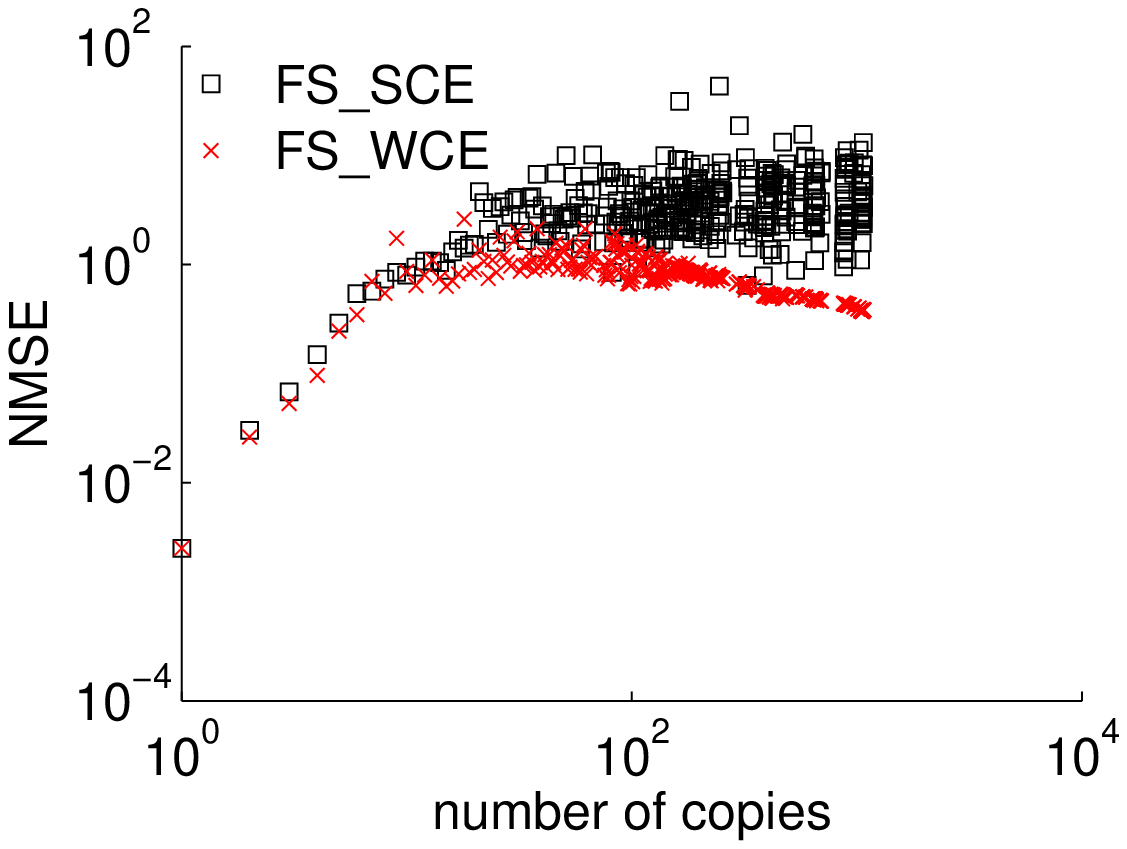}}
\caption{(Xiami) NMSE of content distribution estimates for different estimators and graph sampling methods.}\label{fig:xiamicontent}
\end{figure*}

\begin{figure*}[htb]
\center
\subfigure[CDS I]{
\includegraphics[width=0.23\textwidth]{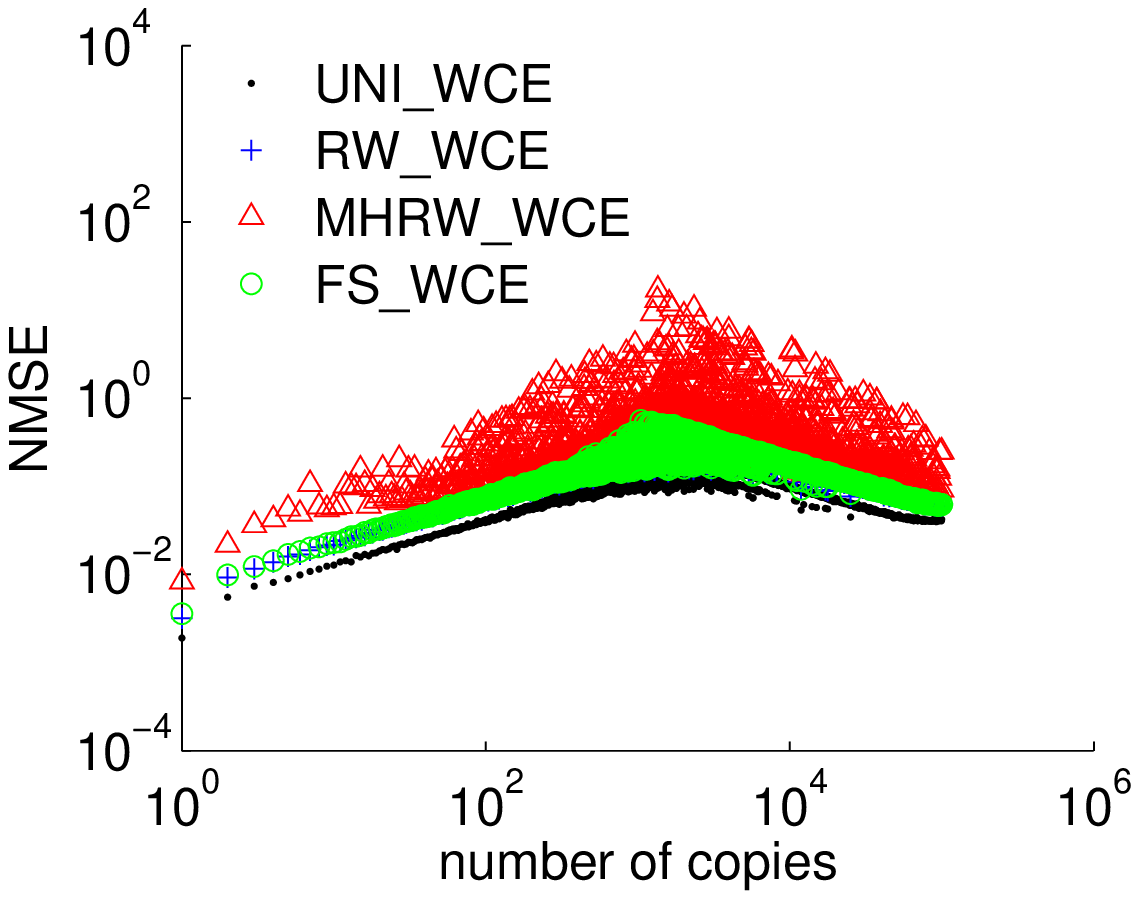}}
\subfigure[CDS II]{
\includegraphics[width=0.23\textwidth]{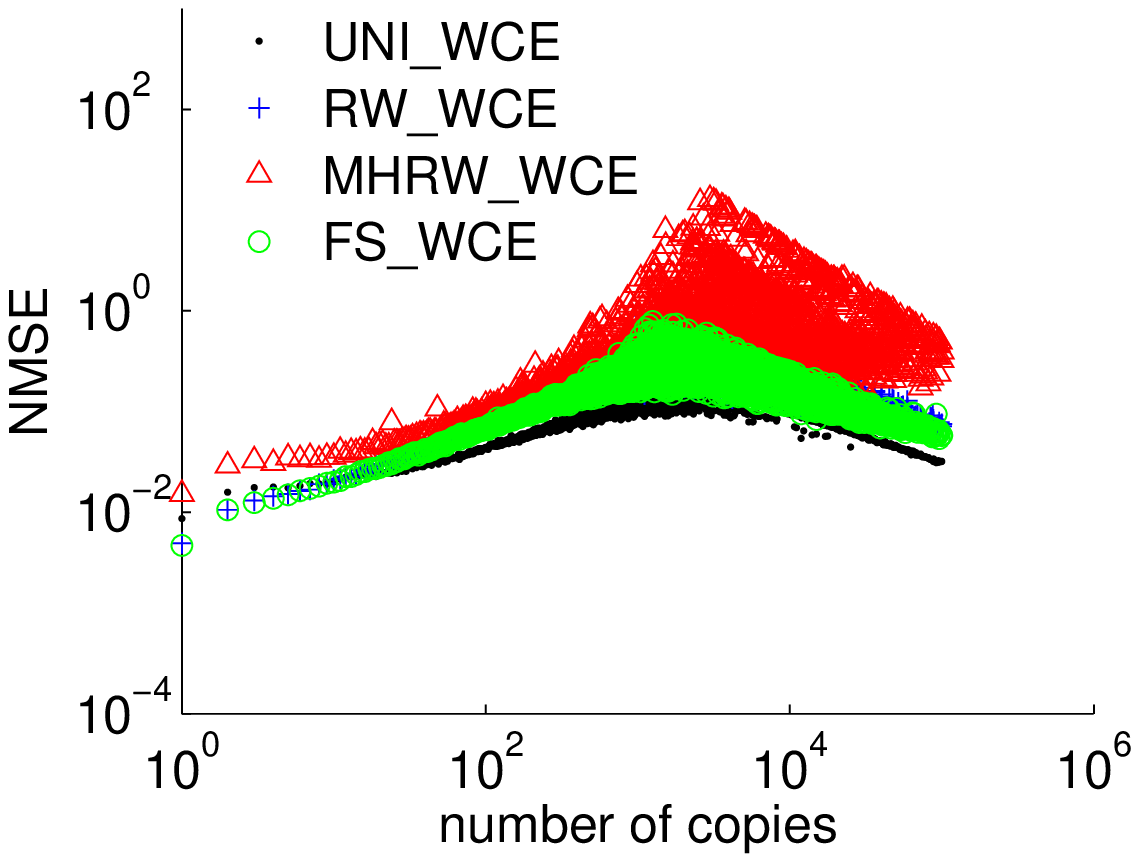}}
\subfigure[CDS III]{
\includegraphics[width=0.23\textwidth]{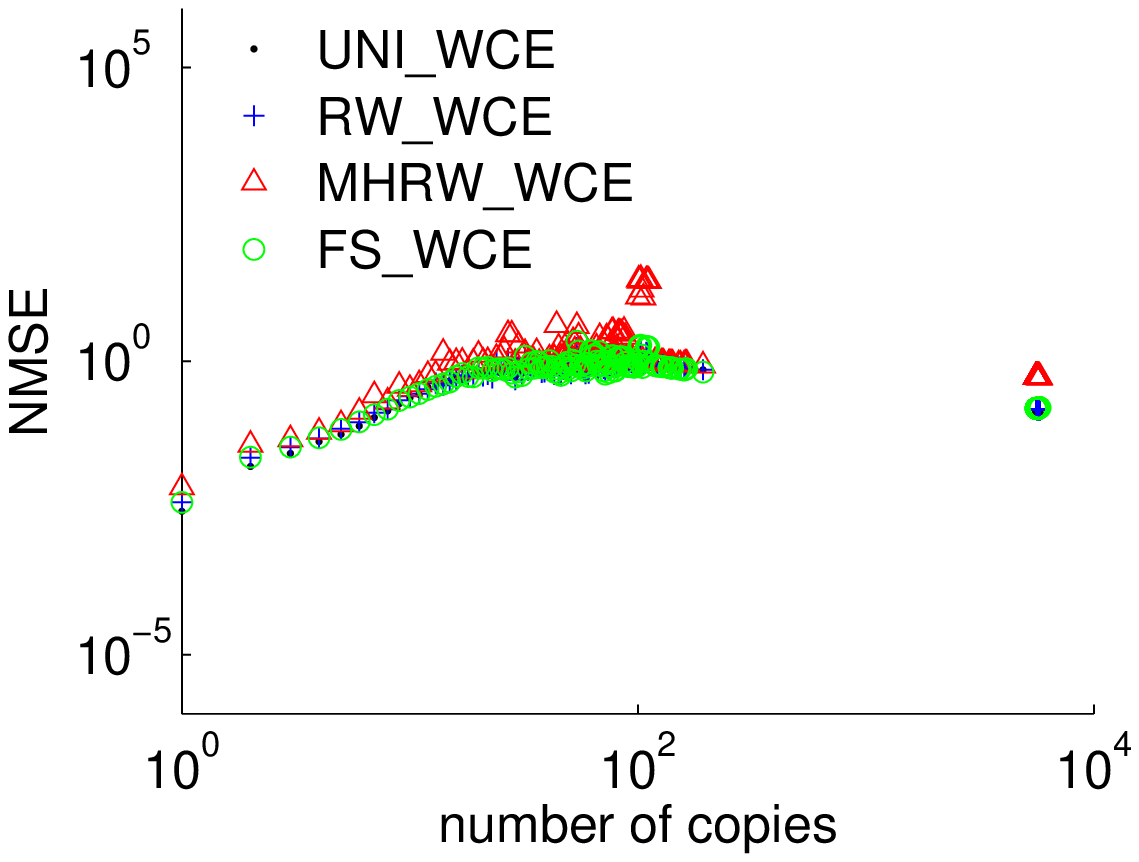}}
\subfigure[CDS IV]{
\includegraphics[width=0.23\textwidth]{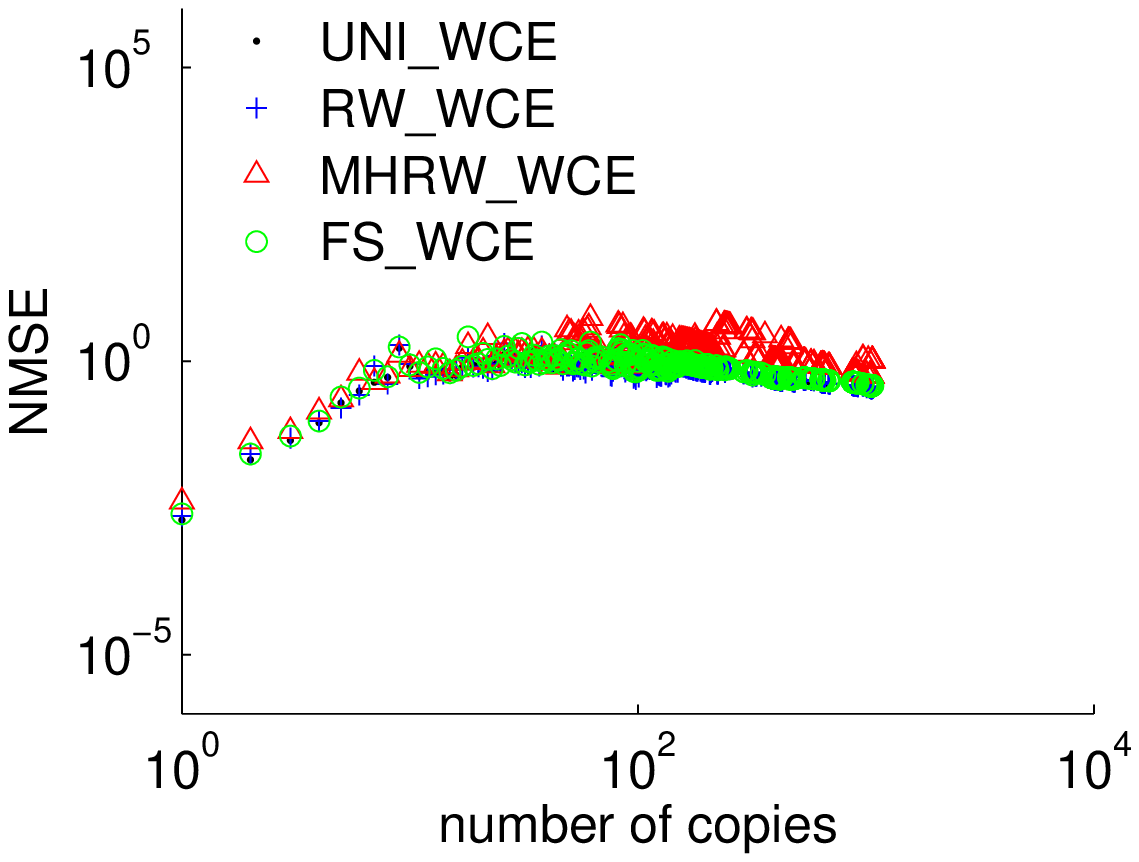}}
\caption{(LCC of Xiami) Compared NMSE of content distribution estimates for WCE using different graph sampling methods.}\label{fig:xiamicontentcmp}
\end{figure*}

In the following experiments, we set $\alpha=1$, $W=10^5$ for CDS I and CDS II, $p_S=0.01$ for CDS III and CDS IV.
We evaluate the performance of SCE and WCE
combined with different graph sampling methods based on
the datasets generated by four different CDS.
Figs.~\ref{fig:xiamicontent} (a)--(d) show the complementary cumulative distribution function (CCDF) of the expectation of content distribution estimates provided by DCE, SCE and WCE, where the graph is Xiami and the graph sampling method is UNI. We find that DCE exhibits large errors, and SCE and WCE are quite accurate.
Similar results are obtained when
we use the other four graph sampling methods described in Section~\ref{sec:struturesampling},
but due to page limits, we omit them here.
Figs.~\ref{fig:xiamicontent} (e)--(t) show the NMSE of SCE and WCE combined with different graph sampling methods for measuring content distribution. The results show that WCE is significantly more accurate than SCE
over most points.
In particular, WCE is almost an order of magnitude more accurate than
SCE for the number of copies larger than 100, and nearly two orders of
magnitude more accurate than SCE for the number of copies larger than 1,000.
Fig.~\ref{fig:xiamicontentcmp} show the compared results for different graph sampling methods
where the graph used is the LCC of Xiami. The results show that UNI is quite accurate and MHRW exhibits large errors for content with a small number of copies. The compared results for WCE show that MHRW is much worse than the other graph sampling methods, while RW and FS have almost the same accuracy.
The results for the graph YouTube are similar and are shown in~\cite{PinghuiContent2012}.

%

\begin{figure}[htb]
\center
\subfigure[\# followers, \# following, and \# recommendations]{
\includegraphics[width=0.4\textwidth]{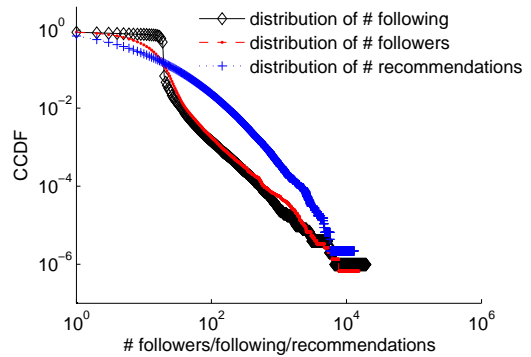}}
\subfigure[Location]{
\includegraphics[width=0.4\textwidth]{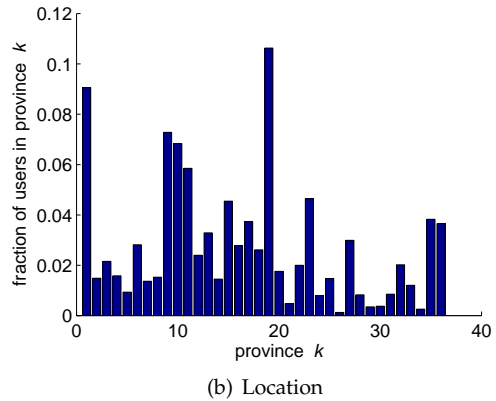}}
\caption{(Xiami) The distributions of users by different labels.}\label{fig:xiamigraphCCDF}
\end{figure}

\begin{figure}[htb]
\center
\subfigure[UNI\_WCE vs. UNI]{
\includegraphics[width=0.23\textwidth]{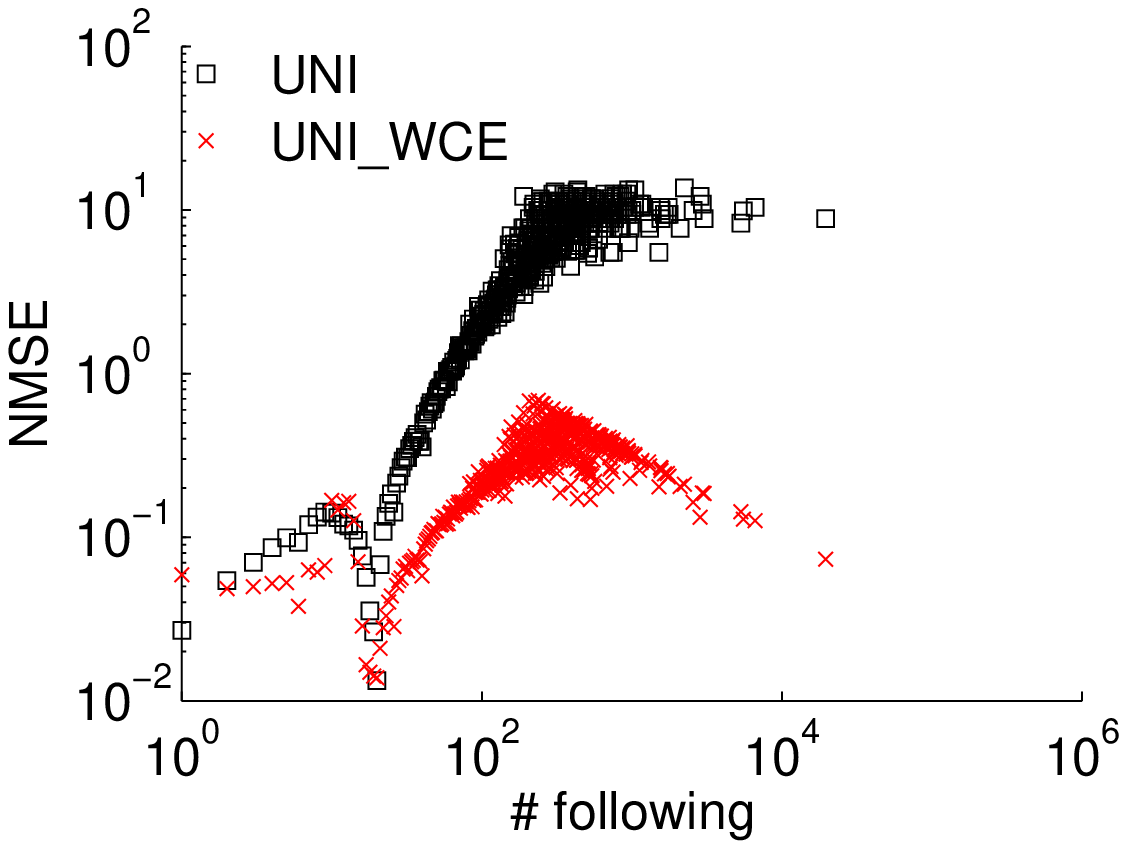}}
\subfigure[RW\_WCE vs. RW]{
\includegraphics[width=0.23\textwidth]{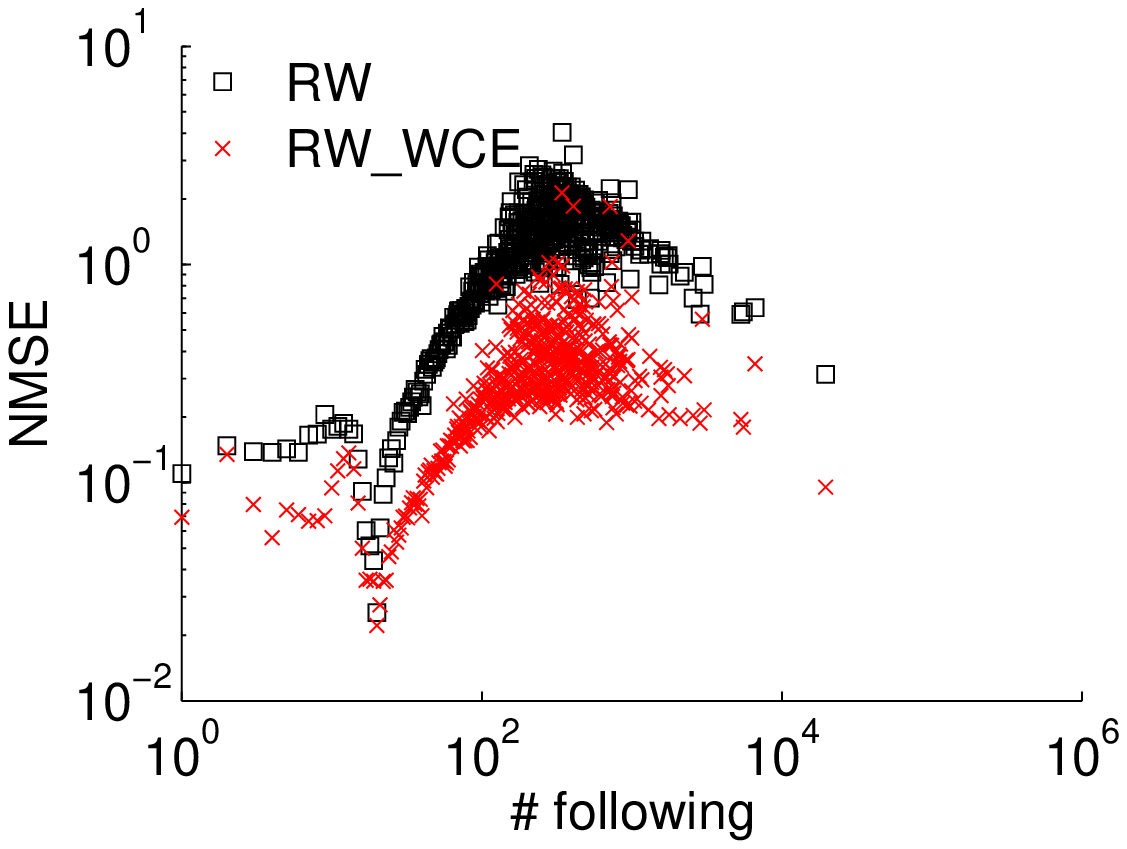}}
\subfigure[MHRW\_WCE vs. MHRW]{
\includegraphics[width=0.23\textwidth]{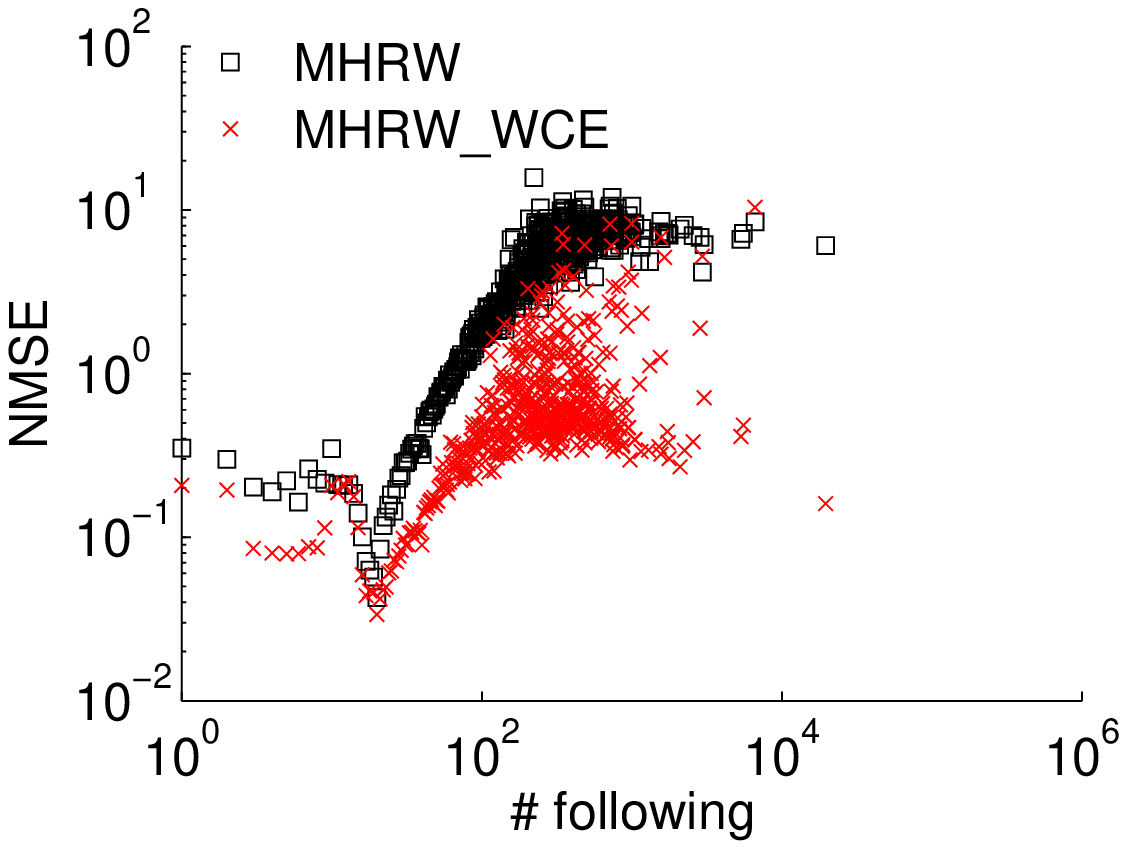}}
\subfigure[FS\_WCE vs. FS]{
\includegraphics[width=0.23\textwidth]{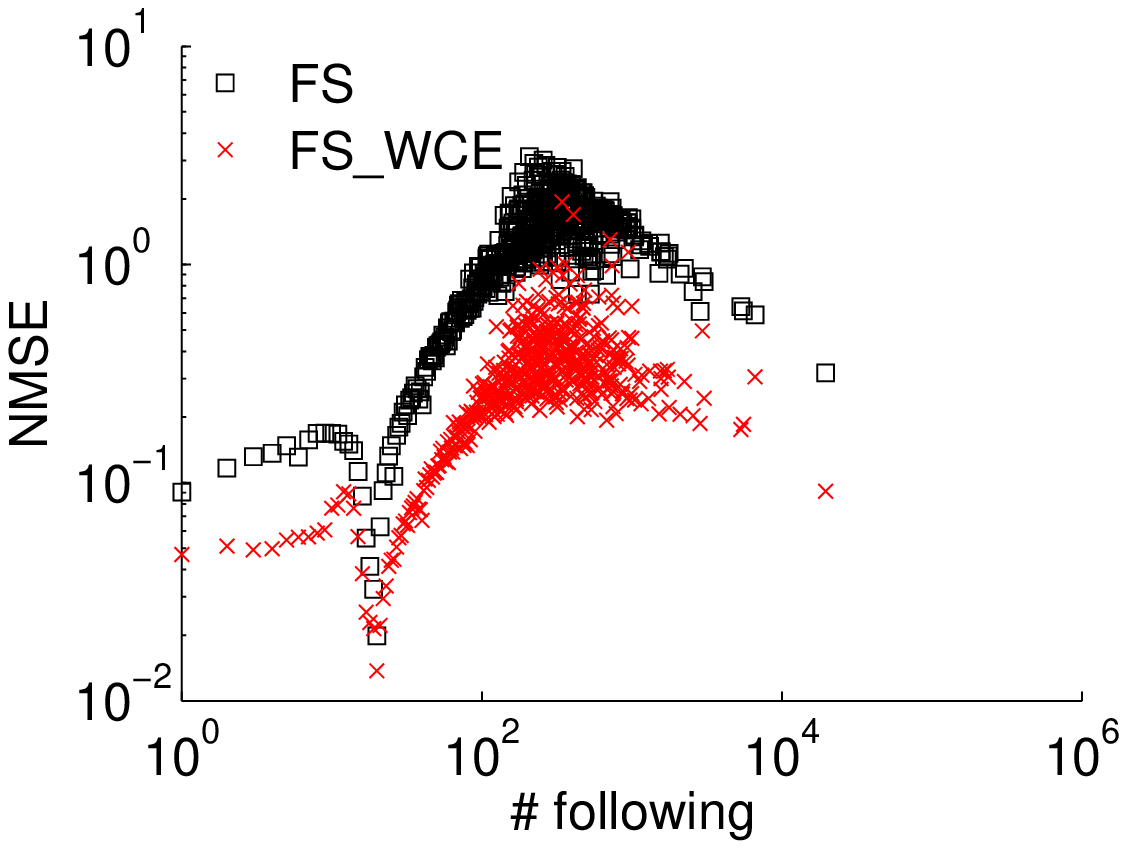}}
\caption{NMSE of following count distribution estimates.}\label{fig:xiamigraphoutdegree}
\end{figure}

\begin{figure}[htb]
\center
\subfigure[UNI\_WCE vs. UNI]{
\includegraphics[width=0.23\textwidth]{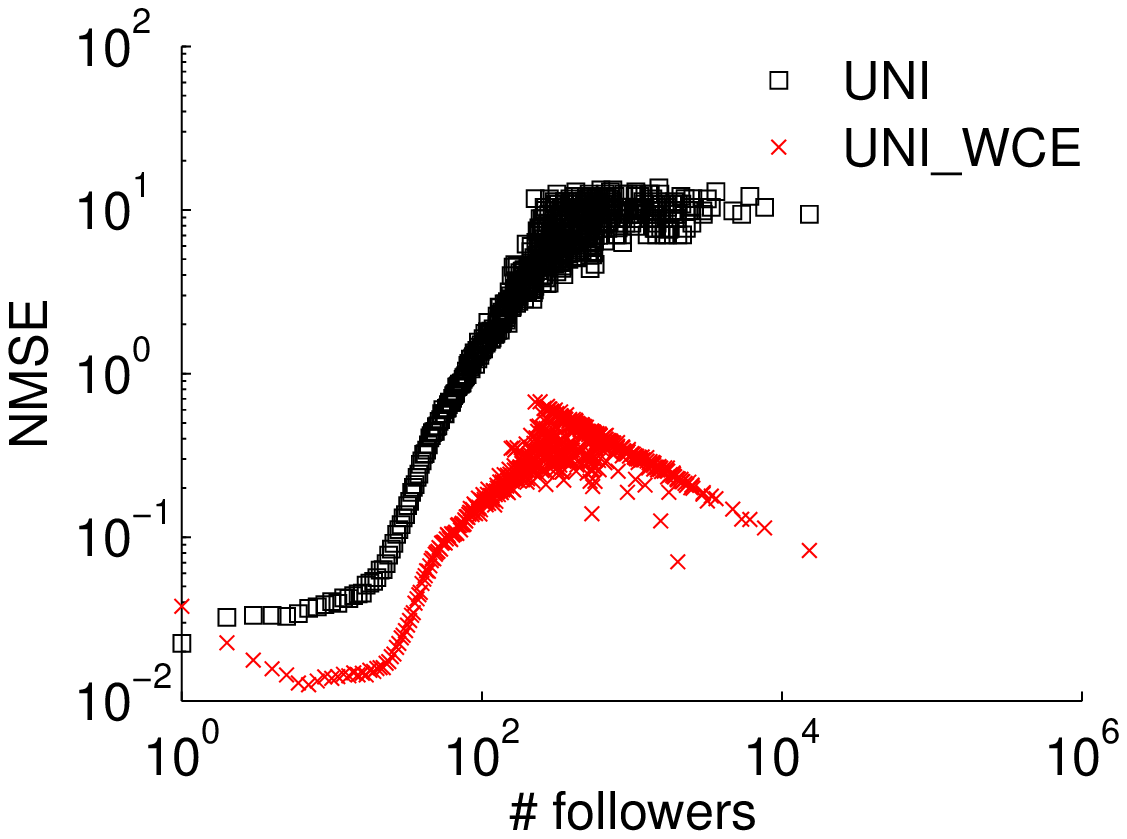}}
\subfigure[RW\_WCE vs. RW]{
\includegraphics[width=0.23\textwidth]{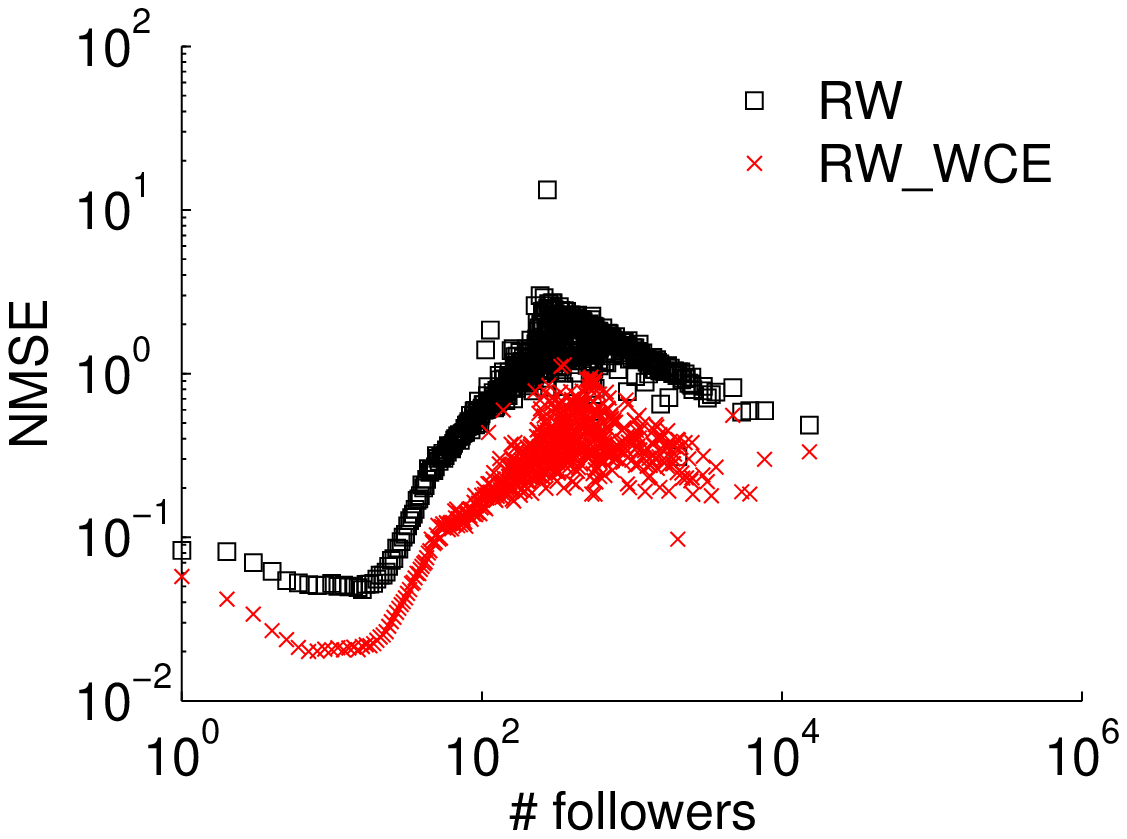}}
\subfigure[MHRW\_WCE vs. MHRW]{
\includegraphics[width=0.23\textwidth]{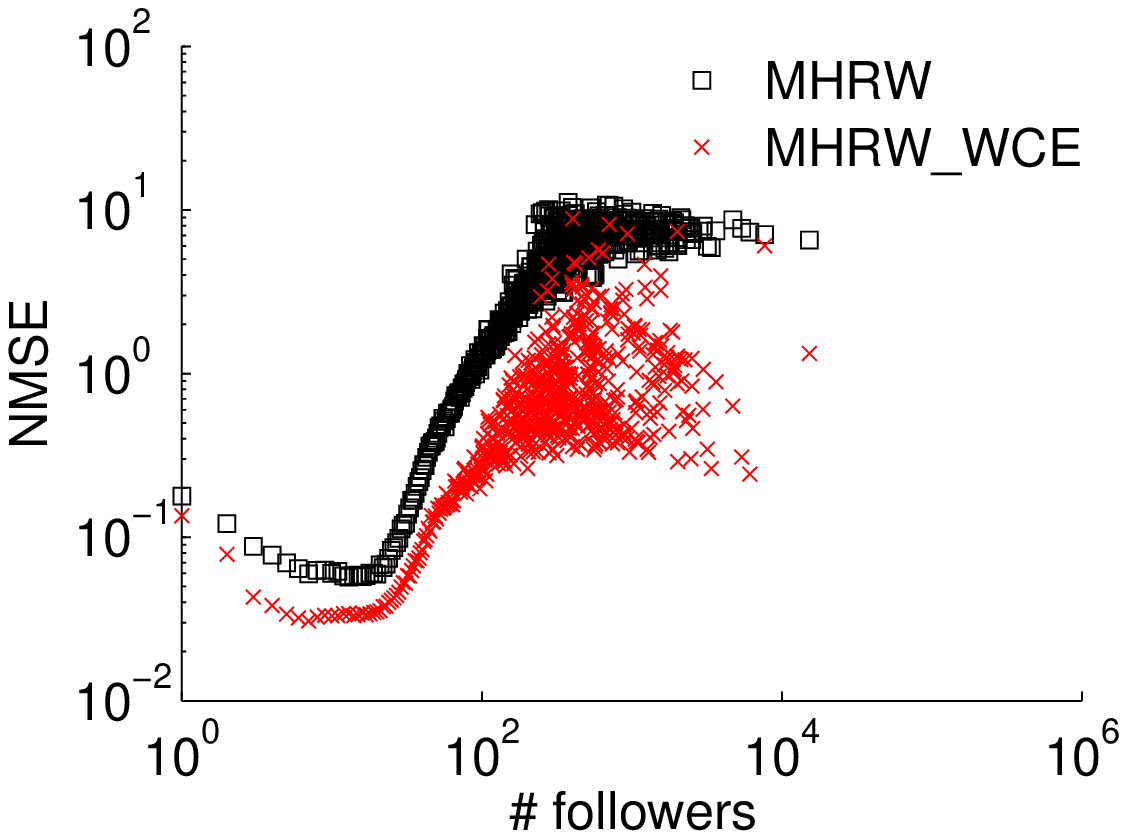}}
\subfigure[FS\_WCE vs. FS]{
\includegraphics[width=0.23\textwidth]{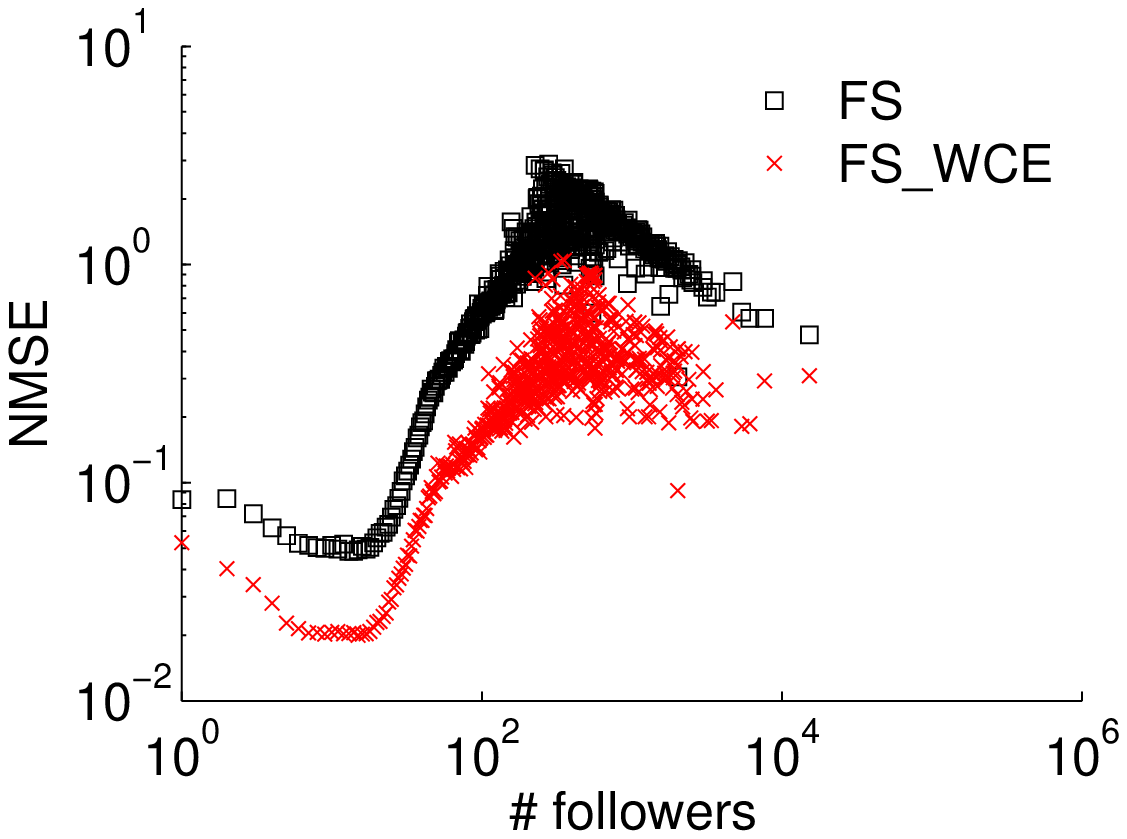}}
\caption{(Xiami) NMSE of follower count distribution estimates. }\label{fig:xiamigraphindegree}
\end{figure}

\begin{figure}[htb]
\center
\subfigure[UNI\_WCE vs. UNI]{
\includegraphics[width=0.23\textwidth]{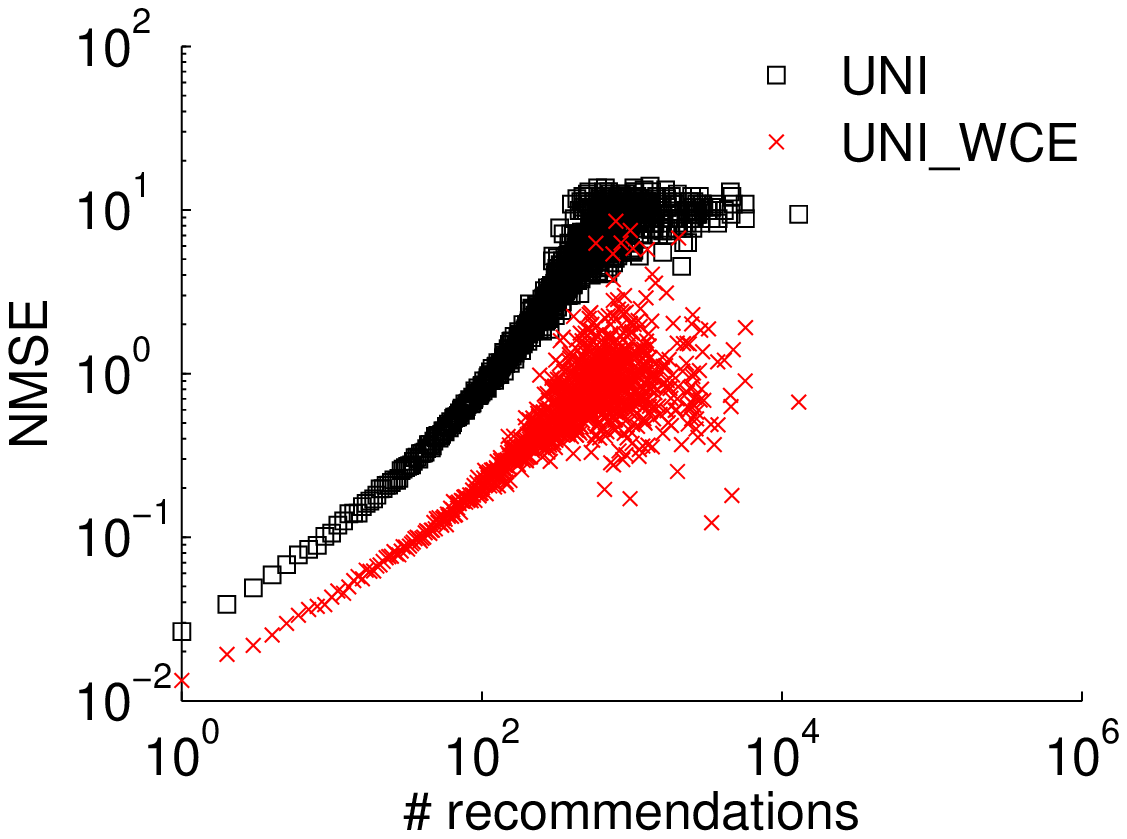}}
\subfigure[RW\_WCE vs. RW]{
\includegraphics[width=0.23\textwidth]{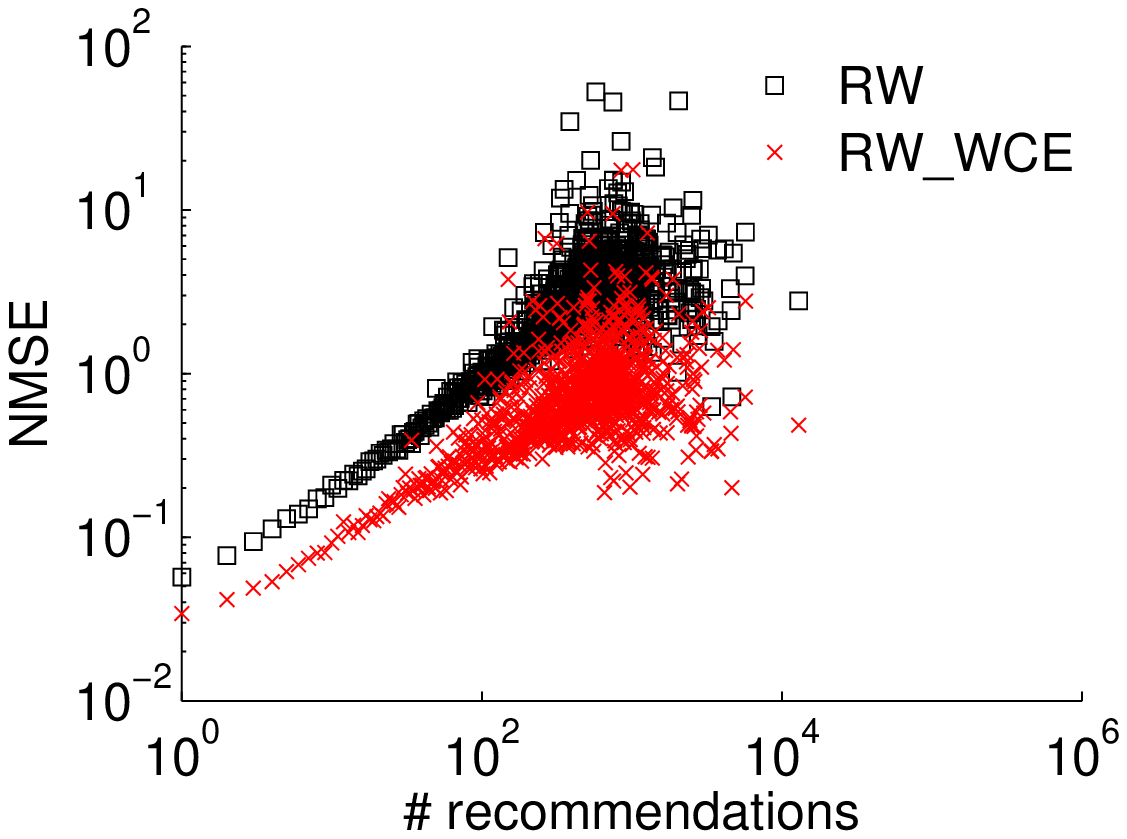}}
\subfigure[MHRW\_WCE vs. MHRW]{
\includegraphics[width=0.23\textwidth]{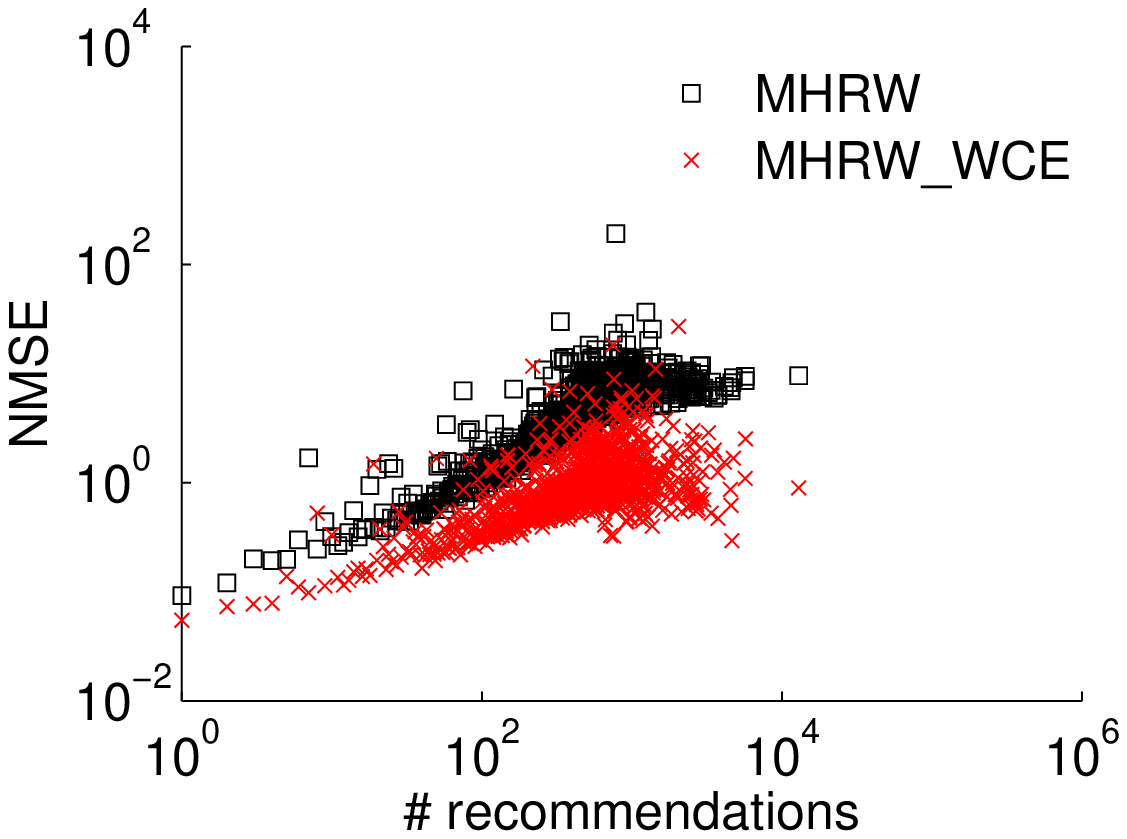}}
\subfigure[FS\_WCE vs. FS]{
\includegraphics[width=0.23\textwidth]{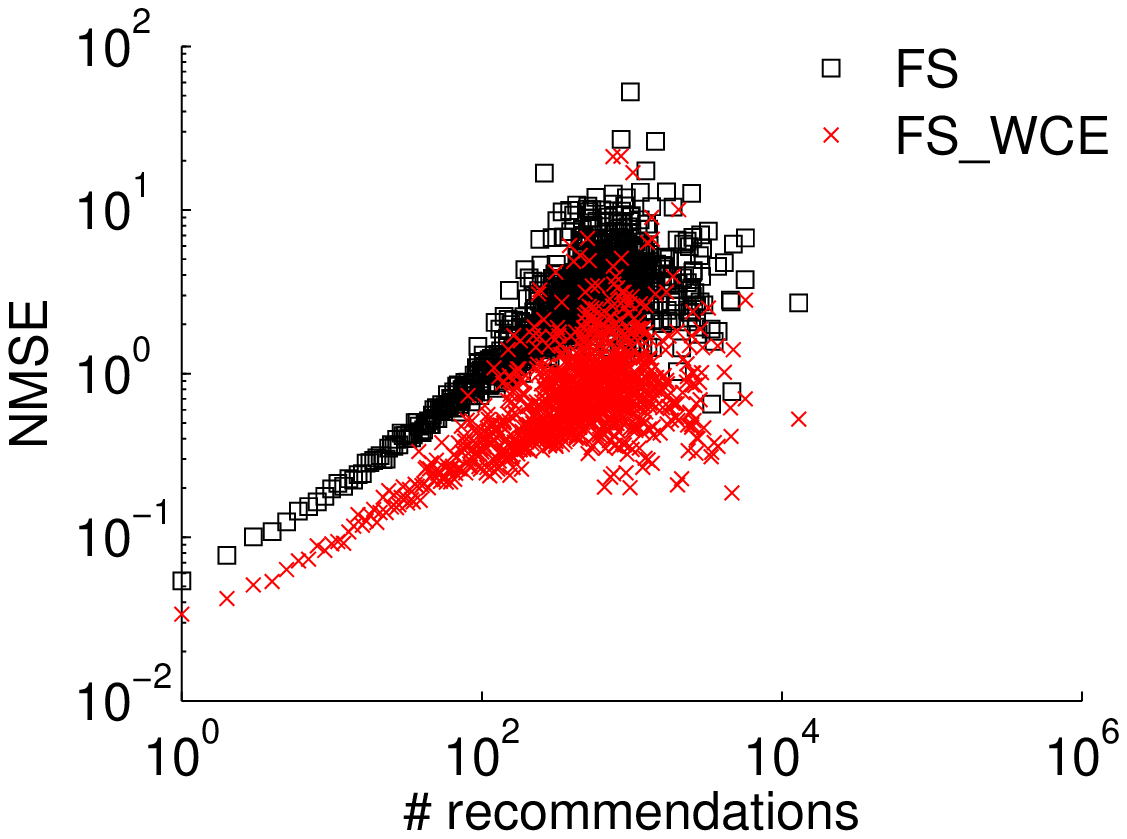}}
\caption{(Xiami) NMSE of recommendation count distribution estimates.}\label{fig:xiamigraphrecommendation}
\end{figure}

\begin{figure}[htb]
\center
\subfigure[UNI\_WCE vs. UNI]{
\includegraphics[width=0.23\textwidth]{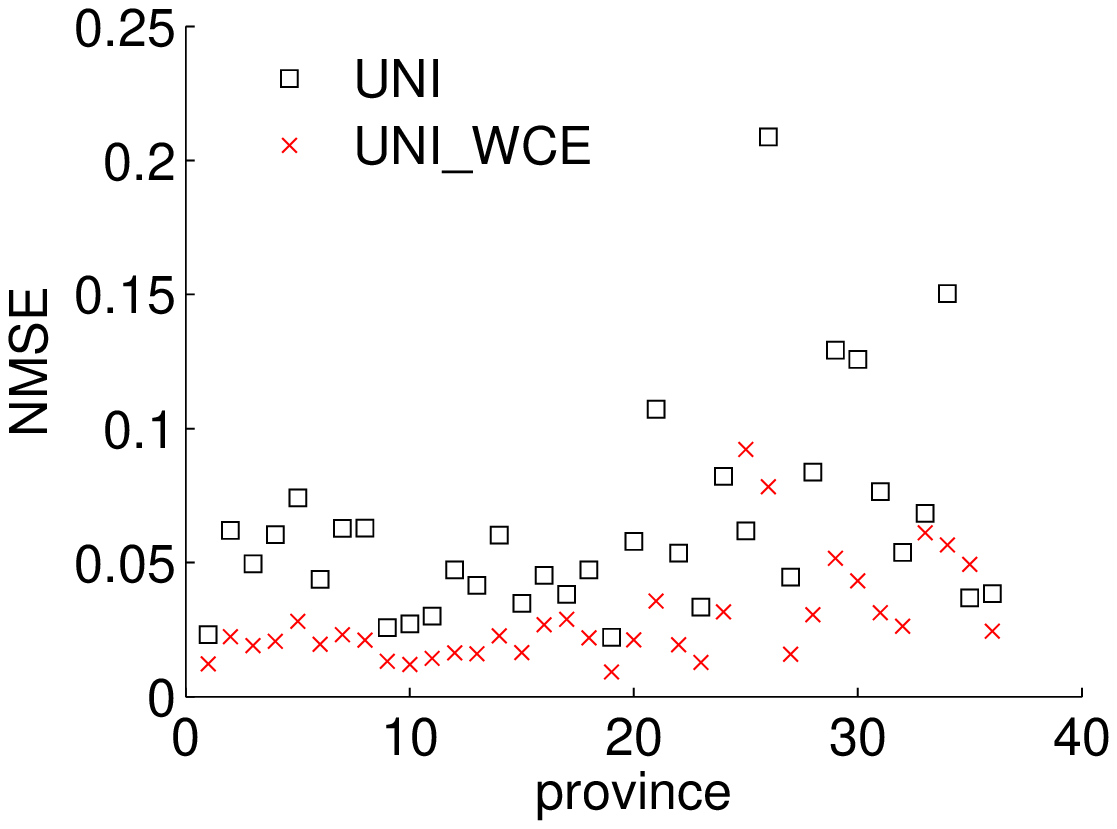}}
\subfigure[RW\_WCE vs. RW]{
\includegraphics[width=0.23\textwidth]{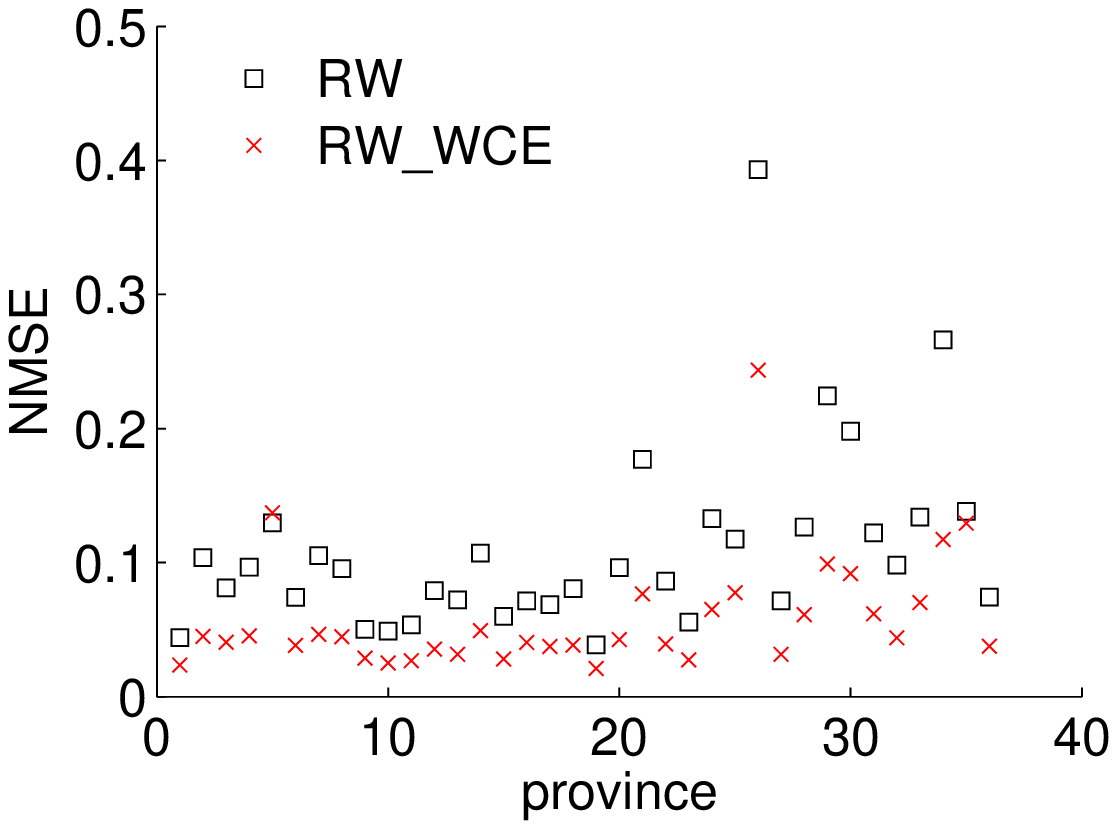}}
\subfigure[MHRW\_WCE vs. MHRW]{
\includegraphics[width=0.23\textwidth]{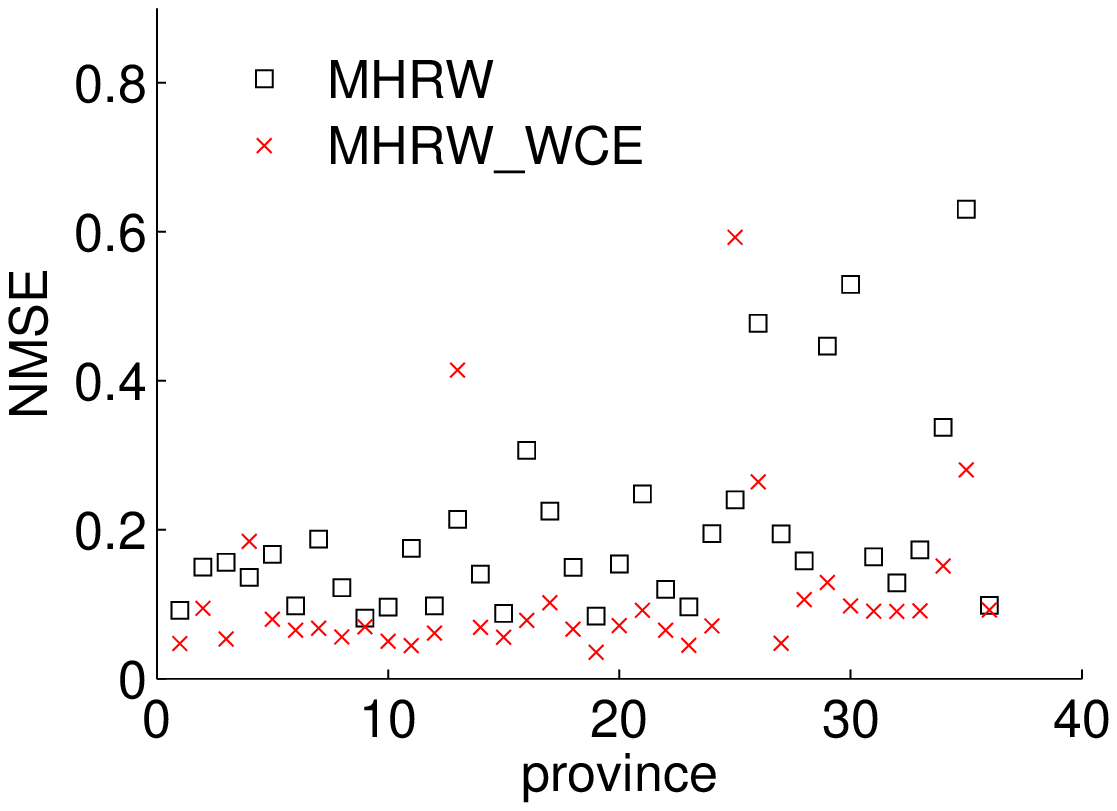}}
\subfigure[FS\_WCE vs. FS]{
\includegraphics[width=0.23\textwidth]{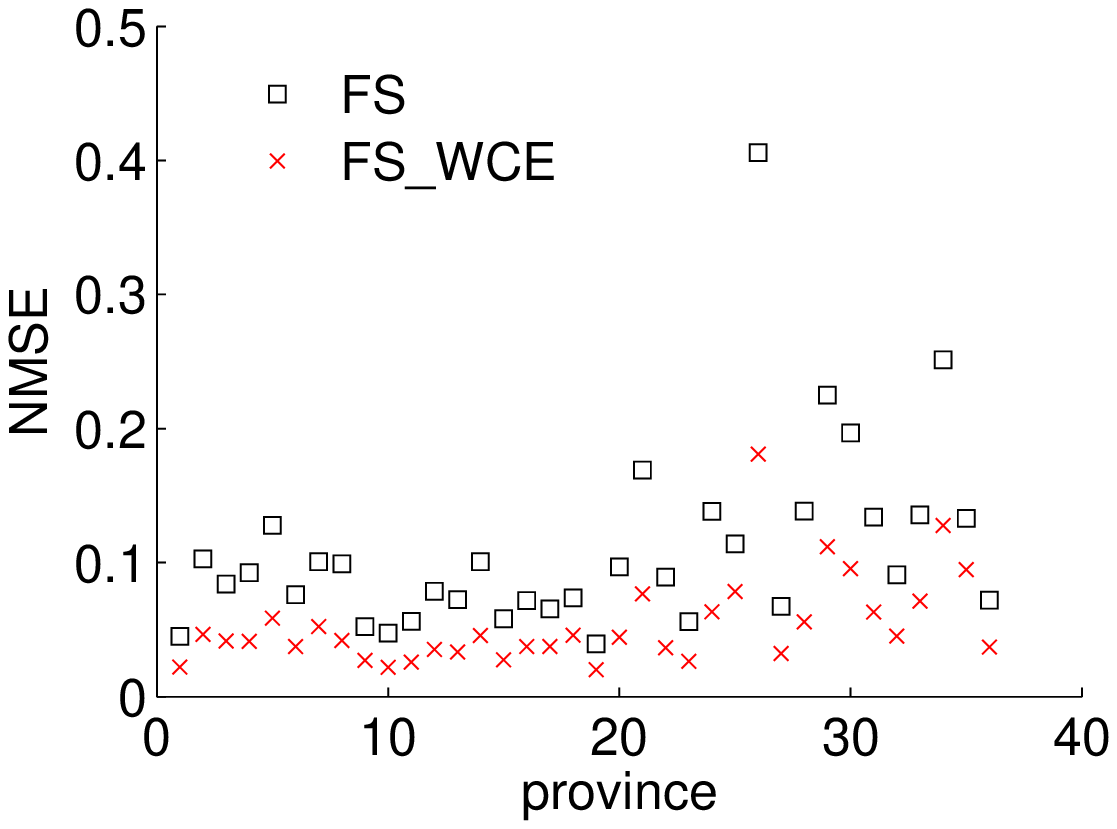}}
\caption{(Xiami) NMSE of location distribution estimates.}\label{fig:xiamigraphlocation}
\end{figure}

\begin{figure}[htb]
\center
\subfigure[\# following]{
\includegraphics[width=0.23\textwidth]{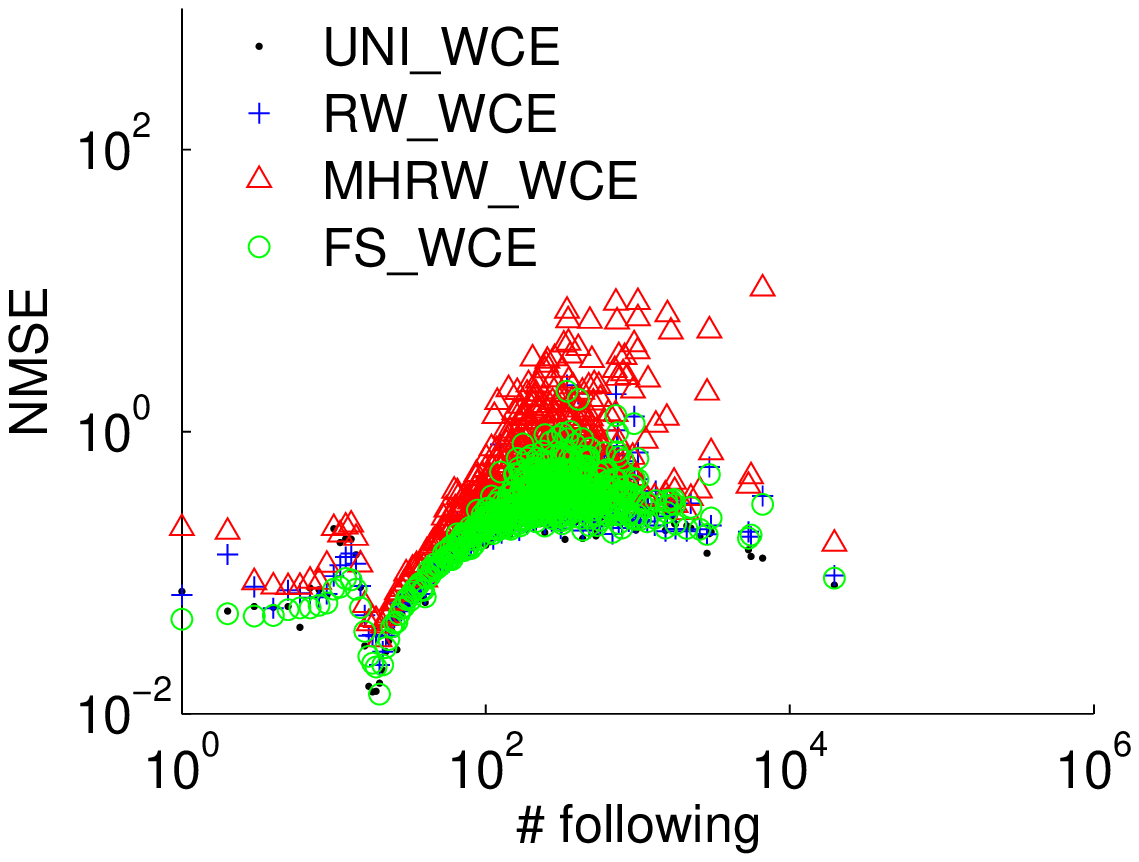}}
\subfigure[\# follower]{
\includegraphics[width=0.23\textwidth]{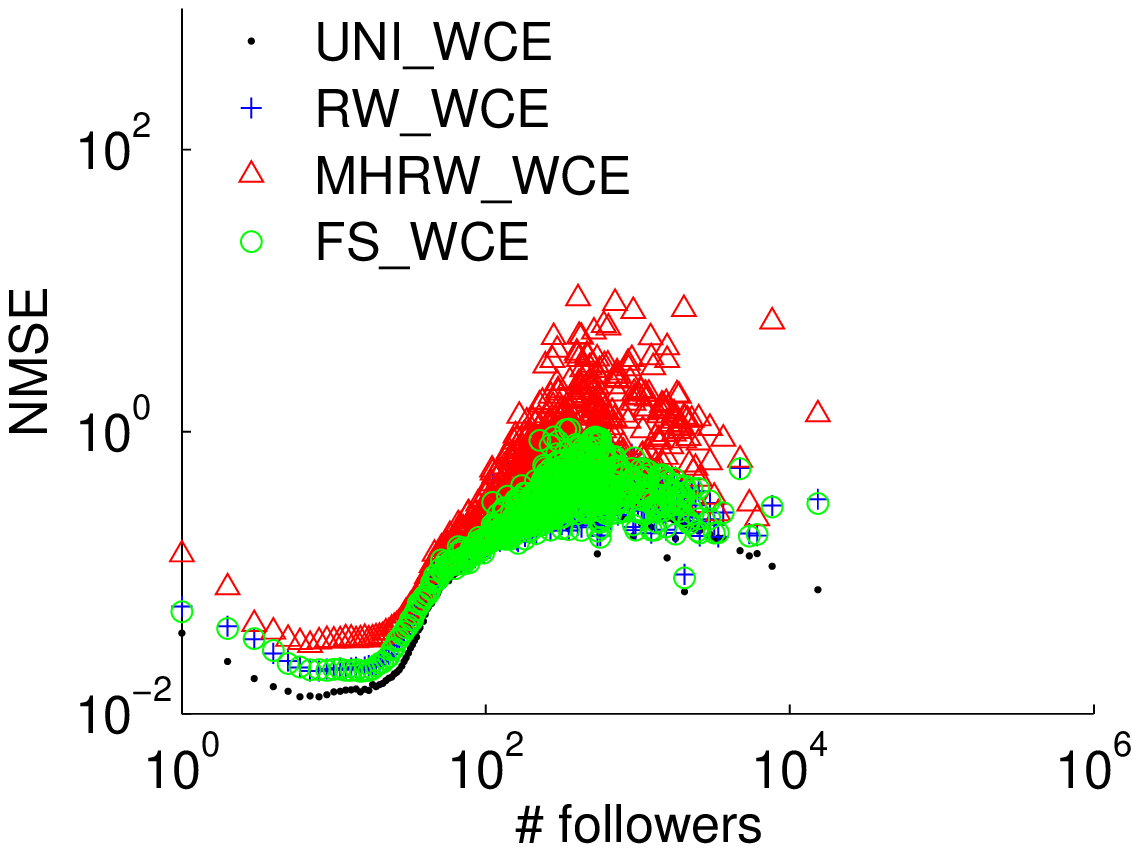}}
\subfigure[\# recommendation]{
\includegraphics[width=0.23\textwidth]{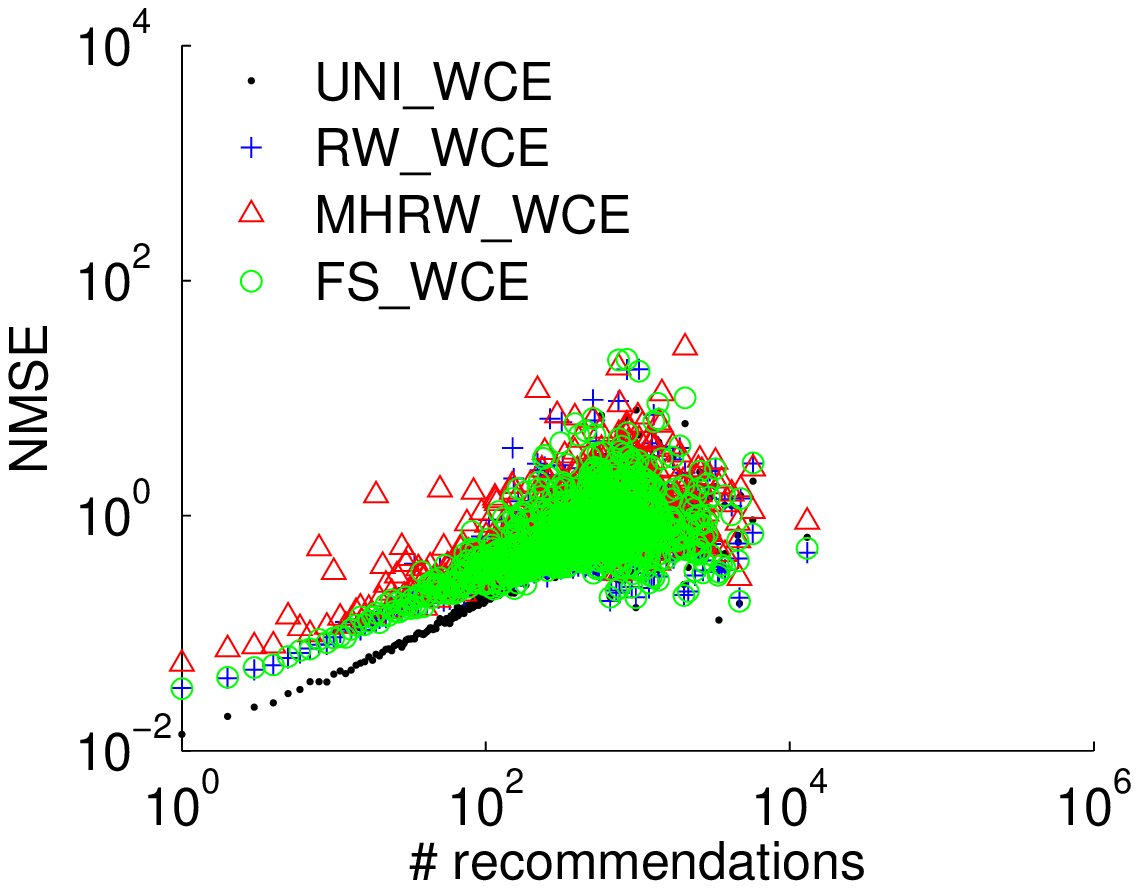}}
\subfigure[Location]{
\includegraphics[width=0.23\textwidth]{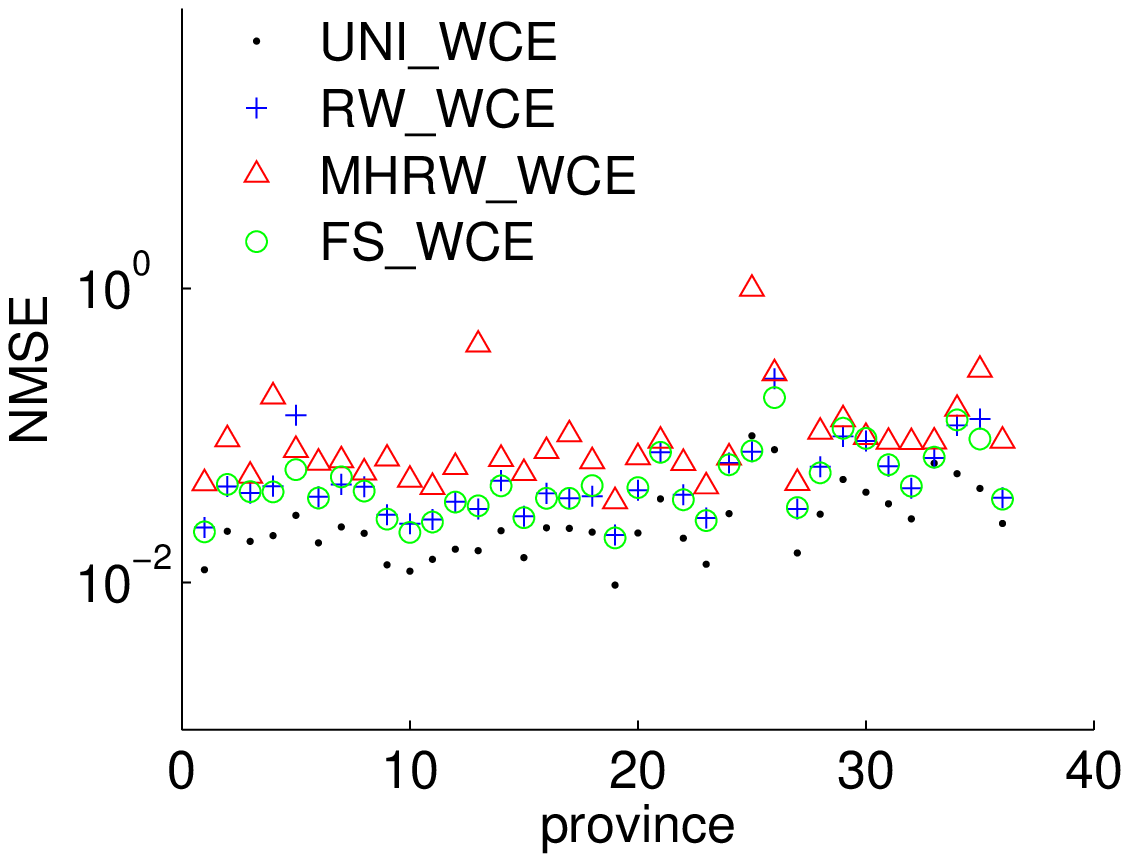}}
\caption{(LCC of Xiami) Compared NMSE of graph label density estimates for WCE using different graph sampling methods.}\label{fig:xiamigraphcmp}
\end{figure}

\begin{figure}[htb]
\center
\subfigure[UNI\_WCE vs. UNI]{
\includegraphics[width=0.23\textwidth]{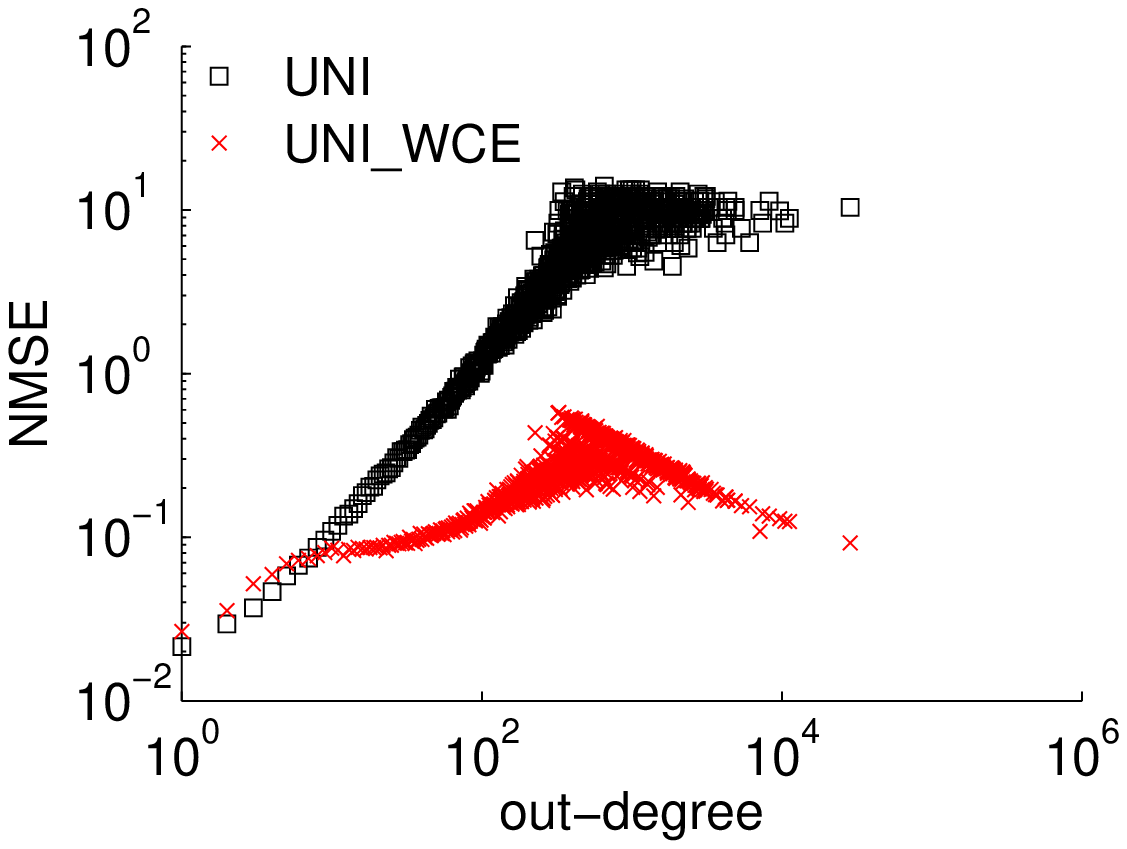}}
\subfigure[RW\_WCE vs. RW]{
\includegraphics[width=0.23\textwidth]{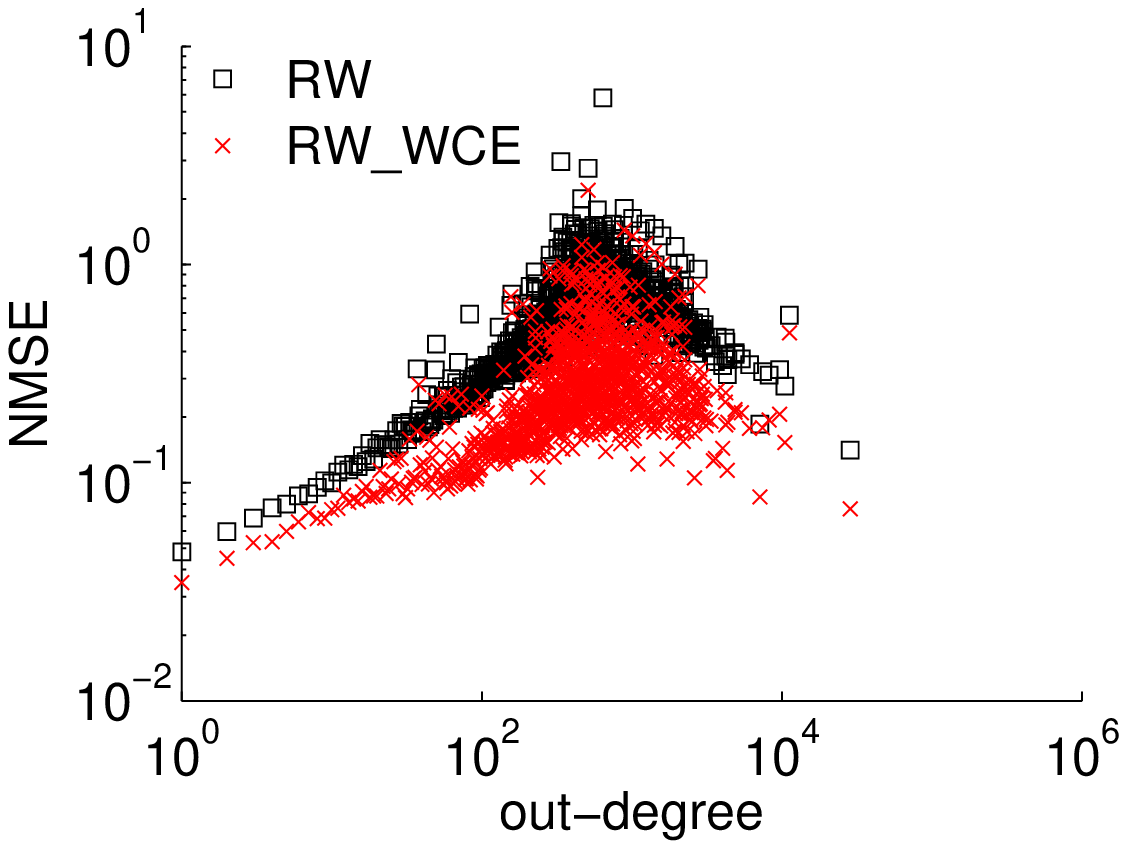}}
\subfigure[MHRW\_WCE vs. MHRW]{
\includegraphics[width=0.23\textwidth]{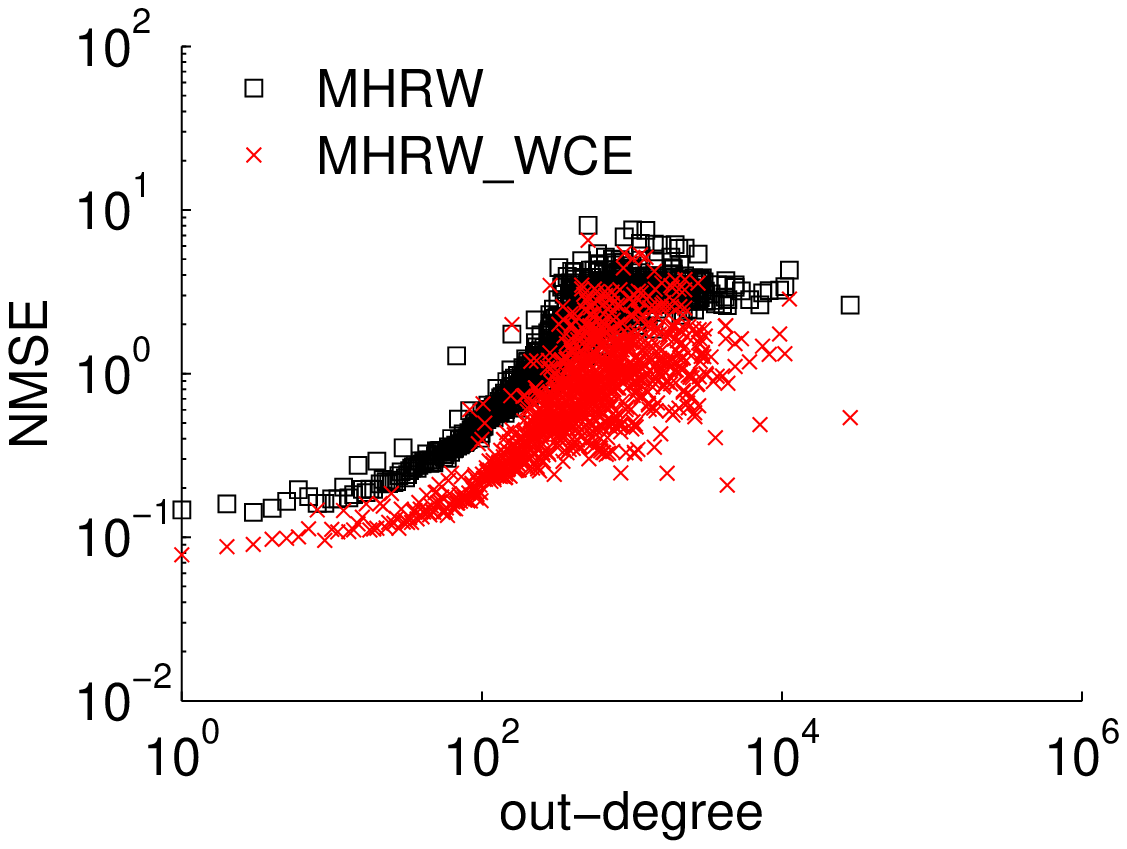}}
\subfigure[FS\_WCE vs. FS]{
\includegraphics[width=0.23\textwidth]{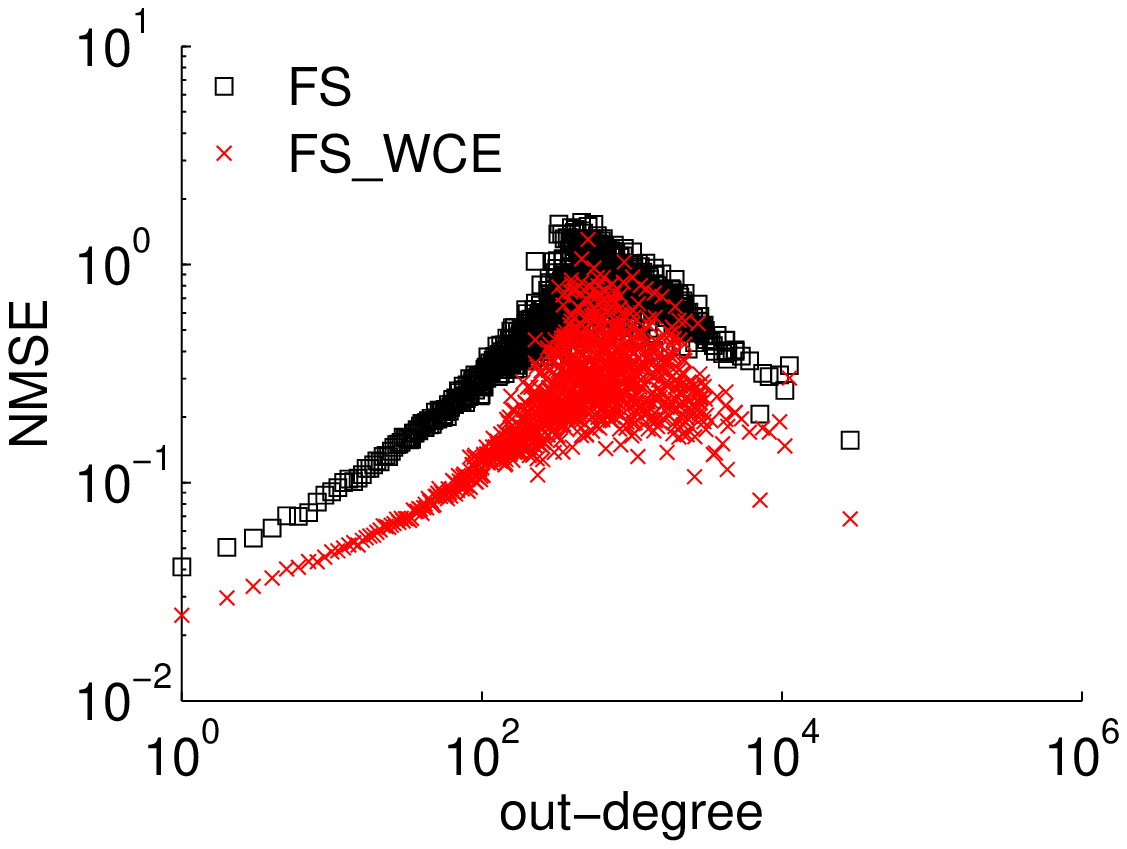}}
\caption{(YouTube) NMSE of out-degree distribution estimates.}\label{fig:youtubegraphoutdegree}
\end{figure}

\begin{figure}[htb]
\center
\subfigure[UNI\_WCE vs. UNI]{
\includegraphics[width=0.23\textwidth]{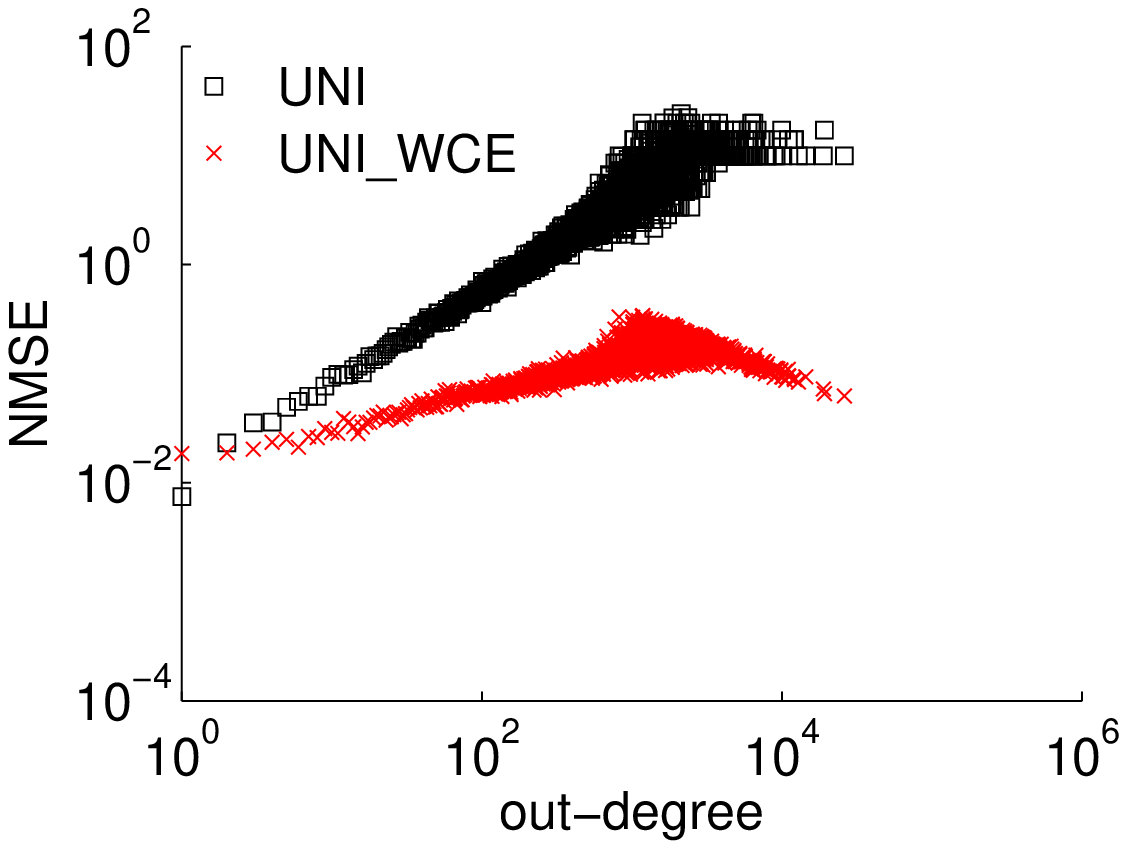}}
\subfigure[RW\_WCE vs. RW]{
\includegraphics[width=0.23\textwidth]{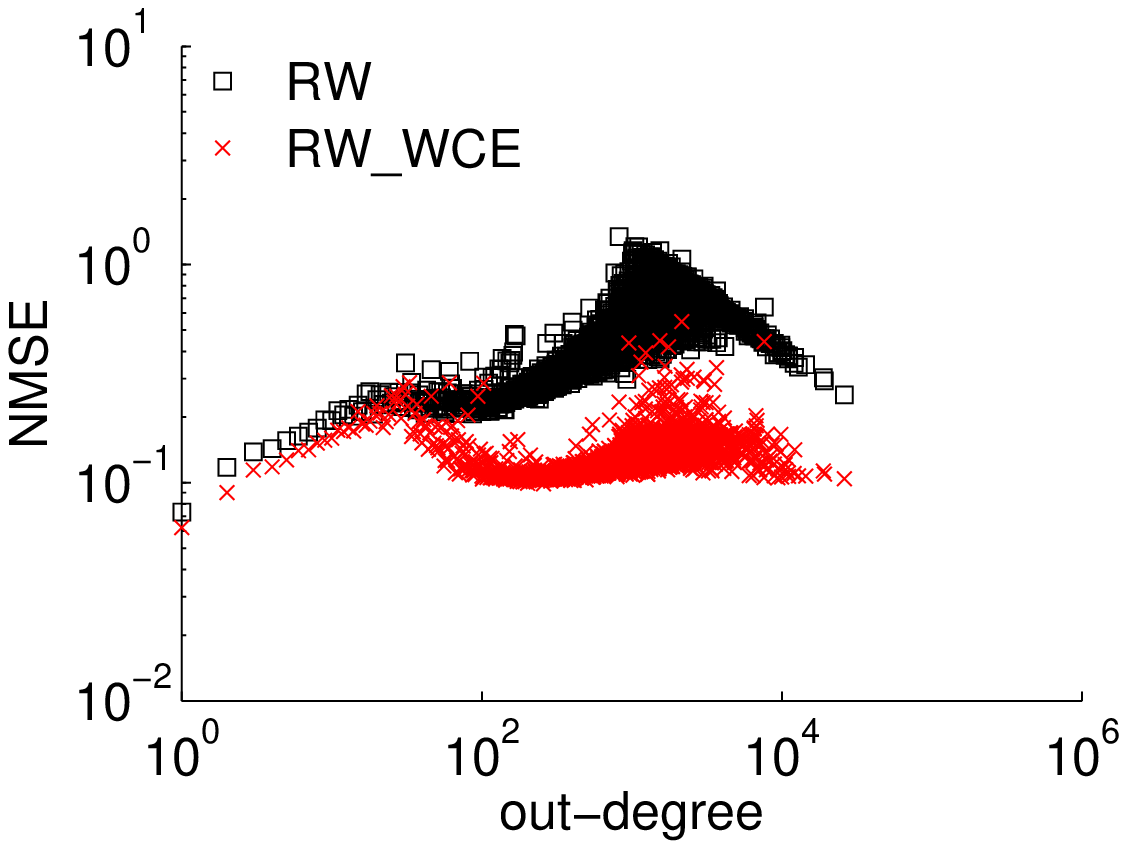}}
\subfigure[MHRW\_WCE vs. MHRW]{
\includegraphics[width=0.23\textwidth]{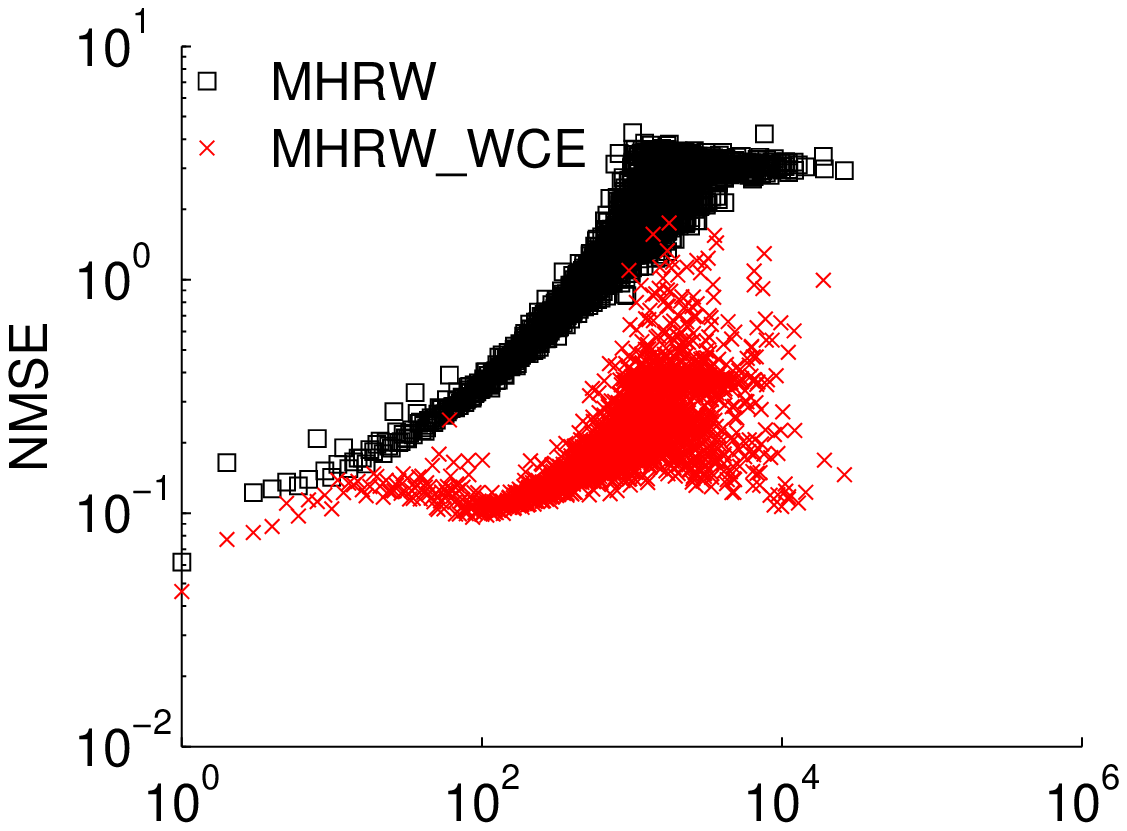}}
\subfigure[FS\_WCE vs. FS]{
\includegraphics[width=0.23\textwidth]{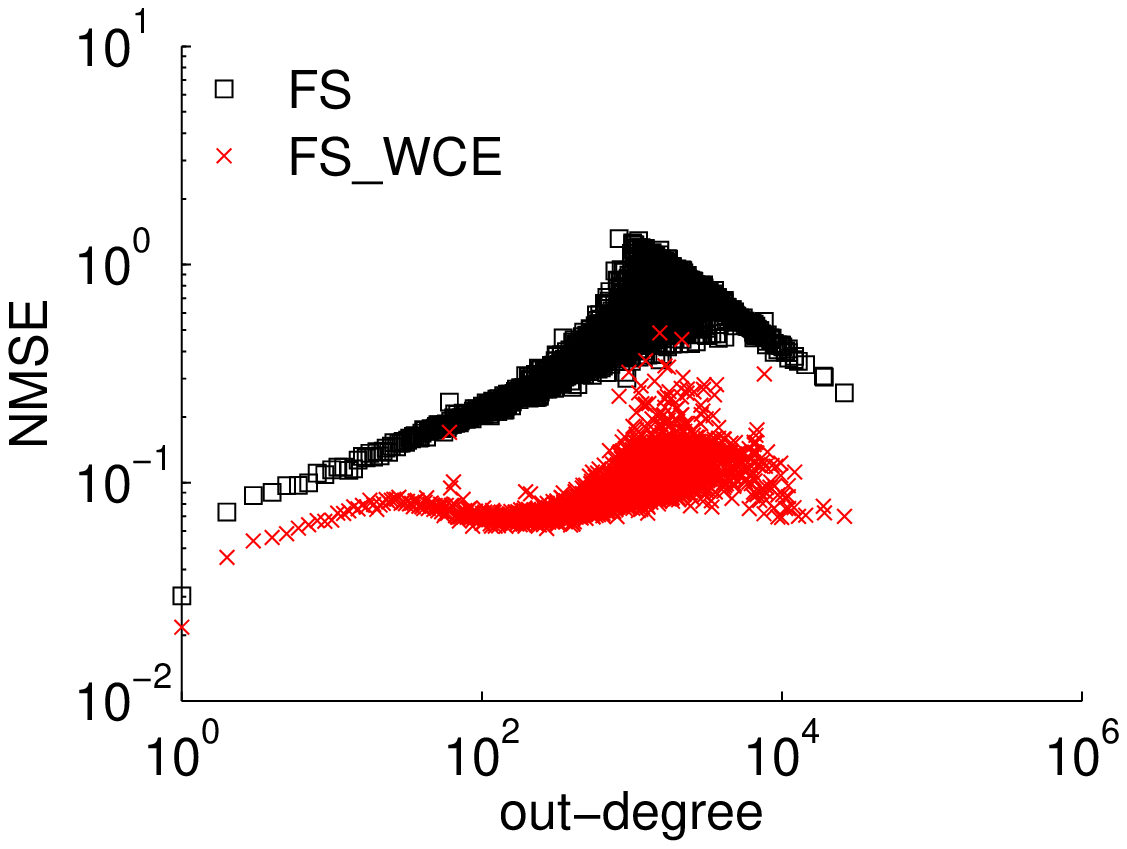}}
\caption{(Flickr) NMSE of out-degree distribution estimates.}\label{fig:flickrgraphoutdegree}
\end{figure}

Fig.~\ref{fig:xiamigraphCCDF} shows the distributions of users in Xiami
using different labels,
where the province numbers and corresponding names are shown in Table~\ref{tab:provinceID}. The
fraction of users with more than $10^4$ followers, following, or recommendations is smaller than $2\times10^{-6}$. The top three popular provinces are Guangdong, Beijing, and Shanghai.
Similar results are also observed for the LCC of Xiami.
Figs.~\ref{fig:xiamigraphoutdegree} to~\ref{fig:xiamigraphlocation} show the results of our new method to estimate these vertex label densities.
The results show that WCE
significantly outperforms the previous methods over almost all points.
This is because WCE uses neighbors' graph property summaries of sampled vertices.
Especially for UNI\_WCE, which is an order of magnitude more accurate than
UNI for follower/following counts larger than 100.
Fig.~\ref{fig:xiamigraphcmp} show the compared results for different graph sampling methods
where the graph used is the LCC of Xiami. The results show that MHRW is quite accurate. RW and FS almost have the same accuracy. For the follower and recommendation count distributions, UNI is more accurate for  follower and recommendation counts with small values. Moreover Figs.~\ref{fig:youtubegraphoutdegree} and~\ref{fig:flickrgraphoutdegree} show the results for estimating out-degree distribution for YouTube and Flickr respectively. We observe that WCE is better than previous methods over almost all points.

\begin{table}[htb]
\begin{center}
\caption{(Xiami) Province numbers and corresponding names.\label{tab:provinceID}}
\begin{tabular}{||lll||}
\hline
1. Beijing&2. Tianjin&3. Hebei\\
\hline
4. Shanxi&5. Inner Mongolia&6. Liaoning\\
\hline
7. Jilin&8. Heilongjiang&9. Shanghai\\
\hline
10. Jiangsu&11. Zhejiang &12. Anhui\\
\hline
13. Fujian&14.  Jiangxi&15. Shandong\\
\hline
16. Henan&17. Hubei&18. Hunan\\
\hline
19. Guangdong&20. Guangxi&21. Hainan\\
\hline
22.  Chongqing&23. Sichuan&24. Guizhou\\
\hline
25. Yunnan&26. Tibet &27. Shannxi\\
\hline
28. Gansu&29. Qinghai&30.  Ningxia\\
\hline
31. Xinjiang&32. Taiwan&33. Hong Kong\\
\hline
34. Macao&35. Null&36. Overseas\\
\hline
\end{tabular}
\end{center}
\end{table}

\section{Applications} \label{sec:application}

\begin{figure}[htb]
\center
\subfigure[\# retweets]{
\includegraphics[width=0.34\textwidth]{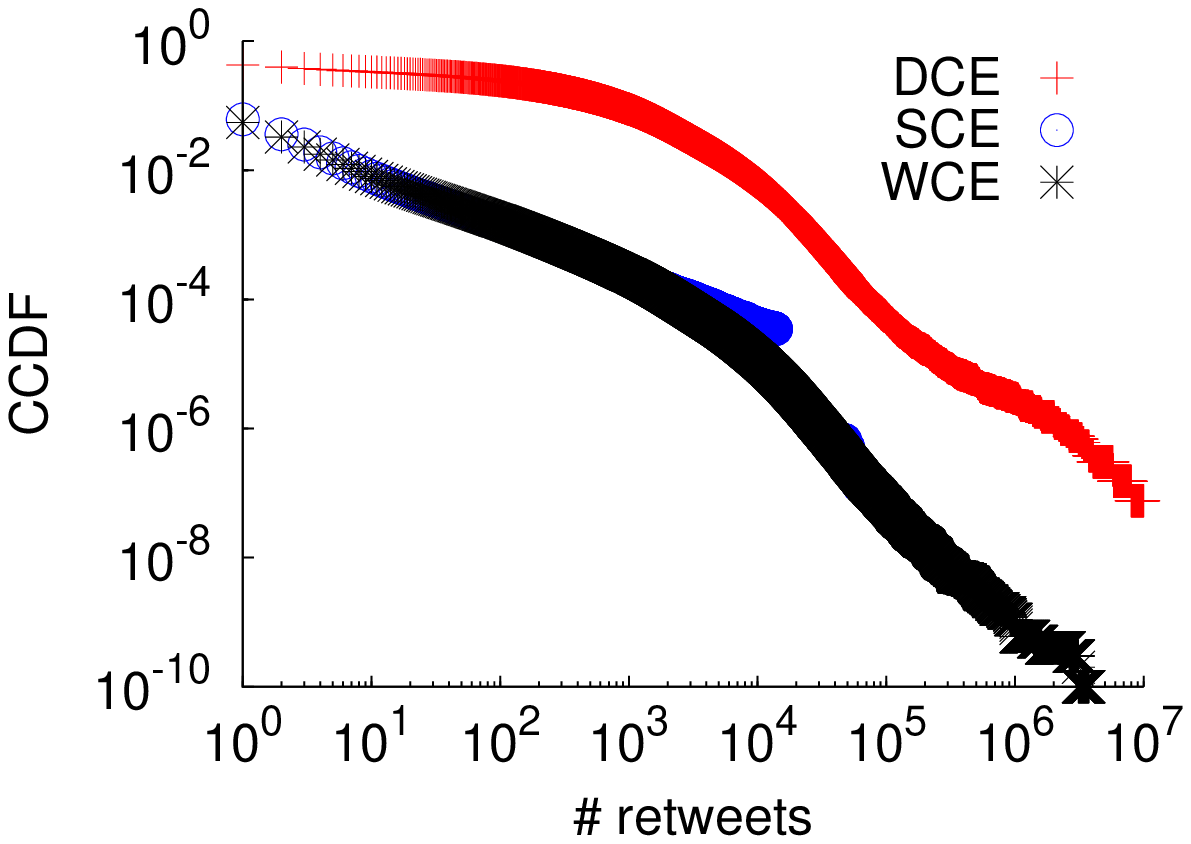}}
\subfigure[\# replies]{
\includegraphics[width=0.34\textwidth]{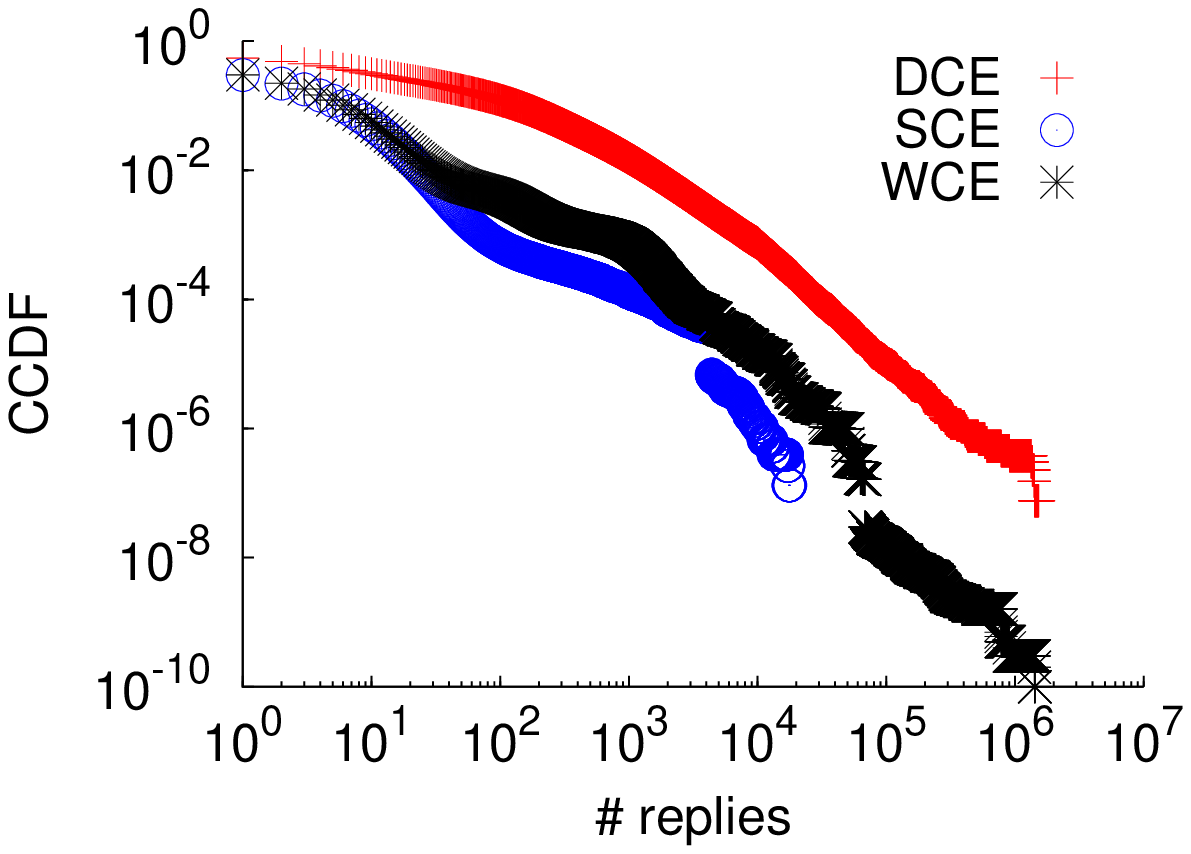}}
\caption{Distributions of tweets by the number of rewteets and replies.}\label{fig:diff_estimators}
\end{figure}

We now apply our methodology to a real OSN to characterize
various content, i.e., average number of retweets or replies per tweet,
types of tweet messages, as well as the associated top rank statistics.
We perform experiments on Sina microblog network. By crawling webpages of
148,313 random accounts selected by UNI, we obtain
19.7 million tweets and retweets. Note that in the following analysis,
tweets refer to the
original tweets. Each tweet or retweet records its original tweet's
information such as the number of retweets and replies.
Fig.~\ref{fig:diff_estimators} shows the results of estimating the
distribution of tweets from retweets and replies using DCE, SCE
and WCE, where the special content is defined as the original tweet for
SCE. The estimates for the average number of retweets and replies per tweet are
shown as Table~\ref{tab:avg_diff_est}.
We observe that the estimates of SCE and WCE are close to each other,
but the estimates obtained by DCE significantly deviate
from SCE and WCE. This is consistent with previous simulation results
which show that DCE introduces large biases.
Furthermore, Fig.~\ref{fig:diff_estimators} shows that the maximum number
of retweets or replies given by SCE is the smallest since it only
uses information of sampled original tweets, and the original tweets
of popular tweets are not always sampled.

\begin{table}[htp]
	\centering
	\caption{Estimates of the average number of rewteets and replies per tweet by different methods}
	\label{tab:avg_diff_est}
	\begin{tabular}{||l|c|c||}
		\hline
		&{\bf Avg. \# retweets}&{\bf Avg. \# replies}\\
		\hline \hline
    {\bf DCE}       & 423                 & 89.8 \\
    {\bf SCE}       & 2.01                   & 3.93  \\
    {\bf WCE}       & 1.60                   & 4.60  \\
		\hline
	\end{tabular}
\vspace{-0.5em}
\end{table}

Let us explore the ``{\em type}'' of tweets.
We classify tweets into three types:\textit{text tweet},
\textit{image tweet}, and \textit{video tweet}.
Table~\ref{tab:avg_diff_tweet} shows their statistics measured by WCE.
We find that 60.1\% are text tweets, 37.6\% are image tweets,
and 2.3\% are video tweets. On average, image and video tweets have
more retweets and replies than text tweets.
Table~\ref{tab:video_sites} shows the statistics of video tweets by
their associated external video source websites.
We find that the top five popular video websites are
youku.com (42.7\%), tudou.com (26.3\%), sina.com (10.0\%), yinyuetai.com (6.2\%)
and 56.com (4.4\%).

\begin{table}[htp]
	\centering
	\caption{(Sina microblog) Statistics of tweets by different categories}
	\label{tab:avg_diff_tweet}
	\begin{tabular}{||l|c|c|c||}
		\hline
	&{\bf Fraction of}&{\bf Avg. \# retweets}&{\bf Avg. \# replies}\\
        &{\bf tweets}     &{\bf per tweet}       &{\bf per tweet}\\
		\hline \hline
        {\bf Text }  &60.1\% & 0.31           & 2.30 \\
        {\bf Image}  &37.6\% & 3.33           & 7.91 \\
        {\bf Video}  &2.3\% & 7.05           & 10.91 \\
		\hline
	\end{tabular}
\vspace{-0.5em}
\end{table}
\begin{table}[htb]
	\centering
	\caption{(Sina microblog) Statistics of video tweets}
	\label{tab:video_sites}
	\begin{tabular}{||l|c|c|c||}
		\hline
	\multirow{3}{*}{\bf Source}&{\bf Fraction of}&{\bf Avg.}&{\bf Avg.}\\
        &{\bf video}&{\bf \# retweets}&{\bf \# replies}\\
        &{\bf tweets}&{\bf per tweet}&{\bf per tweet}\\
		\hline \hline
        {\bf youku.com}                & 42.7\% & 5.57 & 9.68 \\
        {\bf tudou.com}                & 26.3\% & 9.87 & 13.73 \\
        {\bf sina.com }                & 10.0\% & 9.09 & 11.32 \\
        {\bf yinyuetai.com}            & 6.2\%  & 5.06 & 9.36 \\
        {\bf 56.com}                   & 4.4\%  & 11.29 & 26.64 \\
        {\bf ku6.com}                  & 2.9\%  & 11.29 & 13.02 \\
        {\bf sohu.com}                 & 1.7\%  & 4.56 & 1.29 \\
        {\bf kandian.com}              & 1.6\%  & 0.13 & 0.16 \\
        {\bf baomihua.com}             & 1.6\%  & 0.31 & 0.08 \\
        {\bf ifeng.com}                & 0.9\%  & 5.77 & 4.36 \\
		\hline
	\end{tabular}
\vspace{-0.5em}
\end{table}

\section{Related Work} \label{sec:related}

Previous graph sampling work focuses on designing accurate and
efficient sampling methods for measuring graph characteristics,
such as vertex degree distribution~\cite{Stutzbach2009,Rasti2009,
Gjoka2010,Ribeiro2010,Ribeiro2012} and the topology of vertices'
groups~\cite{KurantarXiv2011}.
We summarize previous graph sampling work as follows:
Most previous OSN graph crawling and sampling work
focuses on undirected graph since each vertex in most OSNs
maintains both its incoming and outgoing neighbors, so
it is easy to convert these directed OSNs to their associated
undirected graphs by ignoring the directions of edges.
Breadth-First-Search (BFS), though it is easy to implement but it
introduces a large bias towards high degree vertices, and it is
difficult to remove these biases
in general~\cite{Achlioptas2005, Kurant2010, KurantJSAC2011}.
Random walk (RW) is biased to sample high degree vertices,
however its bias is known and can be
corrected~\cite{Heckathorn2002,Salganik2004}.
Compared with uniform vertex sampling (UNI),
a RW has smaller estimation errors for
high degree vertices, and these vertices
are quite common for many OSNs like Facebook, Myspace and Flickr~\cite{Ribeiro2010}. Furthermore, it is costly to apply
UNI in these networks.
The Metropolis-Hasting RW (MHRW)~\cite{Zhong2006,Stutzbach2009,Gjoka2010}
modifies the RW procedure,
and it aims to sample each vertex with the same probability.
The accuracy of RW and MHRW is compared
in~\cite{Rasti2009,Gjoka2010}.
RW is shown to be consistently more accurate than MHRW.
The mixing time of a RW determines the efficiency of the sampling,
and it is found to be much larger than commonly believed for
many OSNs~\cite{Mohaisen2010}. There are a lot of work on
how to decrease the mixing
time~\cite{Boyd2004, Ribeiro2010,AvrachenkovWAW2010,GjokaJSAC2011,Kurant2011}.
To sample a directed graph with latent incoming links
(e.g. the Web graph and Flickr~\cite{Flickr}), ~\cite{Yossef2008}
and~\cite{Henzinger2000} apply a MHRW over an undirected graph which is
built on-the-fly by adding observed links from the directed graph.
However, these algorithms are biased since the generated undirected graph
may not contain all vertices in the original directed graph.
To address this issue,
Ribeiro et al.~\cite{Ribeiro2012} use a RW with jumps
under the assumption that vertices can be uniformly sampled at
random from directed graphs.
To the best of our knowledge, our work is the first to study the
problem of measuring characteristics of {\em content} distributed over large graphs based on graph sampling techniques.

\section{Conclusions} \label{sec:conclusions}

In this paper, we study the problem of estimating
characteristics of content distributed over large graphs.
The analysis and experiment results show that existing graph sampling methods
are biased to sample content with a large number of copies, and there can be
huge bias in statistics computed by directly using collected content.
To remove this bias, the MLE method is applied. However,
we show that MLE needs to sample most vertices in the graph to
obtain accurate statistics.
To address this challenge, we propose two efficient methods
SCE and WCE using available information in sampled content.
We show that they are asymptotically unbiased. We perform
extensive measurement and experiments, and show that WCE is more accurate
than SCE. Furthermore, we use WCE to estimate graph characteristics
when vertices maintain their neighbors' graph properties.
We carry out experiments to
show that WCE is more accurate than previous sampling methods.

\section*{Acknowledgments}
This work was supported by the NSF grant CNS-1065133 and ARL Cooperative Agreement W911NF-09-2-0053. The views and conclusions contained in this document are those of the authors and should not be interpreted as representing the official policies, either expressed or implied of the NSF, ARL, or the U.S. Government. This work was also supported in part by the NSFC funding 60921003 and 863 Program 2012AA011003 of China.

\ifCLASSOPTIONcaptionsoff
  \newpage
\fi

\bibliographystyle{IEEEtran}
\bibliography{IEEEabrv,samplingcontent}

\begin{IEEEbiography}[{\includegraphics[width=1in,height=1.25in,clip,keepaspectratio]{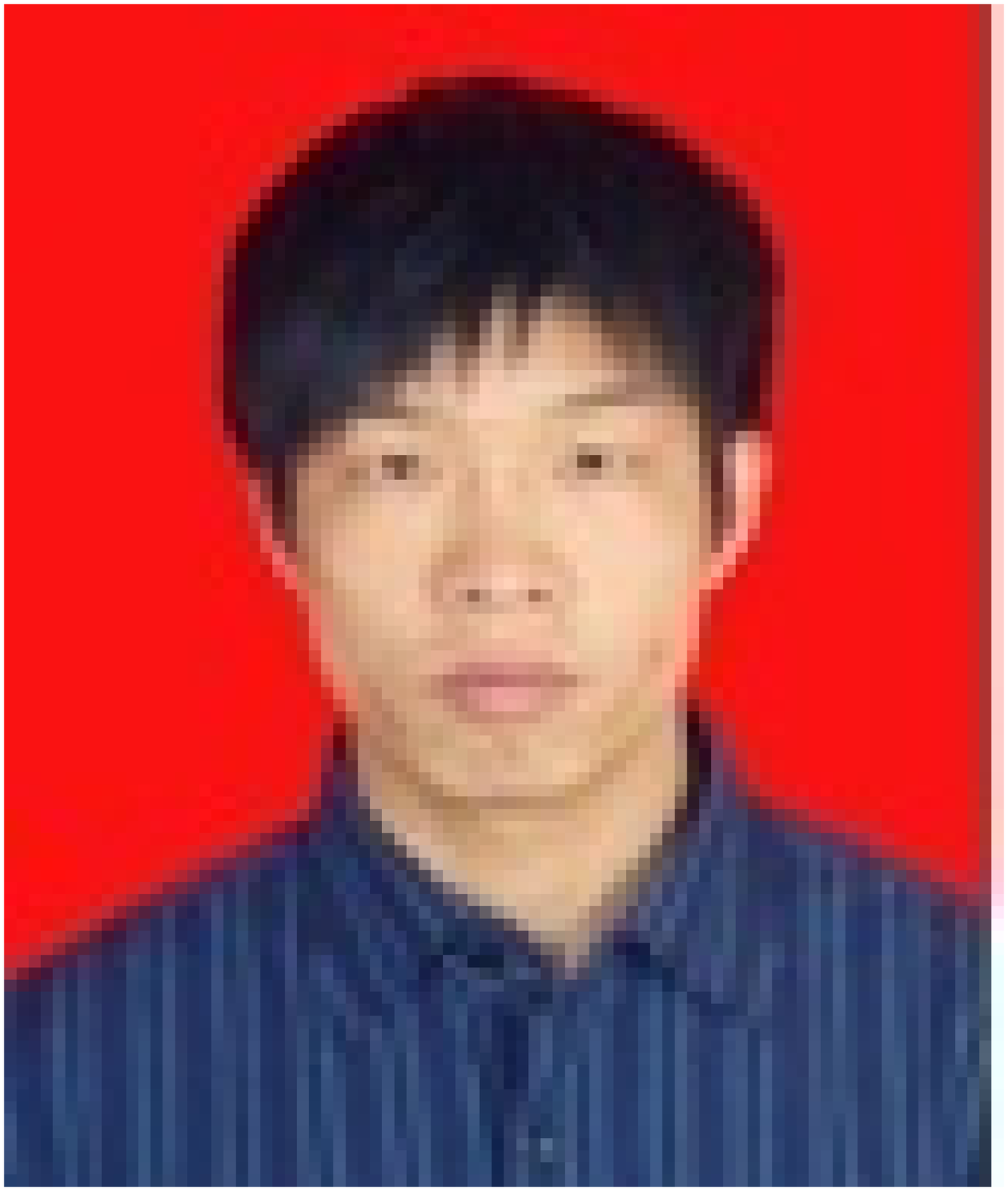}}]{Pinghui Wang}
received the B.S. degree in information engineering and Ph.D degree in  automatic control from Xi'an Jiaotong University, Xi'an, China, in 2006, 2012 respectively. From April 2012 to October 2012, he was a postdoctoral researcher with the Department of Computer Science and Engineering at The Chinese University of Hong Kong. He is currently a postdoctoral researcher with School of Computer Science at McGill University, QC, Canada. His research interests include Internet traffic measurement and modeling, traffic classification, abnormal detection, and online social network measurement.
\end{IEEEbiography}

\begin{IEEEbiography}[{\includegraphics[width=1in,height=1.25in,clip,keepaspectratio]{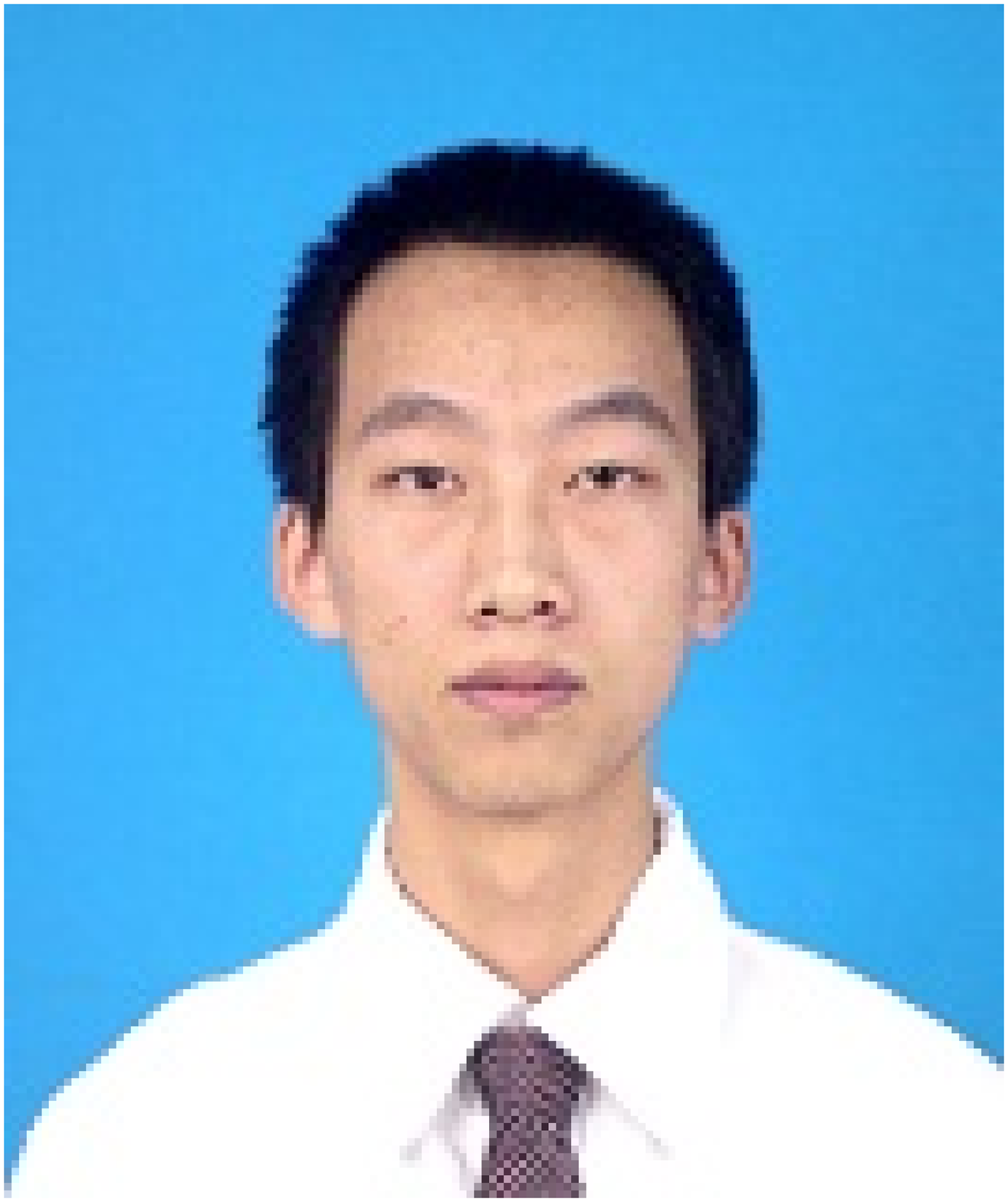}}]{Junzhou Zhao}
received the B.S. degree in automatic control from Xi'an Jiaotong University, Xi'an, China, in 2008. He is currently a Ph.D. candidate with the Systems Engineering Institute and MOE Key Lab for Intelligent Networks and Network
Security, Xi'an Jiaotong University under the supervision of Prof. Xiaohong Guan. His research interests include online social network measurement and modeling.
\end{IEEEbiography}

\begin{IEEEbiography}[{\includegraphics[width=1in,height=1.25in,clip,keepaspectratio]{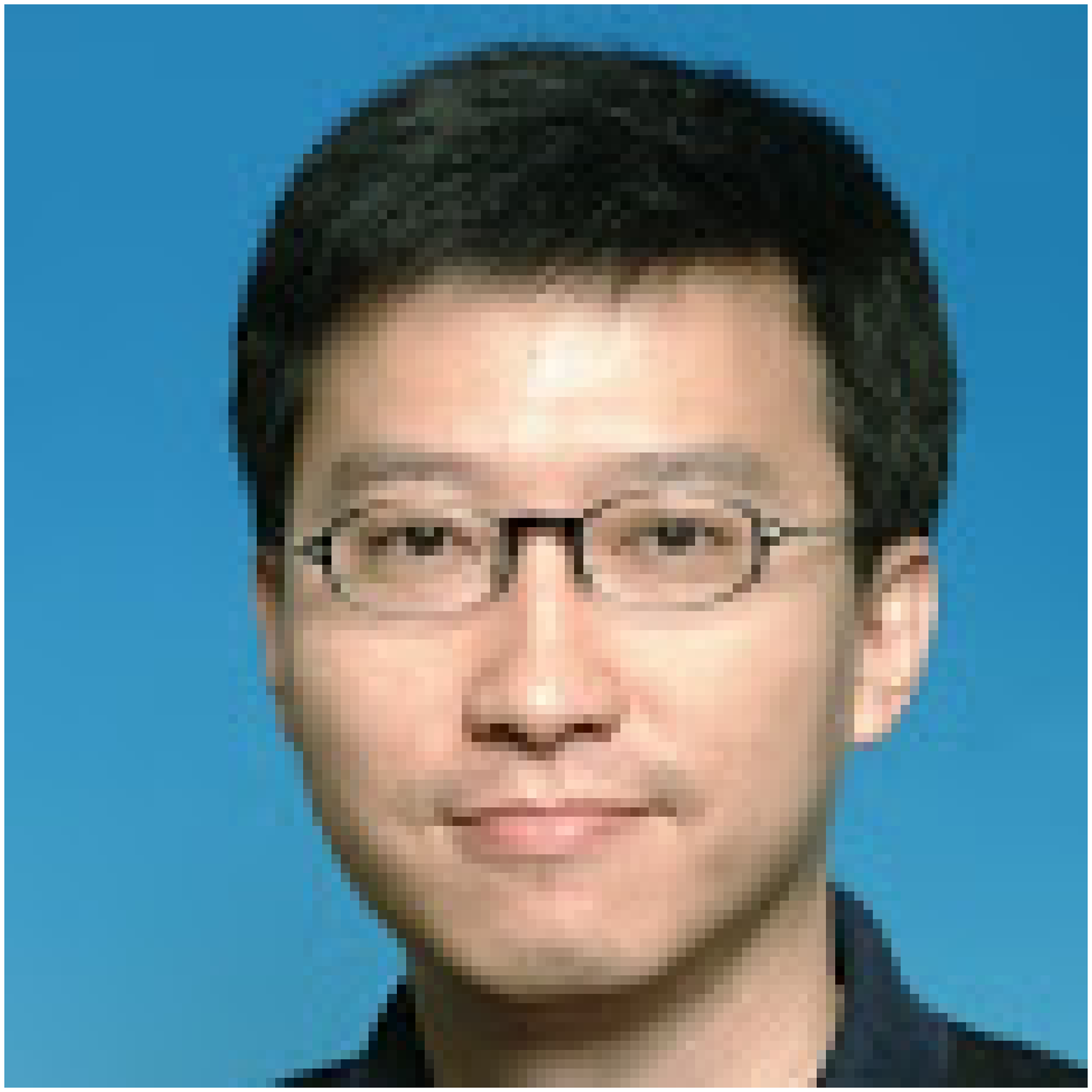}}]{John C.S. Lui}
received the PhD degree in computer science from UCLA. He is currently a
professor in the Department of Computer
Science and Engineering at The Chinese University of Hong Kong. His current research interests
include communication networks, network\/system security (e.g., cloud security, mobile security, etc.), network economics, network sciences
(e.g., online social networks, information spreading, etc.), cloud computing, large-scale distributed systems and performance evaluation theory. He serves in the editorial board of IEEE\/ACM Transactions on Networking, IEEE Transactions on Computers, IEEE Transactions on Parallel and
Distributed Systems, Journal of Performance Evaluation and International Journal of Network Security. He was the chairman of the CSE
Department from 2005-2011. He received various departmental teaching awards and the CUHK Vice-Chancellor¡¯s Exemplary Teaching
Award. He is also a corecipient of the IFIP WG 7.3 Performance 2005 and IEEE\/IFIP NOMS 2006 Best Student Paper Awards. He is an
elected member of the IFIP WG 7.3, fellow of the ACM, fellow of the IEEE, and croucher senior research fellow. His personal interests
include films and general reading.
\end{IEEEbiography}

\begin{IEEEbiography}[{\includegraphics[width=1in,height=1.25in,clip,keepaspectratio]{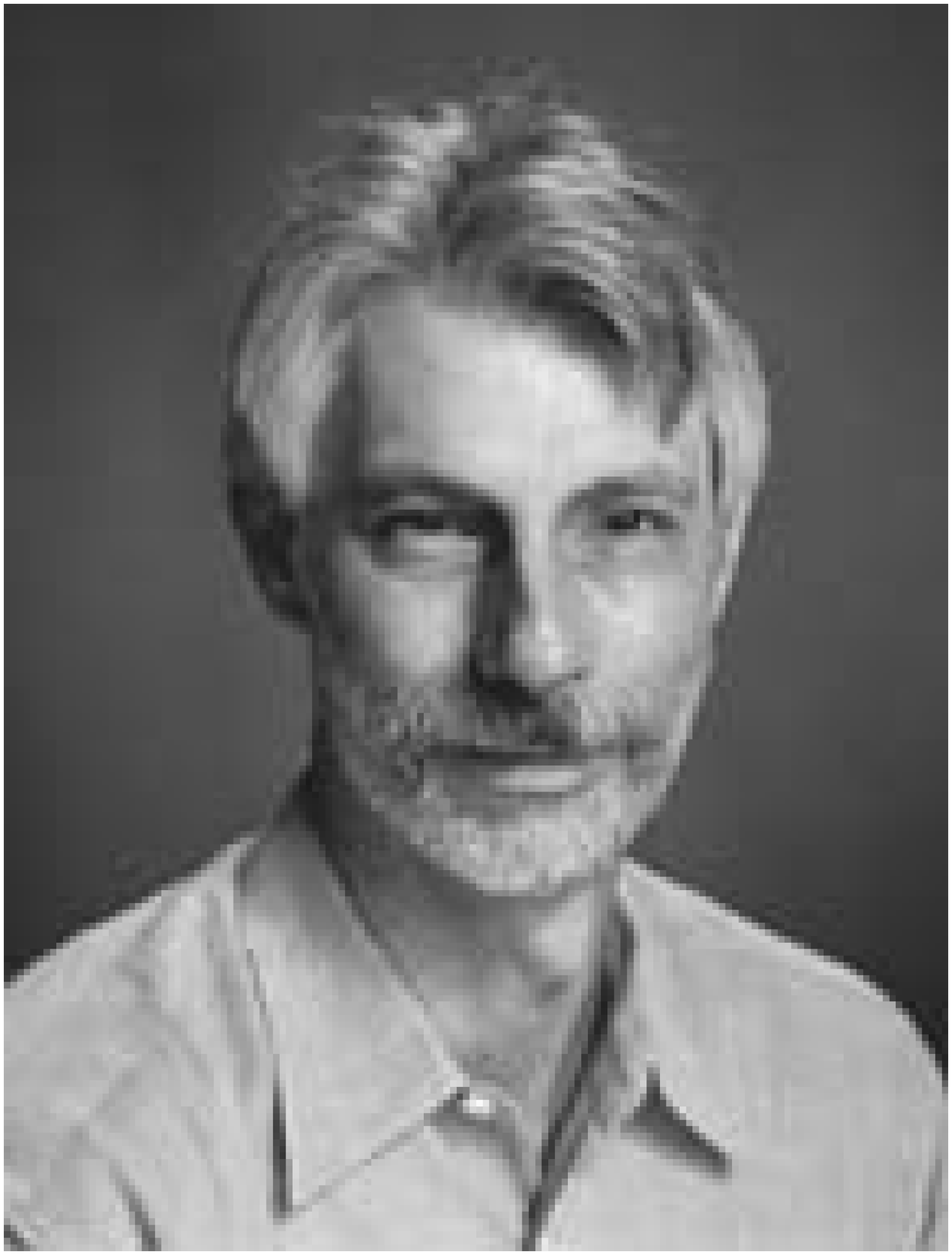}}]{Don Towsley}
holds a B.A. in Physics (1971) and a Ph.D. in Computer Science (1975) from University of Texas. From 1976 to 1985 he was a member of the faculty of the Department of Electrical and Computer Engineering at the University of Massachusetts, Amherst. He is currently a Distinguished Professor at the University of Massachusetts in the Department of Computer Science. He has held visiting positions at IBM T.J. Watson Research Center, Yorktown Heights, NY; Laboratoire MASI, Paris, France; INRIA, Sophia-Antipolis, France; AT\&T Labs-Research, Florham Park, NJ; and Microsoft Research Lab, Cambridge, UK. His research interests include networks and performance evaluation. He currently serves as Editor-in-Chief of IEEE/ACM Transactions on Networking and on the editorial boards of Journal of the ACM, and IEEE Journal on Selected Areas in Communications, and has previously served on numerous other editorial boards. He was Program Co-chair of the joint ACM SIGMETRICS and PERFORMANCE 92 conference and the Performance 2002 conference. He is a member of ACM and ORSA, and Chair of IFIP Working Group 7.3. He has received the 2007 IEEE Koji Kobayashi Award, the 2007 ACM SIGMETRICS Achievement Award, the 1998 IEEE Communications Society William Bennett Best Paper Award, and numerous best conference/workshop paper awards. Last, he has been elected Fellow of both the ACM and IEEE.
\end{IEEEbiography}

\begin{IEEEbiography}[{\includegraphics[width=1in,height=1.25in,clip,keepaspectratio]{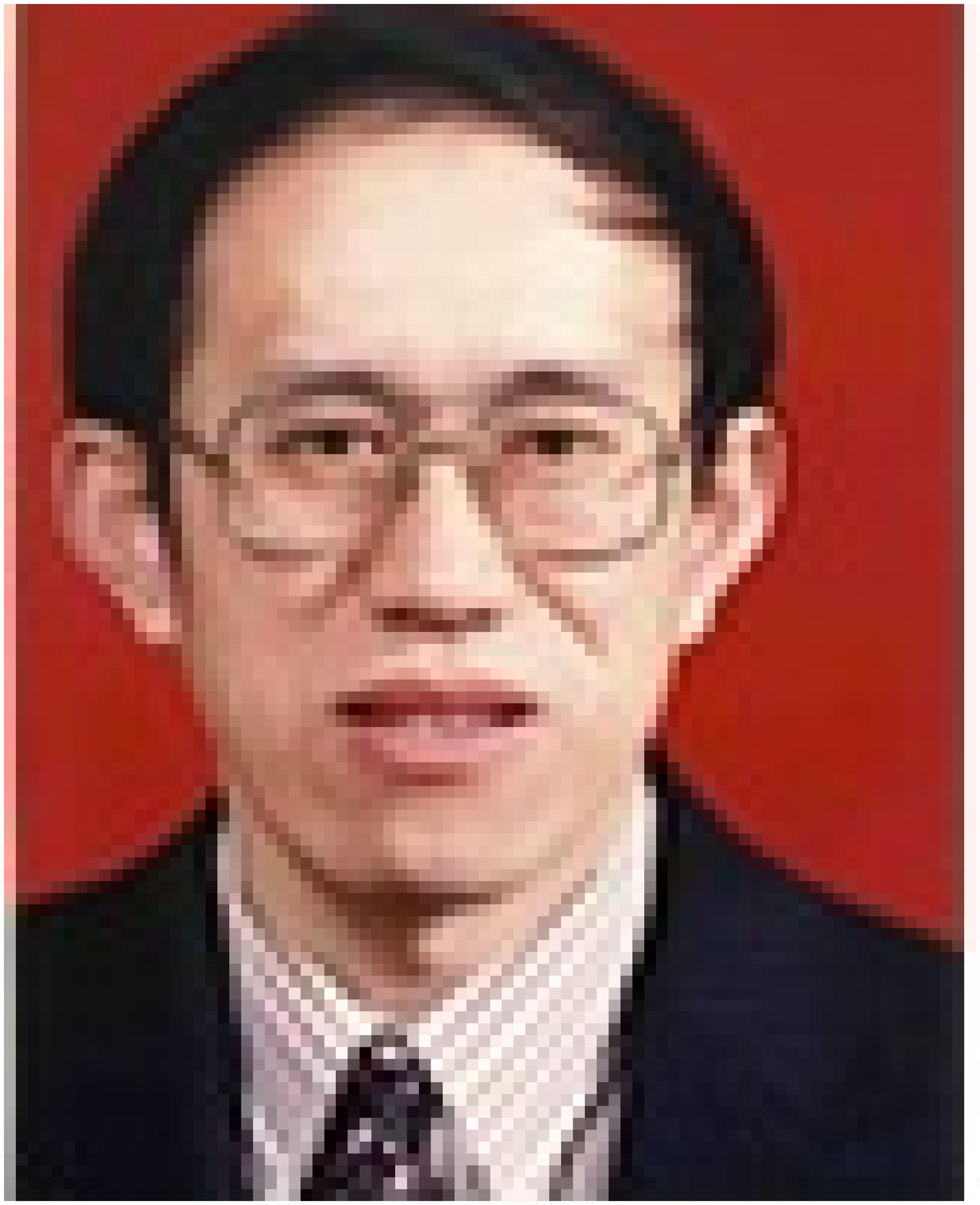}}]{Xiaohong Guan}
received the B.S. and M.S. degrees in automatic control from Tsinghua University, Beijing, China, in 1982 and 1985, respectively, and the Ph.D. degree in electrical engineering from the University of Connecticut, Storrs, US, in 1993.
From 1993 to 1995, he was a consulting engineer at PG\&E. From 1985 to 1988, he was with the Systems Engineering Institute, Xi'an Jiaotong University, Xi'an, China. From January 1999 to February 2000, he was with the Division of Engineering and Applied Science, Harvard University, Cambridge, MA. Since 1995, he has been with the Systems Engineering Institute, Xi'an Jiaotong University, and was appointed Cheung Kong Professor of Systems Engineering in 1999, and dean of the School of Electronic and Information Engineering in 2008. Since 2001 he has been the director of the Center for Intelligent and Networked Systems, Tsinghua University, and served as head of the Department of Automation, 2003-2008. He is an Editor of IEEE Transactions on Power Systems and an Associate Editor of Automata. His research interests include allocation and scheduling of complex networked resources, network security, and sensor networks. He has been elected Fellow of IEEE.
\end{IEEEbiography}

\end{document}